\def\pmb#1{\setbox0=\hbox{#1}%
\kern-.025em\copy0\kern-\wd0
\kern-.05em\copy0\kern-\wd0
\kern-.025em\raise.0433em\box0}

\def    \bOmega {\vec \Omega}
\def      \be       {\begin{equation}}
\def      \ee       {\end{equation}}

\documentclass{elsart}
\usepackage{graphicx}

\usepackage{amssymb}
\newcommand{\Omegabold}{\mbox{{\boldmath $\vec \Omega$}}}
\begin{document}



\begin {frontmatter}
\title{Tracing Magnetic Fields with Aligned Grains}
\author[Alex]{A. Lazarian},
\address[Alex]{University of Wisconsin-Madison,
Astronomy Department, 475 N. Charter St., Madison, WI 53706, e-mail:
lazarian@astro.wisc.edu}



\begin{abstract}
 Magnetic fields play a crucial role in various astrophysical processes, including star
 formation, accretion of matter, transport processes (e.g., transport of heat), and
 cosmic rays. One of the easiest ways to determine the  magnetic field
 direction is via polarization of radiation resulting from extinction on
 or/and emission by aligned dust grains. Reliability of interpretation of the
 polarization maps in terms of magnetic fields depends on how well we understand the
 grain-alignment theory. Explaining what makes grains aligned has been one of the
 big issues of the modern astronomy. Numerous exciting physical effects have been
 discovered in the course of research undertaken in this field. As both the theory and
 observations matured, it became clear that the grain-alignment phenomenon is inherent
 not only in diffuse interstellar medium or molecular clouds but also is a generic
 property of the dust in circumstellar regions, interplanetary space and cometary comae.
 Currently the grain-alignment theory is a predictive one, and its results nicely match
 observations. Among its predictions is a subtle phenomenon of radiative torques. This
 phenomenon, after having stayed in oblivion for many years after its discovery, is
 currently viewed as the most powerful means of alignment. In this article, I shall
 review the basic physical processes involved in grain alignment, and the currently
 known mechanisms of alignment. I shall also discuss possible niches for different
 alignment mechanisms. I shall dwell on the importance of the concept of grain helicity
 for understanding of many properties of grain alignment, and shall demonstrate that
 rather arbitrarily shaped grains exhibit helicity when they interact with gaseous and
 radiative flows.
\end{abstract}
\end{frontmatter}

\section{Introduction}

Magnetic fields are of an utmost importance for most astrophysical systems.
Conducting matter is entrained on magnetic field lines, and magnetic pressure
and tension are very important for its dynamics. For instance, galactic
magnetic fields play key role in many processes, including star formation,
mediating shocks, influencing heat and mass transport, modifying turbulence
etc. Aligned dust grains trace the magnetic field and provide a unique source
of information about magnetic field structure and topology. The new instruments, Sharc II
(Novak et al. 2004), Scuba II (Bastien, Jenness \& Molnar 2005), and an
intended polarimeter for SOFIA open new horizons for studies of astrophysical
magnetic fields via polarimetry.

The enigma that has surrounded
 grain alignment since its discovery in 1949 (Hall 1949; Hiltner 1949)
makes one wonder how reliable is polarimetry as a way of magnetic field
studies. In fact,
for many years grain alignment theory used to have a very limited predictive power
and was an issue of hot debates. Works by great minds like Lyman Spitzer
and Edward Purcell moved the field forward, but the solution looked illusive.
In fact, the reader can see from the review, that a number of key physical processes
have been discovered only recently.
Fig.~1a demonstrates the complexity of grain motion as we understand it now.

The weakness of the theory
 caused a somewhat cynical approach to it among some of the
polarimetry practitioners who preferred to be guided in their work
by the following rule of thumb: {\it All grains are always aligned
and the alignment happens with the longer grain axes perpendicular to
magnetic field.} This simple recipe was shattered, however, by observational
data which indicated that \\
I. Grains of sizes smaller than a critical size are either not aligned
or marginally aligned (Mathis 1986, Kim \& Martin 1995).\\
II. Carbonaceous grains are not aligned, while silicate grains are aligned
(see Mathis 1986).\\
III. A substantial part of small grains
grains deep within molecular clouds are not aligned (Goodman et al. 1995,
Lazarian, Goodman \& Myers 1997, Cho \& Lazarian 2005 and references therein).\\
VI. Grains might be aligned with longer axes parallel to
magnetic fields (Rao et al 1998).

These facts were eloquent enough to persuade even the most sceptical types
that the interpretation
of interstellar polarimetric data does require an adequate
theory. A further boost of the interest to
grain alignment  came from the
search of Cosmic Microwave Background (CMB) polarization (see
Lazarian \& Finkbeiner 2003, for a review). Aligned dust in this case acts as
a source of a ubiquitous
foreground that is necessary to remove from the data. It is clear
that understanding of grain alignment is the key element for such a removal.

While alignment of interstellar dust is a generally accepted fact, the
alignment of dust in conditions other than interstellar has not been fully
appreciated. A common explanation of light polarization from comets or
circumstellar regions is based on light scattering by randomly oriented
particles (see Bastien 1988 for a review). The low efficiency and slow rates
of alignment were sometimes quoted to justify such an approach. However, it
has been proved recently that grain alignment is an efficient and rapid
process. Therefore, we {\it do expect} to have circumstellar, interplanetary,
and cometary dust aligned. Particular interesting in this respect are T-Tauri accretion
disks (see Cho \& Lazarian 2006). Tracing magnetic fields in these
environments with aligned grains opens new exciting avenues for polarimetry.

Taken a broader view, grain alignment is a part of a wider range of
alignment astrophysical processes that can provide the information about
magnetic fields. Molecules aligned in their excited states
can trace magnetic field (Goldreich \& Kylafis 1982), the effect that
was first successfully used in Girart, Crutcher \& Rao (1999) to map
magnetic field in molecular clouds. Atoms and ions with fine and
hyperfine structure can be aligned by
radiation in their ground or metastable states. The magnetic field then
mixes up the states due to the Larmor precession, which allows studies of
interstellar and circumstellar magnetic fields via absorption
and emission lines
(Yan \& Lazarian 2006ab)\footnote{Those studies potentially can restore 3D direction of magnetic fields, compared to the plane-of-sky projection of magnetic field available via
dust polarimetry.}. Making use of several alignment processes is another
avenue for observational studies of astrophysical magnetic fields (see
Lazarian \& Yan 2005).

Getting back to dust, one should mention that in the past the 
 linear starlight polarimetry was used. These days,
far infrared polarimetry of dust
emission has become
a major source of magnetic field structure data in molecular clouds (see Hildebrand 2000).
It is likely that the circular polarimetry may become an important
means of probing magnetic fields in circumstellar
regions and comets.

In this review I shall show that the modern grain alignment
theory allows us for the first time ever make quantitative
{\it predictions} of the polarization degree from various astrophysical objects.
A substantial part of the review is devoted to the
physics of grain alignment, which
 is deep and exciting. Enough to say, its progress 
resulted in a discovery of a number of new solid state physics effects.
The rich physics of grain alignment presents a problem, however,
for its presentation. Therefore I shall describe first the genesis
of ideas that form the basis of the present-day grain alignment theory.
The references to the original papers should help the interested reader
to get the in-depth coverage of the topic. Earlier reviews on the
subject include Lazarian (2003), Roberge (2004),
Lazarian \& Cho (2005) and Vaillancourt (2006). Progress in testing
theory is addressed in Hildebrand et al. (2000), while the exciting
aspects of grain dynamics are covered in Lazarian \& Yan (2004).

Below, in \S 2 I shall show how the properties of polarized radiation are
related to the statistics of aligned grains. In \S 3 I shall discuss the
essential elements of grain dynamics. In \S 4 I shall analyze the main
alignment mechanisms. In \S 5 I shall compare the mechanisms and discuss new
processes related to subsonic mechanical alignment of irregular grains. In \S
6, I shall discuss the observational data that put the grain-alignment theory
to test. An outlook on the prospects of the polarimetric studies of magnetic
fields will be presented in \S 7.

\section{Aligned Grains \& Polarized Radiation}

A practical interest in aligned grains arises from the fact that their
alignment results in polarization of the extinct starlight as well as in
polarization in grain emission. Below we discuss why this happens.

\subsection{Linear Polarized Starlight from Aligned Grains}

For an ensemble of aligned grains the degrees of extinction in the directions
perpendicular and parallel to the direction of alignment are
different\footnote{According to Hildebrand \& Dragovan (1995), the best fit of
the grain properties corresponds to oblate grains with the ratio of axis about
2/3.}. Therefore initially unpolarized starlight acquires polarization
while passing through a volume with aligned grains (see Fig.~2a). If the
extinction in the direction of alignment is $\tau_{\|}$ and in the
perpendicular direction is  $\tau_{\bot}$ one can write the polarization,
$P_{abs}$, by selective extinction
 of grains
as
\begin{equation}
P_{abs}=\frac{e^{-\tau_{\|}}-e^{-\tau_{\bot}}}{e^{-\tau_{\|}}+e^{-\tau_{\bot}}}
\approx -{(\tau_{\|}-\tau_{\bot})}/2~,
\label{Pabs}
\end{equation}
where the latter approximation is valid for $\tau_{\|}-\tau_{\bot}\ll 1$.
To relate the difference of extinctions to the properties of aligned grains
one can take into
account the fact that the extinction is proportional to the product
of the grain density and  their cross sections. If a cloud is composed of
identical aligned grains
$\tau_{\|}$ and $\tau_{\bot}$ are proportional to the number of grains
along the light path times the corresponding cross sections, which
are, respectively,
$C_{\|}$ and $C_{\bot}$.

In reality one has to consider additional complications (like, say, incomplete
grain alignment) and variations in the direction of the alignment axis
relative to the line of sight. (In most cases the alignment axis coincides
with the direction of magnetic field.) To obtain an adequate description, one
can (see Roberge \& Lazarian 1999) consider an electromagnetic wave
propagating along the line of sight (the {\mbox{$\hat{\bf z}^{\bf\rm o}$}}
axis, as on Fig.~1b). The transfer equations for the Stokes parameters depend
on the cross sections  $C_{xo}$ and $C_{yo}$ for linearly polarized waves with
the electric vector,  {\mbox{\boldmath$E$}}, along the {\mbox{$\hat{\bf
x}^{\bf\rm o}$}} and {\mbox{$\hat{\bf y}^{\bf\rm o}$}} axes perpendicular to
{\mbox{$\hat{\bf z}^{\bf\rm o}$}} (Lee \& Draine 1985).

\begin{figure}[h]
\includegraphics[width=2in]{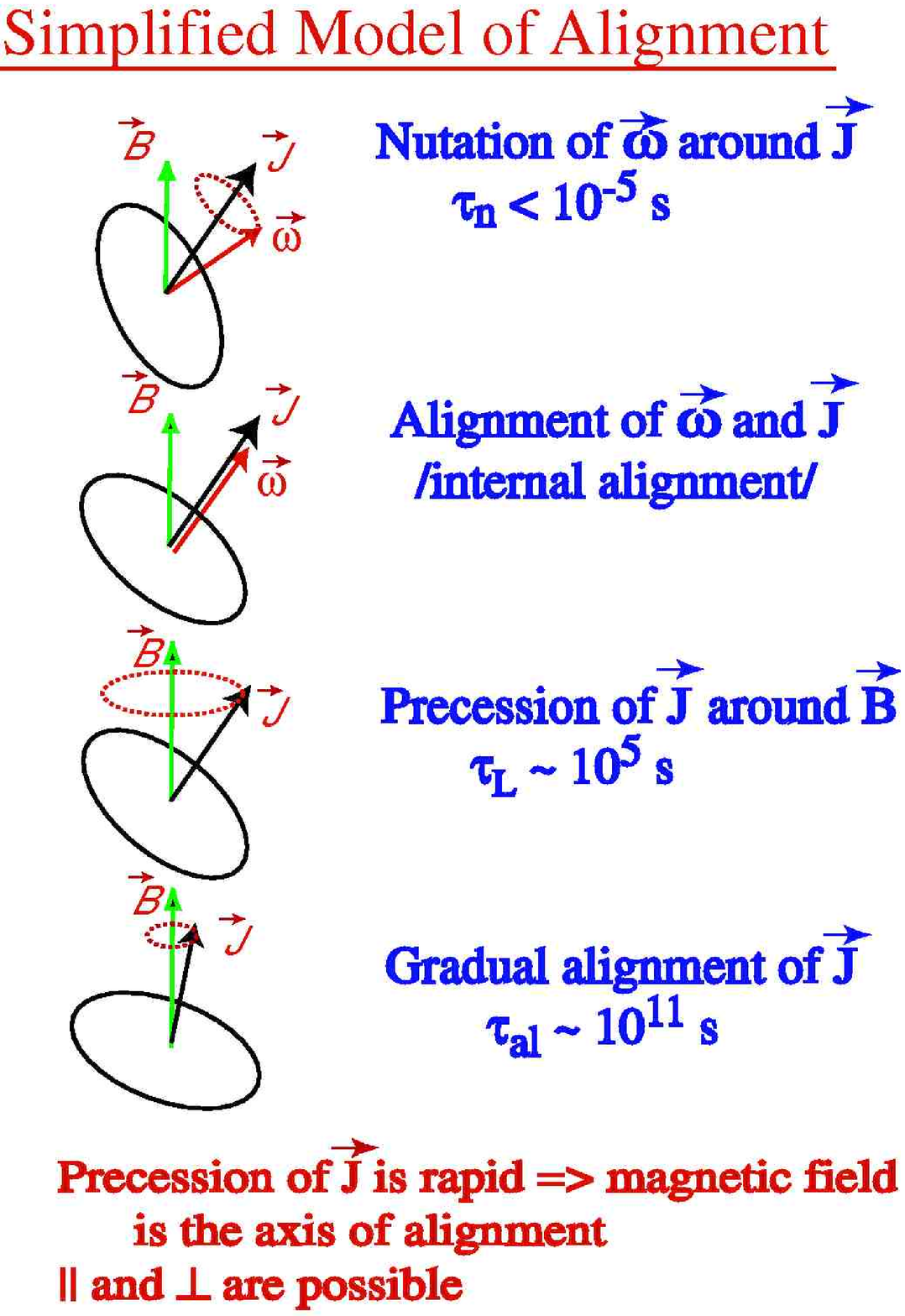}
\hfill
\includegraphics[width=3.in]{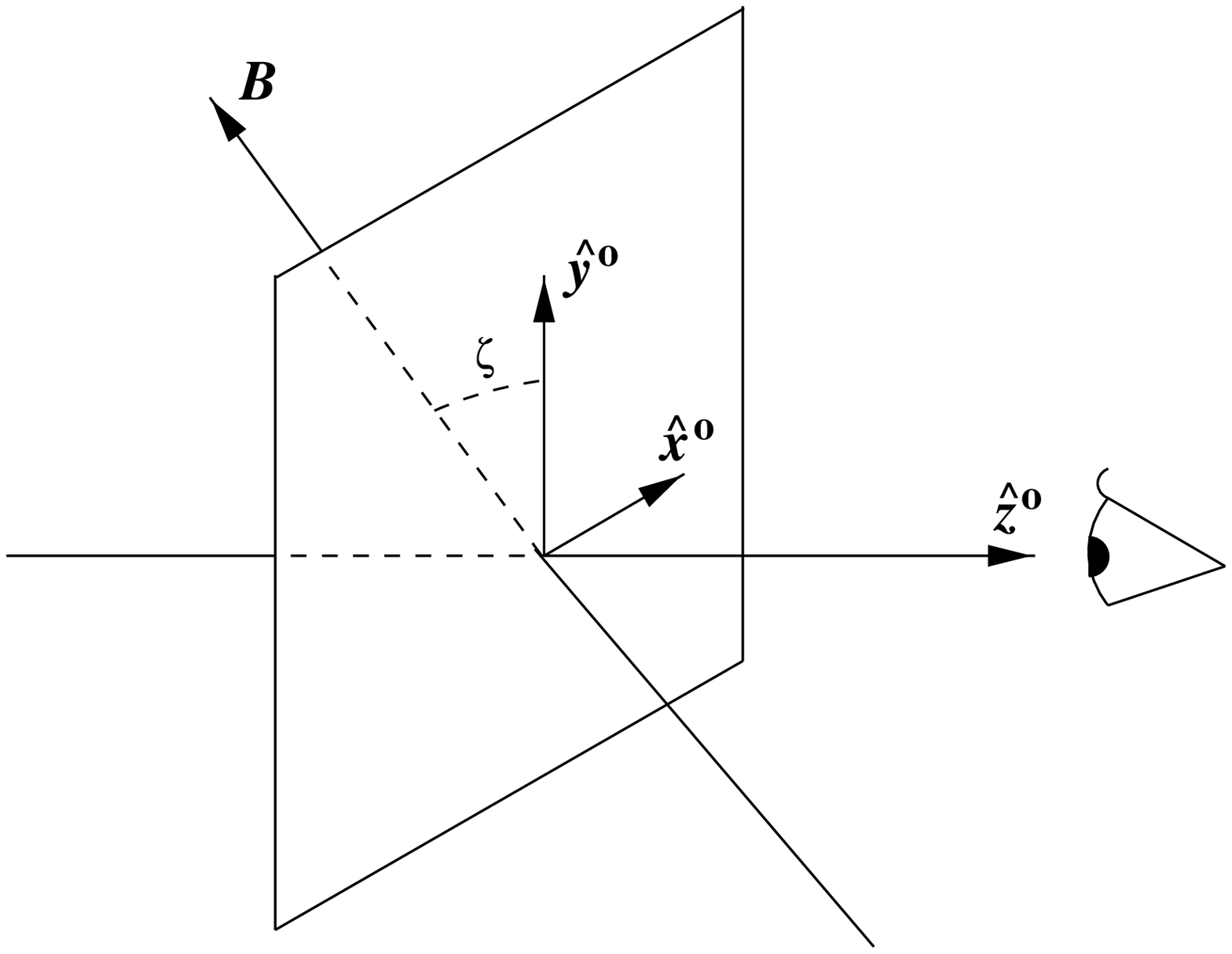}
\caption{\small {\it (a)Left panel}-- Alignment of grains implies several
alignment processes acting simultaneously and covering various timescales.
Internal alignment was introduced by Purcell (1979) and was assumed to be a
slow process. Lazarian \& Draine (1999a) showed that the internal alignment is
$10^6$ times faster if nuclear spins are accounted for.  The time scale of
${\bf J}$ and ${\bf B}$ alignment is given for diffuse interstellar medium. It
is faster in circumstellar regions and for cometary dust.
 {\it (b) Right panel}--
Geometry of observations (after Roberge \& Lazarian 1999).}
\label{fig:2Dspek}
\end{figure}

To calculate  $C_{xo}$ and $C_{yo}$, one transforms the components of
{\mbox{\boldmath$E$}} to the principal axes of the grain, and takes the
appropriately-weighted sum of the cross sections $C_{\|}$ and  $C_{\bot}$ for
{\mbox{\boldmath$E$}} polarized along the grain axes (Fig~1b illustrates the
geometry of observations). When the transformation is carried out and the
resulting expressions are averaged over the precession angles, one finds (see
transformations in Lee \& Draine 1985 for spheroidal grains, and in Efroimsky
2002a for the general case) that the mean cross sections are
\begin{equation}
C_{xo} = C_{avg} + \frac{1}{3}\,R\,\left(C_{\bot}-C_{\|}\right)\,
       \left(1-3\cos^2\zeta\right)~~~,
\label{eq-2_5}
\end{equation}
\begin{equation}
C_{yo} = C_{avg} + \frac{1}{3}\,R\,\left(C_{\bot}-C_{\|}\right)~~~,
\label{eq-2_6}
\end{equation}
$\zeta$ being the angle between the polarization axis and the {\mbox{$\hat{\bf
x}^{\bf\rm o}$}} {\mbox{$\hat{\bf y}^{\bf\rm o}$}} plane, and
$C_{avg}\equiv\left(2 C_{\bot}+ C_{\|}\right)/3$ being the effective cross
section for randomly-oriented grains. To characterize the alignment, we used
in eq.~(\ref{eq-2_6}) the Rayleigh reduction factor (Greenberg 1968) defined
as
 \begin{equation}
 R\equiv \langle G(\cos^2\theta) G(\cos^2\beta)\rangle\;\;\;,
 \label{R}
 \end{equation}
 where angular brackets denote ensemble averaging, $G(x) \equiv 3/2 (x-1/3)$,
 $\theta$ is the angle between the axis of the largest moment of inertia
 (henceforth the axis of maximal inertia) and the magnetic field ${\bf B}$, while
 $\beta$ is the angle between the angular momentum ${\bf J}$ and ${\bf B}$.
 To characterize the alignment with respect to the magnetic field,
 the measures ${Q_X\equiv \langle G(\theta)\rangle}$ and $Q_J\equiv \langle G(\beta)
 \rangle$ are employed. Unfortunately, these statistics are not independent and
 therefore $R$ is not equal to $Q_J Q_X$ (see Lazarian 1998, Roberge \& Lazarian 1999).
 This considerably complicates the description of the alignment process.

\subsection{Polarized Emission from Aligned Grains}
Aligned grains emit polarized radiation (see Fig.~2b). The difference in
$\tau_{\|}$ and $\tau_{\bot}$ for aligned dust results in the emission
polarization:
\begin{equation}
P_{em}=\frac{(1-e^{-\tau_{\|}})-(1-e^{-\tau_{\bot}})}{(1-e^{-\tau_{\|}})+
(1-e^{-\tau_{\bot}})}\approx \frac{\tau_{\|}-\tau_{\bot}}
{\tau_{\|}+\tau_{\bot}}~,
\label{Pem}
\end{equation}
where both optical depths $\tau{\|}$ are $\tau_{\bot}$ were assumed to be
small. Taking into account that both $P_{em}$ and $P_{abs}$ are functions of
the wavelength $\lambda$ and combining eqs.(\ref{Pabs}) and (\ref{Pem}), one
obtains for $\tau=(\tau_{\|}+\tau_{\bot})/2$
\begin{equation}
P_{em}(\lambda) \approx -P_{abs}(\lambda)/\tau(\lambda)~,
\label{Pem}
\end{equation}
which establishes the relation between the polarizations in emission and
absorption. The minus sign in eq~(\ref{Pem}) reflects the fact that emission
and absorption polarizations are orthogonal. This relation enables one to predict the
far infrared polarization of emitted light if the starlight polarization is
measured. This opens interesting prospects of predicting the foreground polarization
arising from emitting dust using the starlight polarization measurements (Cho \& Lazarian
2002, 2003).
 As $P_{abs}$ depends on $R$,
$P_{em}$ also depends on the Rayleigh reduction factor.
\begin{figure}[h]
\includegraphics[width=3.in]{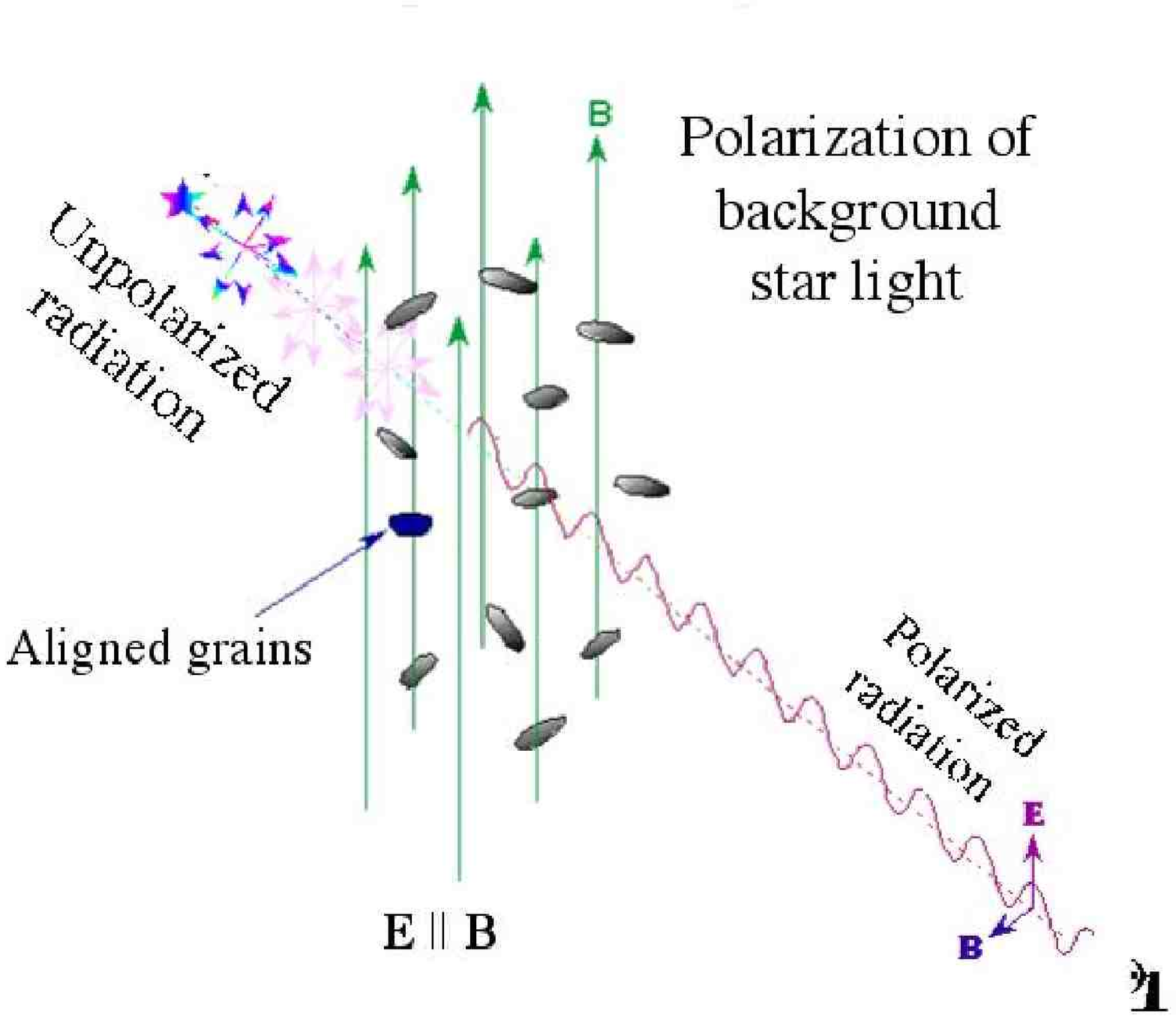}
\hfill
\includegraphics[width=3.in]{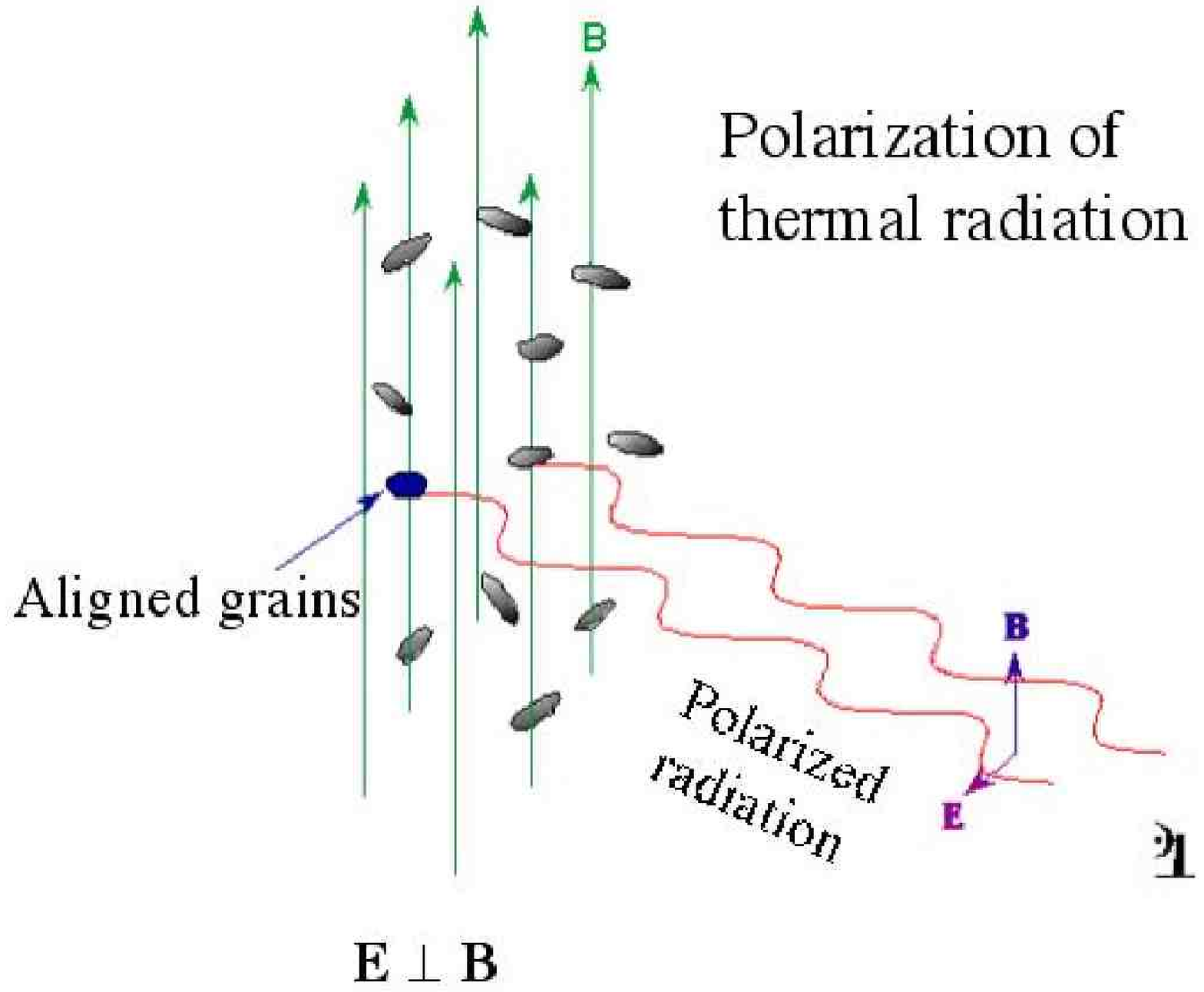}
\caption{
\small
{\it (a)Left panel}-- Polarization of starlight passing through a
cloud of aligned dust grains. The direction of polarization (${\bf E}$)
is parallel to the plane of the sky direction of magnetic field.
 {\it (b) Right panel}-- Polarization of radiation from a optically thin
cloud of aligned dust grains. The direction of polarization (${\bf E}$)
is perpendicular to the plane of the sky direction of magnetic field.
}
\label{fig:2Dspek}
\end{figure}

\subsection{Circular Polarization from Aligned Grains}

One way of obtaining circular polarization is to have a magnetic field that
varies along the line of sight (Martin 1972). Passing through one cloud with
aligned dust the light becomes partially linearly polarized. On passing the
second cloud with dust gets aligned in a different direction. Hence, the light
gets circular polarization. Literature study shows that this effect is well
remembered (see Menard et al 1988), while another process entailing circular
polarization is frequently forgotten. We mean the process of single scattering
of light on aligned particles. An electromagnetic wave interacting with a
single grain coherently excites dipoles parallel and perpendicular to the
grain's long axis. In the presence of adsorption, these dipoles get phase
shift, thus giving rise to circular polarization. This polarization can be
observed from an ensemble of grains if these are aligned. The intensity of
circularly polarized component of radiation emerging via scattering of
radiation with ${\bf k}$ wavenumber on small ($a\ll \lambda$) spheroidal
particles is (Schmidt 1972)
\begin{equation}
V( {\bf e}, {\bf e}_0, {\bf e}_1)=\frac{I_0 k^4}{2 r^2}i(\alpha_{\|}
\alpha^{\ast}_{\bot}-\alpha^{\ast}_{\|}\alpha_{\bot})\left([{\bf e_0}\times
{\bf e}_1] {\bf e}\right)({\bf e}_0 {\bf e})\;\;\;,
\end{equation}
where ${\bf e}_0$ and ${\bf e}_1$ are the unit vectors in the directions of
incident and scattered radiation, ${\bf e}$ is the direction along the aligned
axes of spheroids; $\alpha_{\bot}$ and $\alpha_{\|}$ are the particle
polarizabilities along ${\bf e}$ and perpendicular to it.

The intensity of the circularly polarized radiation scattered in the volume
$\Delta \Gamma({\bf d}, {\bf r})$ at $|{\bf d}|$ from the star at a distance
$|{\bf r}|$ from the observer is (Dolginov \& Mytrophanov 1978)
\begin{equation}
\Delta V ({\bf d}, {\bf r})=\frac{L_{\star} n_{\rm dust}\sigma_{V}}{6\pi |{\bf d}|^4
|{\bf r}||{\bf d}-{\bf r}|^2}R \left([{\bf d}\times {\bf r}] h\right)
({\bf d}{\bf r})\Delta \Gamma({\bf d}, {\bf r})~~~,
\label{circular}
\end{equation}
where $L_{\star}$ is the stellar luminosity, $n_{\rm dust}$ is the number of
dust grains per a unit volume, and $\sigma_V$ is the cross section for
producing circular polarization, which for small grains is
$\sigma_V=i/(2k^4)(\alpha_{\|}\alpha^{\ast}_{\bot}-\alpha^{\ast}_{\|}\alpha_{\bot})$.
According to Dolginov \& Mytrophanov (1978) circular polarization arising from
single scattering on aligned grains can be as high as several percent for
metallic or graphite particles, which is much more than one may expect from
varying magnetic field direction along the line of sight (Martin 1971). In the
latter case, the linear polarization produced by one layer of aligned grains
passes through another layer where alignment direction is different. If on
passing through a single layer, the linear polarization degree is $p$, then
passing through two layers produces circular polarization that does not exceed
$p^2$.

\section{Grain Dynamics: Never Ending Story}
Grain dynamics is really rich, as it involves an abundant variety of effects.
We provide a brief over-review of this exciting field of research.

\subsection{Wobbling Grains}
 To produce the observed starlight polarization, grains must be aligned, with their
 {\it long axes} perpendicular to magnetic field. According to eq. (\ref{R}) this
 involves alignment not only of the grains' angular momenta ${\bf J}$ with respect to
 the external magnetic field ${\bf B}$, but also the alignment of the grains'
 long axes with respect to ${\bf J}$. Jones \& Spitzer (1967) assumed a Maxwellian
 distribution of the angular momentum, distribution that favored the preferential
 alignment of ${\bf J}$ with the axis of a maximal moment of inertia (henceforth,
 maximal inertia, using Purcell's terminology). As we already mentioned in \S 3.2,
 Purcell (1979, henceforth P79) later considered grains rotating much faster than the
 thermal velocities and showed that the internal dissipation of energy in a grain will
 make grains spin about the axis of maximal inertia.

Indeed, it is intuitively clear that a tumbling and precessing grain should,
due to the internal dissipation, tend to get into the state of minimal energy,
i.e., to spin about the axis of maximal inertia. P79 discussed two possible
causes of internal dissipation -- one due to the well known inelastic
relaxation (see also Lazarian \& Efroimsky 1999), another due to the mechanism
that he discovered and termed ``Barnett relaxation".

We would remind to the reader that the Barnett effect is inverse to the
Einstein-de Haas effect. The essence of the Einstein-de Haas effect is that a
paramagnetic body acquires rotation during remagnetizations, when the flipping
electrons transfer to the lattice their spin angular momentum. The essence of
the Barnett effect is that the rotating body shares its angular momentum with
the electron subsystem, thus causing magnetization. The magnetization is
directed along the {\it grain's angular velocity},  and the value of the
Barnett-induced magnetic moment is $\mu\approx 10^{-19}\Omega_{(5)}$~erg
gauss$^{-1}$ (where $\Omega_{(5)}\equiv \Omega({\rm
s}^{-1})/10^5$)\footnote{Therefore the Larmor precession has a period
$\tau_L\sim 10^6 B_{5}^{-1}$~s (where $B_{5}\equiv B/(10^{-5}~{\rm G})$), and the
magnetic field defines the axis of alignment (see also \S 5.4)}.

Into the grain-alignment theory, the Barnett effect was introduced by Dolginov
\& Mytrophanov (1976), who noticed that the magnetization of rotating grains
due to this effect far exceeds the one arising from their typical charge. By
itself, this was a big advance in understanding the grain dynamics. Moreover,
it induced Purcell to think about the relaxation that this magnetization could
cause.

The Barnett relaxation process may be easily understood. We know that a freely
rotating grain preserves the direction of ${\bf J}$, while the
(body-axes-related) angular velocity precesses about ${\bf J}$ (see Fig.~3a).
The ``Barnett equivalent magnetic field'', i.e. the equivalent external
magnetic field that would cause the same magnetization of the grain material,
is $H_{BE}=5.6 \times10^{-3} \Omega_{(5)}$~G. Due to the precession of the
angular velocity, the co-directed ``Barnett equivalent magnetic field''
precesses in the grain axes. This causes remagnetization accompanied by the
inevitable dissipation.

\begin{figure}
\includegraphics[width=3.2in]{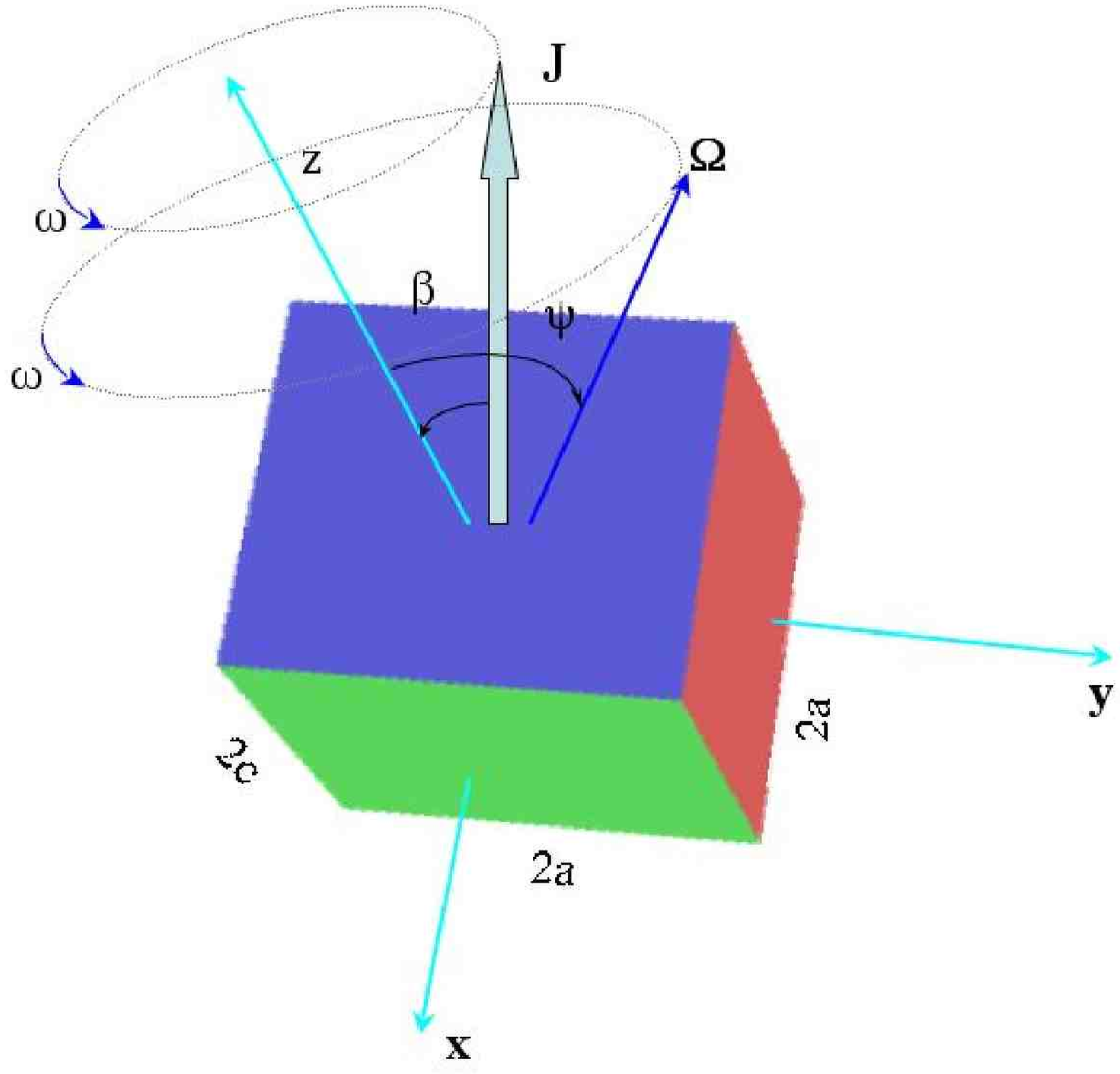}
\hfill
\includegraphics[width=2.3in]{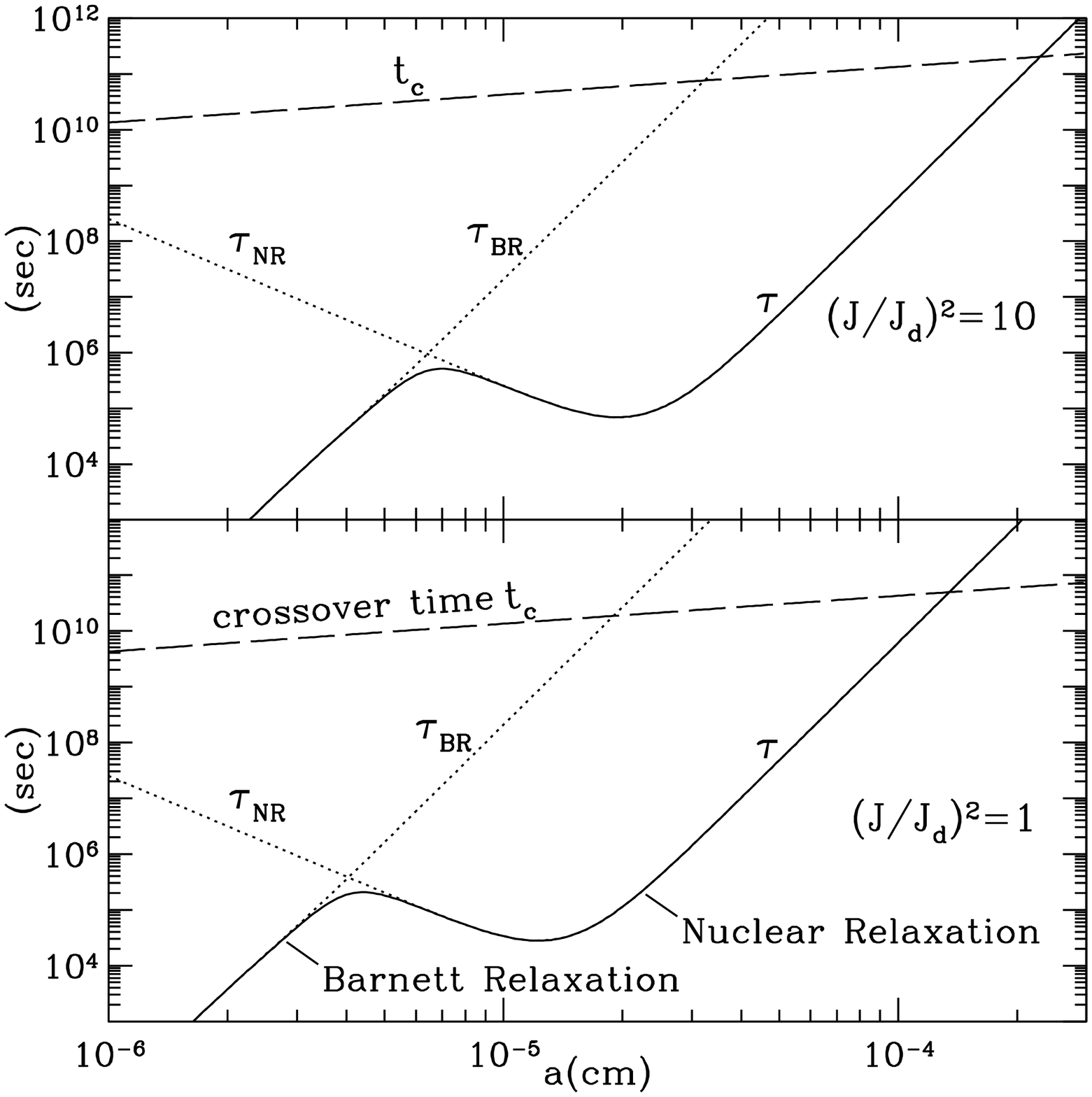}
\caption{\small {\it (a) Left panel}-- Grain dynamics as seen in the grain
frame of reference. The Barnett magnetization is directed along $\bOmega$, and
it causes a gradual grain remagnetization as the $\bOmega$ precesses in the
grain axes. {\it (b) Right panel}-- Time scale for the internal alignment due
to nuclear and Barnett relaxation processes. $J_d=(kT_{dust} I_{\bot} I_{\|}/(I_{\|}-I_{\bot}))^{1/2}$.
Also shown the ``crossover time'' $t_c=J/\dot{J_{\|}}$, where the torques are
due to the H$_2$ formation, with a density of active sites $10^{13}$
cm$^{-2}$. From Lazarian \& Draine (1999a). }
\end{figure}

The Barnett relaxation takes place over the time scale of $t_{Bar}\approx
4\times 10^7 \Omega_{(5)}^{-2}$~sec, which is very short compared to the time
$t_{gas}$ over which randomization through gas-grain collisions takes place.
As a result, models of interstellar-dust polarization developed since 1979
have often assumed that the Barnett dissipation aligns $\bf J~ {\it
perfectly}\/$ with the major axis of inertia. However, Lazarian (1994,
henceforth L94) showed that this approximation is invalid if the grains rotate
with thermal kinetic energies: thermal fluctuations in the Barnett
magnetization will excite rotation about all 3 of the body axes, preventing
perfect alignment unless either the rotation velocity  is suprathermal
($\Omega\gg \Omega_{thermal}$) or the grain's material temperature is zero.

Following Lazarian \& Roberge (1997, henceforth LR97), consider an oblate
grain (see Fig.~3a) with an angular momentum $J$. Its  energy can be written
as
\begin{equation}
E(\beta)=\frac{J^2}{I_{\|}}\left(1+\sin^2\beta(h-1)\right)~~~,
\label{1}
\end{equation}
 where $h=I_{\|}/I_{\bot}$ is the ratio of the maximal to minimal moments of grain inertia. Internal forces cannot change the angular
momentum, but it is evident from eq.(\ref{1}) that the energy can be
decreased by aligning the axis of maximal inertia along ${\bf J}$,
i.e. by decreasing $\beta$.
However, whatever the efficiency of internal relaxation,
in the presence of thermal fluctuations the grain energy
as a function of $\beta$ should have a Boltzmann distribution, i.e.
 $\exp(-E(\beta)/kT_{grain})$, where $T_{grain}$ is the grain
temperature, rather than the $\delta$-function distribution assumed in the
literature thitherto. The quantitative analysis offered in LR97 allowed many
further theoretical advances.

As the numbers of parallel and antiparallel spins become different, the body
develops magnetization, even if the unpaired spins are nuclear spins. The relation between $\Omega$ and the strength of the
``Barnett-equivalent'' magnetic field
$H_{\rm BE}^{\rm (n)}$
(P79)
that would cause the same
nuclear
magnetization (in a non-rotating body) is given by 
\be
 {\bf H}_{\rm BE}^{(\rm
n)}=\frac{\hbar}{g_{\rm n}\mu_{\rm N}}\bOmega~~~, 
\label{H} 
\ee
 where $g_{\rm
n}$ is the so-called nuclear $g$-factor (see Morrish 1980), and $\mu_{\rm
N}\equiv e\hbar/2m_{\rm p}c$ is the nuclear magneton, which is equal to the
Bohr magneton multiplied by the electron to proton mass ratio, $m_{\rm
e}/m_{\rm p}$.

The nuclear magnetization was mentioned in P79 as an subdominant effect that
can induce Larmor precession. The same paper discussed the Barnett relaxation,
but did not address a possible effect of the nuclear spins on the  internal
relaxation. Presumably, this was due to the fact that the nuclear moments
induce the magnetization of grains that is $m_{\rm e}/m_{\rm p}$ smaller that
the magnetization by electrons.

The {\it nuclear relaxation} was considered by Lazarian \& Draine (1999a, further on LD99a).
Surprisingly and rather counter-intuitively, the effect happened to be very
strong. Indeed, a striking feature of eq.~(\ref{H}) is that the Barnett-equivalent
magnetic field is inversely proportional to the species' magnetic moment. As
grain tumbles, the magnetization changes in the grain's body coordinates, and
this causes paramagnetic relaxation. This relaxation is proportional to
$\chi_N^{\prime\prime} (\Omega) H^2_{BE}$ (where $\chi_N^{\prime\prime}$ is
the imaginary part of the nuclear contribution to the susceptibility) and is
approximately $10^6$ times faster for nuclear moments than for their electron
counterparts (see Fig.~3b).

In terms of parameters involved, our arguments may be summarized as follows. The
Barnett equivalent field ${\bf H}_{BE}$ is $\sim 1/\mu$, while the
paramagnetic relaxation is proportional (for sufficiently slow rotation) to
$H^{2}_{BE}$, which means that the relaxation rate is proportional to
$1/\mu^2$. As $\mu\sim 1/m$, the heavier the species to align along $\bOmega$
the higher the relaxation rate.

Curiously enough, while the Barnett effect is {\it reduced} for nuclear spins
by a factor of $\sim m_{\rm e}/m_{\rm p}$, the relaxation {\it increases} by a
factor of $\sim (m_{\rm p}/m_{\rm e})^2$. Therefore it would be {\it
incorrect} to identify this relaxation as a modification of the Barnett
relaxation for nuclear spins. This is a separate relaxation process. In terms
of its domain of applicability it is limited by the spin-spin relaxation rate.
Indeed, the nuclear spins precess in the field of their neighbors,
which is approximately $\sim 3.8 n_n \mu_n$ (van Vleck 1937), where $\mu_n$ is
the magnetic moment of the nuclei, $n_n$ is the density of the nuclei. For 
hydrogen nuclei $\mu_n\approx 2.7 \mu_N$, for $^{29}Si$ $\mu_n\approx 0.5\mu_N$ (see
Robinson 1991). The rate of precession
in such a field is $\tau^{-1}_{nn}\sim \hbar/(3.8 g_n n_n\mu_n)$, where $g_n$ is the 
nuclear $g$-factor, which is, for instance, $\sim -0.6$ for $^{29}Si$. According to
LD99a the interaction of nuclei in the interstellar grains with electrons induce a
nuclei-electron relaxation rate $\tau^{-1}_{ne}$ which is comparable with $\tau^{-1}_{nn}$
and the actual spin-spin relaxation rate $\tau^{-1}_n$ is the sum of the two. If grain rotational
frequency $\omega$ exceeds the rate of spin-spin relaxation, the internal nuclear dissipation
rate $t^{-1}_{nucl}$ gets suppressed by a factor $[1+(\omega \tau_n)^2]$ (Draine \& Lazarian 1998b). 
This explains why
for small fast rotating grains the Barnett relaxation may be more efficient than the
nuclear one (see Fig.~3).

However, the nuclear relaxation dominates the Barnett one for grains
larger than $5\times 10^{-6}$~cm, the range that includes most of the aligned
interstellar grains. In general, for several relaxation processes acting
simultaneously, the overall internal relaxation rate is $t_{relax, tot}^{-1}=\Sigma
t_{relax, i}^{-1}$.

\subsection{Grains that are Swiftly Rotating, Flipping, and Thermally Trapped}

All the studies undertaken prior to 1979, with a notable exception of Dolginov
\& Mytrophanov (1976) that we shall discuss separately, assumed the Brownian
grain rotation with the effective temperature equal to the mean of the grain
and gas temperatures (see Jones \& Spitzer 1967). The greater complexity of
grain rotation was appreciated only later. Purcell (1975; 1979) realized that
grains may rotate at a much faster rate resulting from systematic torques. P79
 identified three separate systematic torque
mechanisms: inelastic scattering of impinging atoms when gas and grain
temperatures differ, photoelectric emission, and H$_2$ formation on grain
surfaces (see Fig.~4a). Below we shall refer to the latter as "Purcell's
torques". These were shown to dominate the other two for typical conditions in
the diffuse ISM (P79).  The existence of systematic H$_2$ torques is expected
due to the random distribution over the grain surface of catalytic sites of
H$_2$ formation, since each active site acts as a minute thruster emitting
newly-formed H$_2$ molecules. The arguments of P79 in favor of suprathermal
rotation were so clear and compelling that other researchers were immediately
convinced that the interstellar granules in diffuse clouds must rotate
suprathermally.

\begin{figure}
\includegraphics[width=2.3in]{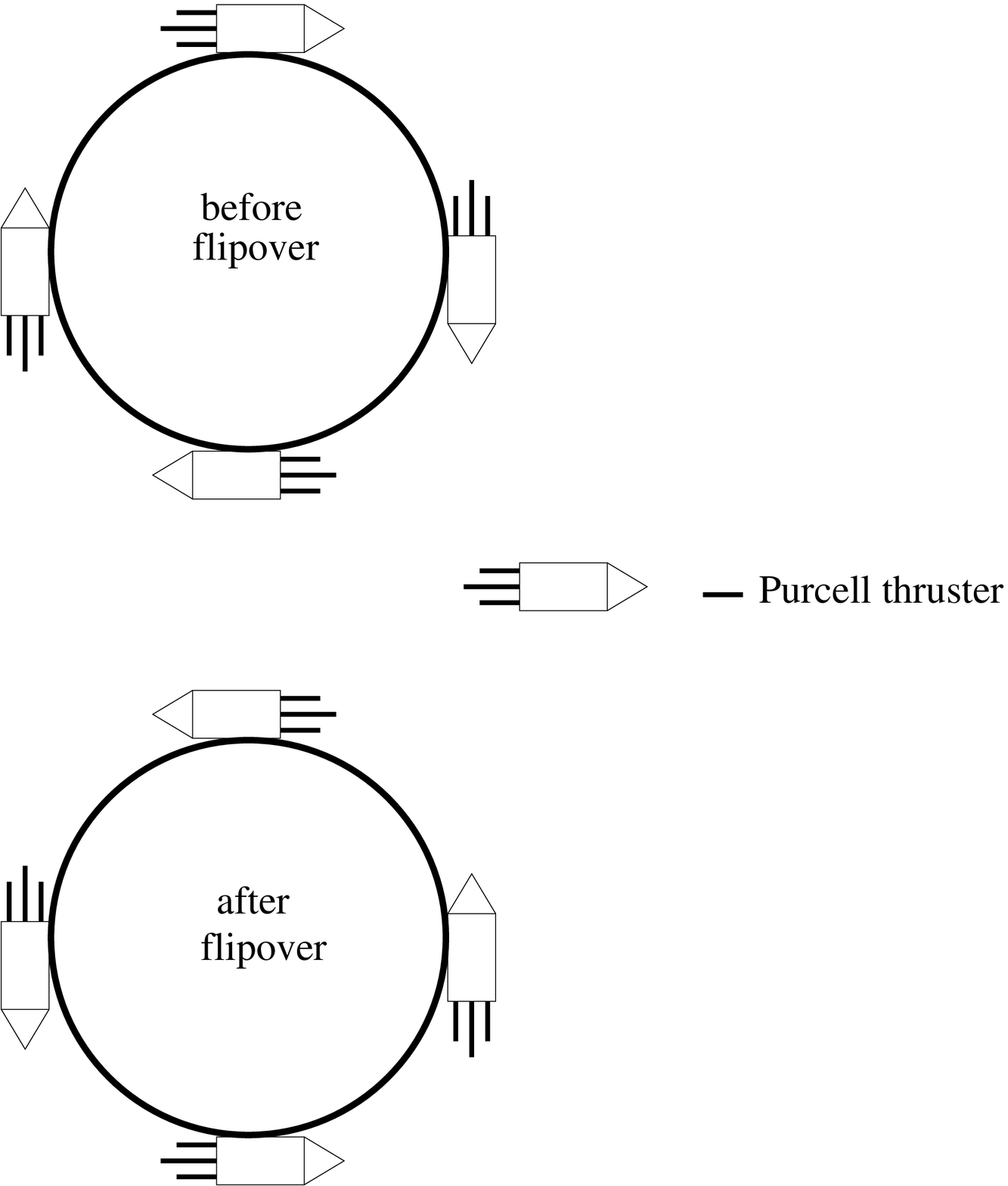}
\hfill
\includegraphics[width=2.0in]{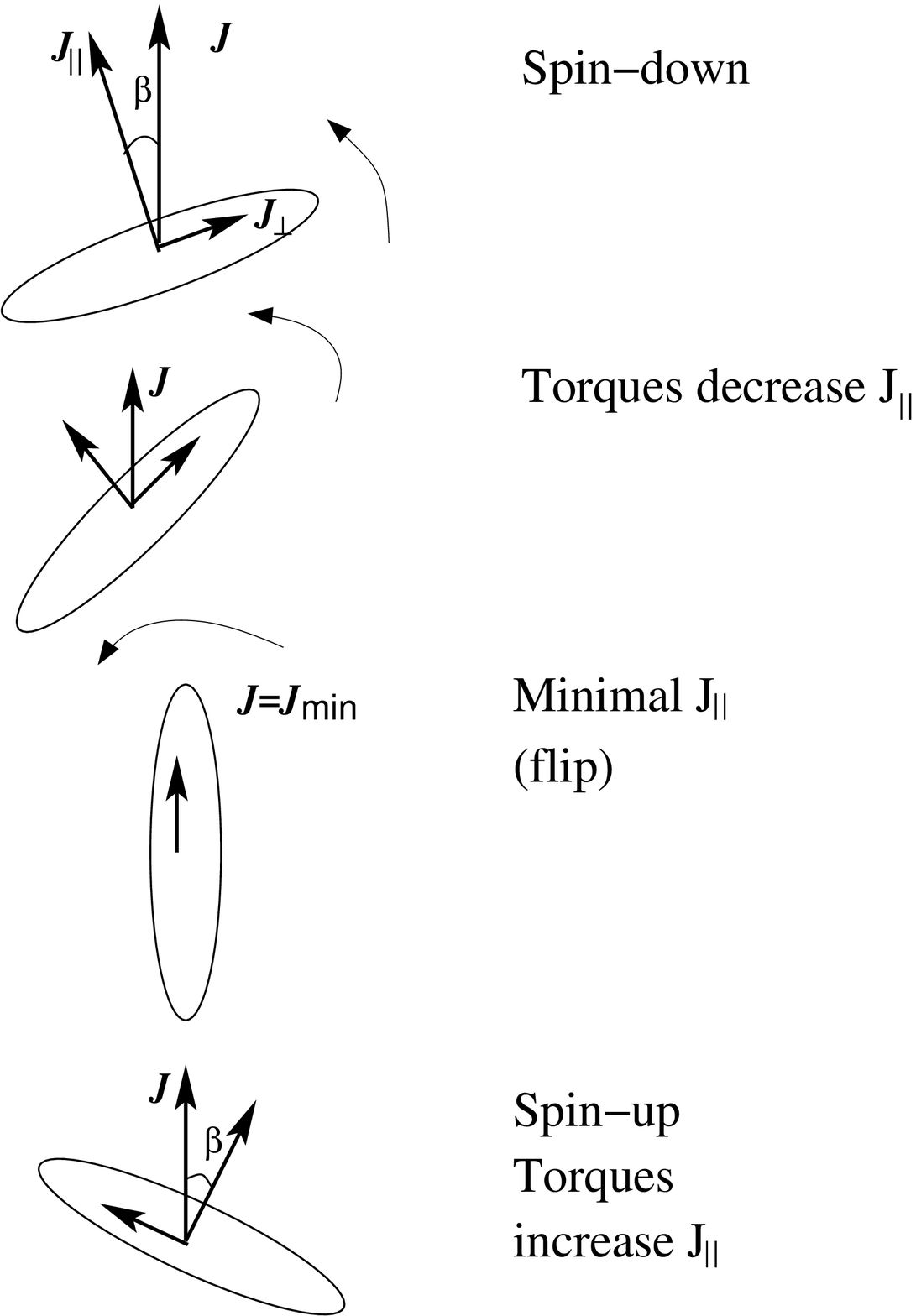}
\caption{ \small {\it (a) Left panel}-- A grain acted upon by Purcell's
torques before and after a flipover event. As the grain flips, the direction
of torques alters. The H$_2$ formation sites act as thrusters. {\it (b) Right
panel}-- A regular crossover event as described by Spitzer \& McGlynn (1979).
The systematic torques nullify the amplitude of the ${\bf J}$ component
parallel to the axis of maximal inertia, while preserving the other component,
$J_{\bot}$. If $J_{\bot}$ is small then the grain is susceptible to
randomization during crossovers. The direction of ${\bf J}$ is preserved in
the absence of random bombardment.}
\end{figure}

P79  considered changes of the grain surface properties and noted that those
should stochastically change the direction (in body-coordinates) of the
systematic torques. Spitzer \& McGlynn (1979, henceforth SM79) developed a
theory of such {\it crossovers}. During a crossover, the grain slows down,
flips, and thereafter is accelerated again (see Fig.~4b).

From the viewpoint of the grain-alignment theory, the important question is
whether or not a grain gets randomized during a crossover. If the value of the
angular momentum is small during the crossover, the grains are susceptible to
randomization arising from atomic bombardment. The original calculations in
SM79 obtained only marginal correlation between the values of the angular
momentum before and after a crossover, but their analysis disregarded thermal
fluctuations within the grain material. Indeed, if the alignment of ${\bf J}$
with the axis of maximal inertia is perfect, all the time through the
crossover the absolute value of $|{\bf J}|$ passes through zero during the
crossover. Therefore gas collisions and recoils from nascent $H_2$ molecules
would {\it completely} randomize the final direction of ${\bf J}$ during the
crossover. Thermal fluctuations partially decouple ${\bf J}$ from the
axis of maximal inertia (see \S 3.1). As a result, the minimal value of $|{\bf
J}|$ during a crossover is equal to the component of ${\bf J}$ perpendicular
to the axis of maximal inertia. This value for moderately oblate grains
is approximately $J_{d}\approx
(2kT_{dust} I_{\|})^{1/2}$, and the randomization during a crossover decreases
(Lazarian \& Draine 1997, henceforth LD97). LD97 obtained a high degree of
correlation between the angular-momentum directions before and after the
crossover for grains larger than the critical radius $a_{c, Bar} \approx
1.5\times10^{-5}$cm. This is the radius for which the time for internal
dissipation of the rotational kinetic energy is equal to the duration of a
crossover.

As nuclear relaxation is faster than the Barnett one for grains larger than 
$5\times 10^{-6}$~cm (see Fig. 3), the actual grain critical size $a_c$ gets
larger than $10^{-4}$~cm. In view of this, the results of LD97 study are
related only to very large grains, e.g. grains inside molecular clouds or
accretion disks.

\begin{figure}
\includegraphics[width=2.7in]{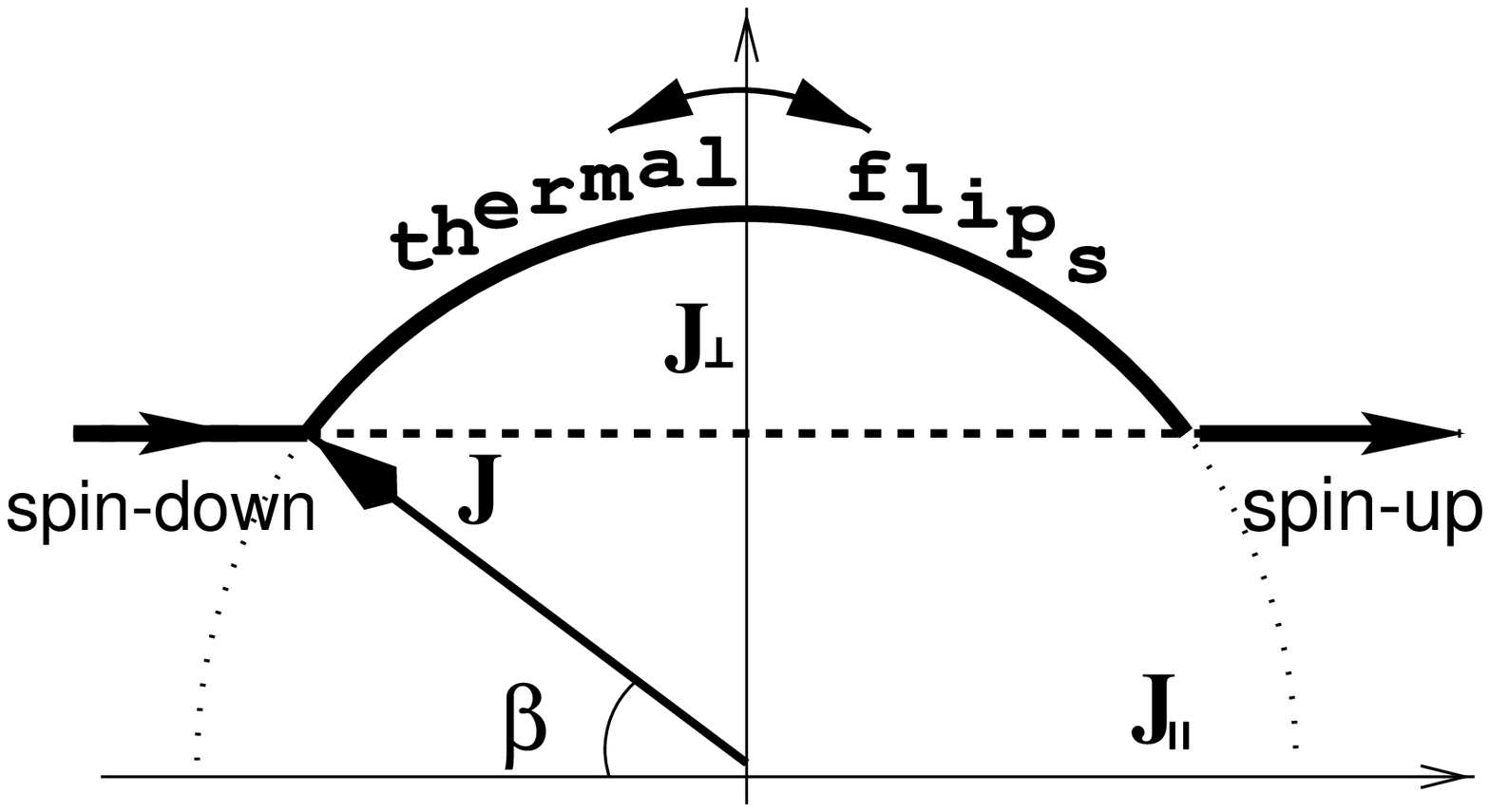}
\hfill
\includegraphics[width=2.6in]{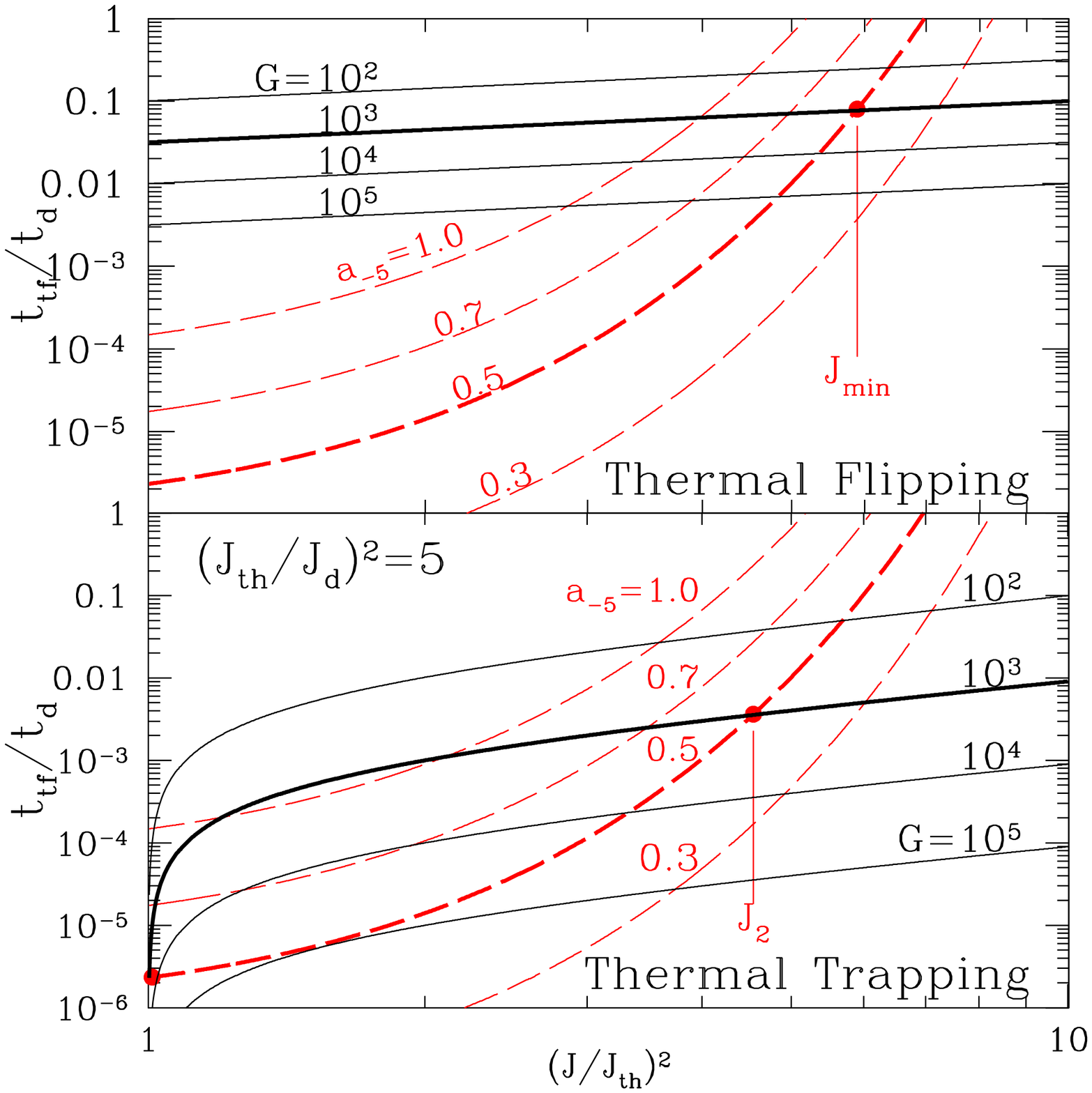}

\caption{
\small
{\it (a) Left panel}--
Grain trajectory on the $J_\perp$ -- $J_\parallel$ plane, where
        $J_\perp$ and $J_\parallel$ are components of $\bf J$
        perpendicular or parallel
        to the grain's principal axis of largest moment of inertia.
        The solid trajectory shows a ``thermal flip'', while the broken line
        shows the ``regular'' crossover which would occur in the absence of
        a thermal flip. {\it (b) Right panel}--
       Top: Thermal flipping to damping ratio as a function of
       $J/J_{thermal}$ for grains of given size
    [broken lines, labeled by $a_{-5}\equiv a(cm)/10^{-5}$] and
    for grains with a given value of systematic torques
        [solid lines, labeled by $G$].
    Dot shows $J_{\rm min}= \dot{J}\cdot t_{tf}$
        for flipping-assisted crossover of $a_{-5}=0.5$
    grain with $G=10^3$.
    Bottom: Thermal trapping for grains of given size [broken lines,
    labeled by $a_{-5}$], and given value of torques
    [solid lines, labeled by $G$].  From Lazarian \& Draine (1999b).}
\end{figure}

What would happen for grains that are smaller than $a_c$? The SM79 theory
prescribed that the granules should follow the phase-space trajectory along
which $J_{\bot}$ is approximately constant while the component of ${\bf J}$
parallel to the axis of maximal inertia $J_{\|}$ changes sign. Later, though,
Lazarian \& Draine (1999b, henceforth LD99b) demonstrated that in reality the
grains undergo flipovers (see Fig.~5a) during which the absolute value ${\bf J}$
does not change. If these flipovers repeat, the grains get ``thermally
trapped'' (LD99b and Fig.~5b). This process can be understood in the following
way. For sufficiently small $|J|$, the rate of flipping $t_{tf}^{-1}$ becomes
large. Purcell's torques change sign as grain flips, and they cannot
efficiently spin the grain up. As a result, a substantial part of grains
smaller than $a_{cr}$ cannot  rotate at high rates predicted by P79, even in
spite of the presence of systematic torques that are fixed in the body axes
(LD99a). A more elaborate study of the phenomenon in Roberge \& Ford
(preprint; see also Roberge 2004) supports this conclusion.

While the thermal trapping limits the range of grain sizes which can be spun
up by Purcell's torques, a natural question arises: do the astrophysical
grains rotate suprathermally?

Earlier than Purcell, Dolginov (1972) and Dolginov \& Mytrophanov (1976)
identified radiative torques as the way of spinning up a subset of the
interstellar grains. Unlike Harwit (1971), who addressed the issue of
interaction of symmetric, e.g. spheroidal, grains with a radiative flow,
Dolginov and Mytrophanov considered ``twisted grains'' that can be
characterized by some {\it helicity}. They noticed that ``helical'' grains
would scatter differently the left- and right-polarized light, for which
reason an ordinary unpolarized light would spin them up.
        The subset of the ``helical'' grains was not properly identified,
and the later researchers could assume that it is limited to special
shapes/materials. One way or another, this ground-breaking work did not make
much impact to the field until Draine \& Weingartner (1996, henceforth DW96)
numerically showed that grains of rather arbitrary irregular shapes get spun
up efficiently.

DW96 and Draine \& Weingartner (1997, henceforth DW97) demonstrated that
radiative torques can be separated into isotropic and anisotropic parts. While
the isotropic torques that are fixed in body coordinates are averaged out
similarly to the Purcell torques, the anisotropic torques do not change sign when the grain flips.
If those spin-up grains are fast enough to avoid constant flipping, the Purcell torques
can also act on a grain in a regular way. Do all grains get spun up efficiently
by anisotropic radiative torques? While DW97 provide arguments in favor
of the positive answer, it should be mentioned that
they treated crossovers in a crude way, i.e.,
as singularities at which the grain does not
flip, while the direction of $\bf J$ changes to the opposite one. This is
different from the crossover prescriptions in SM79 and
Lazarian \& Draine (1997). On the contrary, the study in Weingartner \& Draine
(2003, henceforth WD03), that accounted for thermal wobbling
of grains (LR97, LD99b), indicated that only a fraction of grains rotates
suprathermally when acted upon by anisotropic radiative torques. Lazarian \&
Hoang (2006, henceforth LH07) showed that the same effect is also present when
thermal wobbling is absent, but a more rigorous treatment of crossovers is applied.
In fact, LH06 showed that at $T_{grain}\rightarrow 0$ and no gaseous
bombardment most grains undergo multiple crossovers and get settled
with $J\rightarrow 0$. For finite $T_{grain}$, the same subset of
grains settles with $J\sim J_{d}$ in accordance with the findings in WD03. 
The effective temperature of grain rotation increases to approximately
$T_{gas}>T_{grain}$  when gaseous bombardment is present (Hoang \& Lazarian 2007).

This presents an unexpected twist in the theory of radiative torques.
Interestingly enough, for most grains their alignment by radiative torques is a way
to {\it minimize} their rotational velocity. Therefore most grains in the
diffuse interstellar gas, contrary to the common belief, {\it do not rotate
suprathermally}.
In addition, essentially none of the small grains (i.e.  ones with
$a<5\times 10^{-6}$~cm), rotate suprathermally as the radiative torques are
too weak to spun up the grains of size much less than the wavelength\footnote{In
the vicinity of stars with UV excess smaller grains can be spun up
as well.}.
On the contrary, grains deep within starless molecular
clouds were usually assumed to rotate
thermally. However,
Cho \& Lazarian (2005) showed that the radiative torques efficiency increases
with the grain size. Therefore some fraction
of large grains will rotate suprathermally even in dark cores of molecular
clouds. As we explain further, rapid rotation is not a necessary
requirement for the
efficient alignment, if radiative torques are concerned.

\subsection{Grains Zooming in Space}

Grains can stream through ambient gas. One of the
processes to induce such streaming
was suggested by Gold (1952) who considered 
penetration of grains from one cloud to another as the clouds
collide. Later, though, Davis (1955) showed that the applicability realm of
the process is quite limited.

A more standard way of driving grain-gas motion is by radiation pressure
(see Purcell 1969).
Grains are exposed to various forces in anisotropic radiation fields.
Apart from
radiation pressure, grains are subjected to forces due to the asymmetric
photon-stimulated ejection of electrons. A detailed discussion can
be found in Weingartner \& Draine (2001).  They
 demonstrated that the emission caused 
force is comparable to the one arising from the usual radiation pressure,
provided that
the grain potential is low and the radiation spectrum is hard. 
Another photon-stimulated ejection process showing up in the picture
 is photodesorption of atoms absorbed
on grain surface. The force due to photodesorption of atoms
is comparable to the radiation and photoelectric ones (Draine 2003). 
However, none of these forces is expected to induce a
supersonic
grain drift under the typical interstellar conditions.

A residual imbalance arises from the difference of
the number of catalytic active sites for H\( _{2} \) formation on
upper and lower grain surfaces (P79). The nascent H\( _{2} \) molecules leave
the active sites with kinetic energy \( E \), and the grain experiences
a push in the opposite direction. The uncompensated
force is parallel to the spin direction as the other components
of force are averaged out due to the grain's fast rotation. Applying
the best-guess values\footnote{The number of H$_{2}$ formation sites is highly uncertain. It may
also depend on the interplay of the processes of photodesorption and
poisoning (Lazarian 1995b; 1995c).}  adopted in LD97, Lazarian \& Yan (2002)
got the ``optimistic'' velocity
$v\simeq330(10^{-5}$cm$/a)^{1/2}$cm/s
for the Cold Neutral Medium (CNM) and $v\simeq 370(10^{-5}$cm$/a)^{0.7}$cm/s for the Warm
Neutral Medium (WNM), provided that grains do not flip (see \S 3.2). In
dark clouds, a similar effect arising from variations of the
accommodation coefficient can induce translational motion of grains.

Turbulence is another driver for grain drift with respect to gas.
It is generally accepted that the interstellar medium is turbulent (see
Elmegreen \& Scalo 2002). Turbulence has been invoked by a number of authors
(see Weidenschilling \& Ruzmaikina 1994, Lazarian \& Yan 2002 and references
therein) to induce grain motion relative to the gas.
In hydrodynamic turbulence, the grain motions are caused by the frictional
interaction with the gas. At large scales, grains are coupled with the
ambient gas, and the fluctuating gas motions mostly cause
an overall advection of the grains with the gas (Draine 1985). At
small scales, grains are decoupled. The largest velocity difference
occurs on the largest scale at which the grains are still decoupled. Thus
the characteristic velocity of a grain with respect to the gas corresponds
to the velocity dispersion of the turbulence on the scales corresponding to eddies with turnover
time equal to $t_{drag}$
(Draine \& Salpeter 1979). Using the Kolmogorov scaling
relation $v_{k}\propto k^{-1/3}$, Draine (1985) obtained the largest
velocity dispersion in hydrodynamic turbulence
$v\simeq V(t_{drag}/\tau_{max})^{1/2}$, where $\tau_{max}$ is the eddy turnover time
 at the injection scale.

A complication, though, comes from the fact that most astrophysical
fluids are magnetized. Therefore
 magnetohydrodynamic (MHD) turbulence should be used to characterize
interstellar turbulence. This was attempted first in
L94. A more quantitative approach was
adopted in Lazarian \& Yan (2002)
and Yan \& Lazarian (2003, henceforth YL03). There, in accordance with the
simulations
in Cho \& Lazarian (2002), the MHD turbulence was decomposed into an Alfven,
slow and fast modes. The particular scalings of the modes were applied,
i.e., Goldreich \& Sridhar (1995) scaling for Alfven and slow modes, and
acoustic turbulence scaling for fast modes. Moreover, in YL03 we considered
a gyro-resonance between the fluctuating magnetic field and charged
grains, and thus identified a new mechanism of grain acceleration.

Specifically, the resonance condition that the Doppler-shifted frequency of the
wave in the grain's guiding center
rest frame $\omega_{gc}=\omega-k_{\parallel}v\mu$ is a multiple of
the particle gyrofrequency $\Omega_g$: $\omega-k_{\parallel}v\mu=n\Omega_g$,
($n=0,\pm1,\pm2...$). Basically, there are two main types
of resonant interactions: gyroresonance acceleration and the transit one.
The transit acceleration ($n=0$) requires longitudinal motions that are present only
for compressible modes. As the dispersion relation for fast waves is
$\omega=kV_{f}>kV_{A}$, it is clear that it is applicable only to the super-Alfvenic
(for a low $\beta$ medium, i.e. with magnetic pressure higher than the thermal one, as
$\beta\equiv P_{gas}/P_{mag}$) or supersonic (for a high $\beta$ medium)
grains. For low speed grains that we deal with, gyroresonance is the
dominant MHD interaction.
The calculation by YL03 showed that
grains gain the maximum velocities perpendicular
to the magnetic field, so the averaged $\mu$ decreases.
This is understandable since the electric field accelerating
the grain is perpendicular to the magnetic field.

The results of the theory
were applied to various idealized phases of the interstellar
medium in Yan, Lazarian \& Draine (2004).
In Fig.~6, we show the velocity of grain
as a function of the grain size in CNM.

\begin{figure}
\includegraphics[width=2.5in]{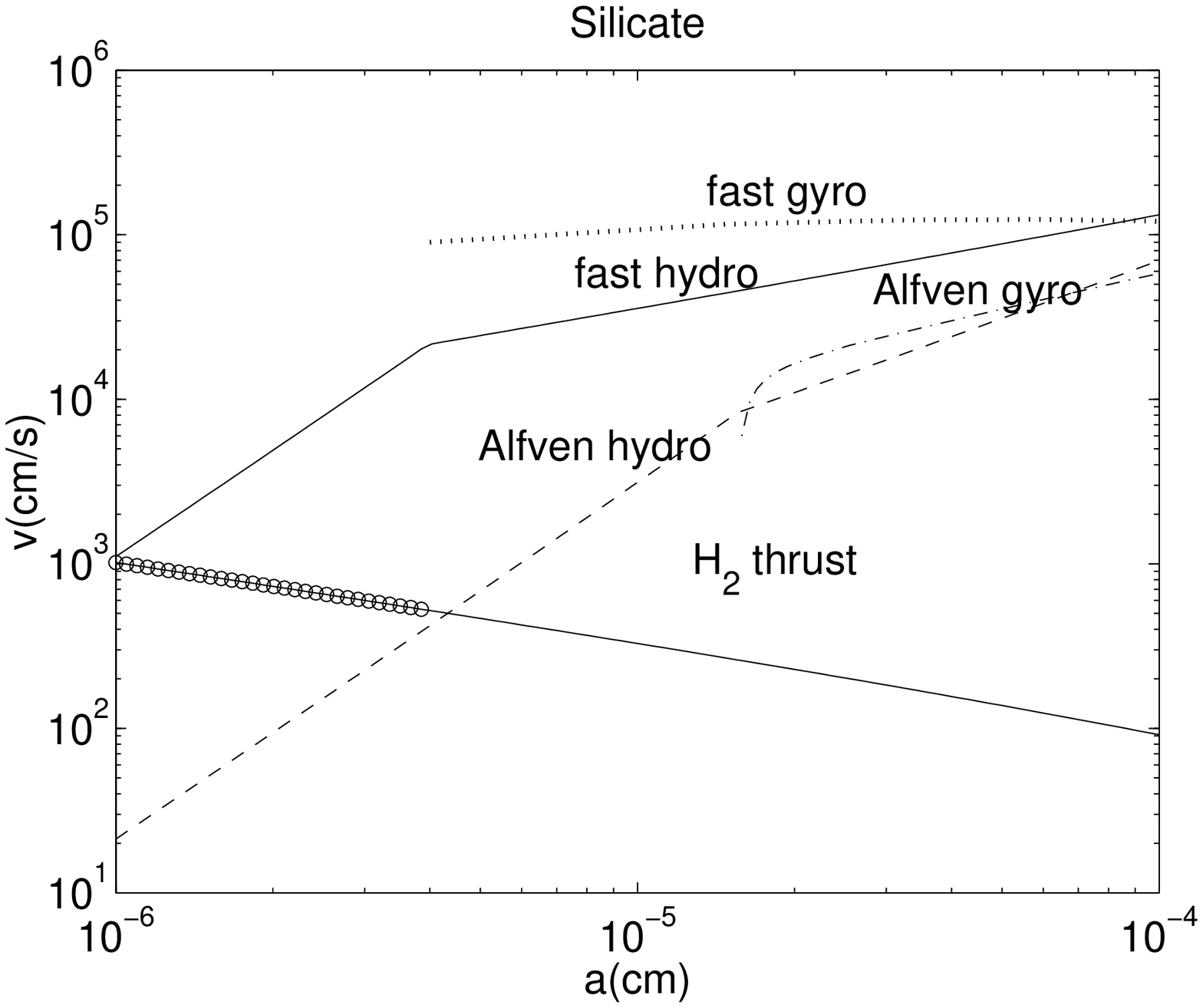}
\hfill
\includegraphics[width=2.5in]{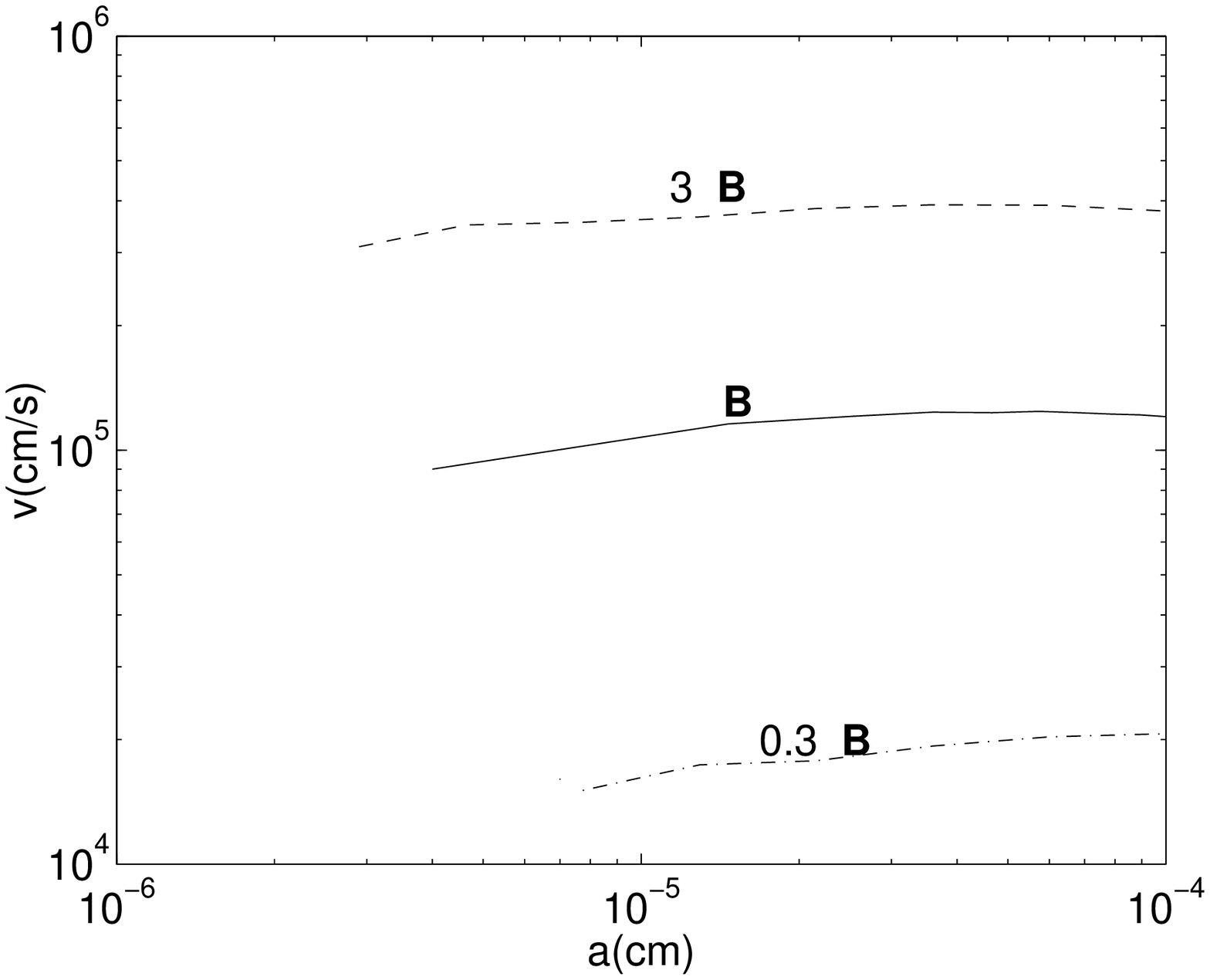}

\caption{
\small
{\it (a) Left panel}.--
Relative velocities as functions of grain radius 
for silicate grains in the Cold Neutral Medium. The dotted
line represents the gyroresonance with fast modes. The dash-dot line
refers to the gyroresonance with Alfv\'{e}n modes. The cutoff is due to
viscous damping. From Yan \& Lazarian (2003). {\it (b) Right panel}.-- Grain velocities
in CNM
 gained from gyroresonance for different magnetic field
strengths. From Yan, Lazarian \& Draine (2004).}
\end{figure}

The acceleration by gyroresonance in both MC and DC are not so efficient
as in other media. This happens in MC and DC because the time for the gyroresonant
acceleration, $t_{drag}$, are much shorter that in the WNM. In MC and DC, due to high density,
the drag time  is less than the gyro-period
for grains larger than $10^{-5}$~cm.

For molecular clouds Roberge \& Hanany (1990) and
Roberge, Hanany \& Messinger (1995)
considered ambipolar diffusion\footnote{A similar process was considered by Roberge  \& Desch (1990) for
 molecular accretion disks.}.
They demonstrated that this diffusion entails supersonic relative 
drift. The action of the mechanism is expected to be localized, however.

To finish our brief discussion of grain motion in magnetized medium
consider magnetized shocks.
The basic idea is that the weakly charged grains are like ions with
high mass to charge ratio (Epstein 1980). Thus they can easily diffuse
farther back
upstream of the shock and be accelerated more efficiently to suprathermal
energies. Nevertheless, the shock acceleration is inefficient for low
speed grains. The reason for this is that the efficiency of the shock acceleration
depends on the scattering rate, which is determined by the stochastic
interaction with the turbulence. For low speed particles, the scattering
rate is lower than the rate
of momentum diffusion. In this case, the stochastic acceleration
 by turbulence happens faster than dust acceleration
by shocks (YL03).

\section{Grain Alignment Theory: Major Mechanisms}

\subsection{Tough Problem}

We have seen in the previous sections that both linear and circular
polarizations depend on the degree of grain alignment given by the
Rayleigh reduction
factor (see Eq.~(\ref{R})). Therefore it is the goal of the grain alignment theory
to determine this factor. Table~1 shows that the wide range of different time
 scales involved makes
the brute force numerical approach doomed.

A number of different mechanisms that produce grain alignment has been
developed by now. Dealing
with a particular situation one has to identify the dominant alignment process.
Therefore it is essential to understand different mechanisms.

The history of grain alignment is really exciting. A real constellation
of illustrious scholars, e.g. L. Spitzer and E. Purcell contributed to the field. Our earlier
discussion of the complex dynamics of a grain explains why the grain alignment
theory still requires theoretical efforts.
Note, that most of the effects we discussed in the previous section
were {\it discovered in the process of work on grain alignment}.

A drama of ideas in historic perspective is presented in Lazarian (2003).
It was shown there that the work on grain alignment can be subdivided
into a number stages, such that at the end of each the researchers believed that the
theory was adequate. However, higher quality observational data made it
clear that more work was required.
\begin{table}[h]
\caption{Time-scales relevant for grain alignment}
\begin{displaymath}
\begin{array}{rrrrrr} \hline\hline\\
\multicolumn{1}{c}{\rm Symbol} & \multicolumn{1}{c}{\rm Meaning}  & \multicolumn{1}{c}{\rm Definition} &
\multicolumn{1}{c}{\rm Value~~(s)} \\[1mm]
\hline\\
{\rm t_{rot}}&{\rm thermal~rotational~period}&{2\pi/\Omega}
&{6\times 10^{-5}\hat{T}_{rot}^{-1/2}a_{-5}^{5/2}s^{-2}}\\[1mm]
{\rm t_{Bar}}&{\rm Barnett~relaxation~time}&
{\frac{\gamma_e^{2}I_{\|}^{3}}{VK_{e}h^{2}(h-1)J^{2}}}&{9.84\times 10^{6}(\frac{\hat{\rho}^{2}}{\hat{K_{e}}\hat{T}_{d}})f_1(s) a_{-5}^{7}}(\frac{J_{d}}{J})^{2}F(\tau_{el})\\[1mm]
{\rm t_{nucl}}&{\rm
  nuclear~relaxation~time}&{(\frac{\gamma_n}{\gamma_e})^2(\frac{K_e}{K_n}) t_{Bar}}
  &{21.35 \hat{\rho}^{2} a_{-5}^{7}f_{1}(s)\hat{g}_{n}^{4}\hat{\mu}_n^{-2}}(\frac{J_{d}}{J})^{2}F(\tau_n)\\[1mm]
{\rm t_{tf}}&{\rm thermal~flipping~time}&&{~~~~t_{B,nucl}{\rm exp}(0.5[\frac{J^{2}}{J_{d}^{2}}-1])}\\[1mm]
{\rm t_{c}}&{\rm crossover~time}&{\frac{2 J_{d,\bot}}{L_{z}^{b}}~~~~} & {1.6\times 10^{9} (\frac{\hat{\rho}
\hat{T_d}\hat{\alpha}}{\hat{W}\hat{\zeta}^2\hat{n}^2 \hat{T}_{g}})^{1/2} f_2(s) a_{-5}^{1/2}} \\[1mm]
{\rm t_{L}} &{\rm Larmor~precession~time}& {\frac{2\pi\mu_{d} I_{\|}}{\chi^{'}VB}}&{1.1\times
   10^{6}(\frac{\hat{\rho}\hat{T_{d}}}{\hat{\chi}\hat{B}})a_{-5}^{2}}s^{2}\\[1mm]
{\rm t_{RT}}&{\rm Radiative~precession~time}&{\frac{2\pi}{|d\phi/dt|}}&{\frac{3\times 10^{7}}{\hat{Q}_{e3}}\hat{\rho} b_{-5}^{1/2}(\frac{1}{\hat{\lambda}\hat{u}_{rad}})} \\[1mm]
{\rm t_{gas}}&{\rm gas~damping~time}& {\frac{4I_{\|}}{nmv_{th} b^{4}}}& {4.6\times
  10^{12}(\frac{\hat{\rho}_{s}}{\hat{n}\hat{T}_{g}^{1/2}}) sb_{-5}}\\[1mm]
{\rm t_{E}}&{\rm electric~precession~time}& {\frac{2\pi}{\Omega_{E}}}&{0.2 \times 10^{11} p^{-1} \hat{E}^{-1}\hat\rho \hat{\omega}{a_{-5}}}\\[1mm]
{\rm t_{DG}}&{\rm paramagnetic~damping~time}&{\frac{2\rho a^{2}}{5K(\omega)T_{2}B^{2}}}&{10^{13}\hat{B}^{-1}\hat{K}^{-1}a_{-5}^{2}s^2}\\[1mm]
\hline
\end{array}
\end{displaymath}
{\tiny Notations:\\
$a$: minor axis~ ~~~~~~~~~~~~~~~~~~~~~~~~~~~~~~~~~~~~~~~~~~~~~~~~~~~~~~~~~~~~~~~$b$: major axis\\
$a_{-5}=a/10^{-5} cm$~~~~~~~~~~~~~~~~~~~~~~~~~~~~~~~~~~~~~~~~~~~~~~~~~~~~~~~~~~~~$s=a/b < 1$: ratio of axes\\
$h=I_{\|}/I_{\perp}$: ratio of moment inertia~~~~~~~~~~~~~~~~~~~~~~~~~~~~~~~~~~~~$\hat{\rho}=\rho/3 gcm^{-3}$: normalized grain density\\
$\hat{T}_{g}=T_{g}/85 K$: normalized gas
temperature~~~~~~~~~~~~~~~~~~~~~~~~~~$\hat{T}_{d}=T_{d}/15 K$: normalized dust
temperature\\
$T_{rot}=(T_{g}+T_{d})/2$: rotation temperature\\
$\hat{n}=n/20 cm^{-3}$: normalized gas
density~~~~~~~~~~~~~~~~~~~~~~~~~~~~~~$\hat{B}=B/5 \mu
\mbox{G}$: normalized magnetic field\\
$\chi^{'}=10^{-3}\hat{\chi}/\hat{T_{d}}$: real part of magnetic
susceptibility~~~~~~~~~~~~$\hat{K}_{e}=K_{e}/10^{-13}  F^{-1}(\tau_e)$\\
$K_{e,n}\omega$: imaginary part of magnetic susceptibility by electron and nuclear spin\\
$\mu_d$: grain magnetic
moment~~~~~~~~~~~~~~~~~~~~~~~~~~~~~~~~~~~~~~~~~~~~~~$\gamma_e=\frac{g_{e}\mu_{B}}{\hbar}$: magnetogyric ratio for electron\\
$\gamma_{n}=\frac{g_{n}\mu_{n}}{\hbar}$: magnetogyric ratio nuclei\\
$\hat{\mu}_n=\mu_n/\mu_N$: normalized magnetic moment of nucleus~~~~~~~~~~$\mu_N=e\hbar/2m_pc=5.05\times 10^{-24}$~ergs G$^{-1}$\\
$J_{d}=(\frac{I_{\|}I_{\perp}kT_{d}}{I_{\|}-I_{\perp}})^{1/2}$: grain angular 
momentum at  $T=T_{d}$ ~~~~~~~$J_{therm}$: grain angular momentum at $T=T_{gas}$\\
$t_{B,nucl}^{-1}=t_{B}^{-1}+t_{nucl}^{-1}$: total nuclear relaxation time~~~~~~~~~~~~ can also include
inelastic relaxation\\
$\hat{u}_{rad}=u_{rad}/u_{ISRF}$: energy density of radiation
field~~~~~~~~~~~$\hat{\lambda}=\overline{\lambda}/1.2\mu m$: wavelength of radiation field\\
$\hat{Q}_{e3}=\mbox{Q}_{\Gamma}.\mbox{e}_{3}/10^{-2}$: third component of radiative torques~~~~~~~$E=\hat{E}/10^{-5} Vcm^{-1}$: electric field\\
$p=10^{-15} \hat{U}a_{-5}\hat{\kappa_{e}}$: electric dipole moment~~~~~~~~~~~~~~~~~~~~~~~~~~$\hat{\kappa_{e}}=\kappa_{e}/10^{-2}$: electric constant\\
$\hat{U}=U/0.3 V$: normalized voltage~~~~~~~~~~~~~~~~~~~~~~~~~~~~~~~~~~~~~~~$\hat{\omega}=\omega/10^{5}
rad~s^{-1}$: angular velocity\\
$L_{z}^{b}$: magnitude of $H_{2}$
torque~~~~~~~~~~~~~~~~~~~~~~~~~~~~~~~~~~~~~~~~~~~~~$\hat{\zeta}=\zeta/0.2$
fraction of absorbed atoms\\
$\hat{W}=W/0.2$: kinetic energy of H$_2$~~~~~~~~~~~~~~~~~~~~~~~~~~~~~~~~~~~~~$\hat{\alpha}=\alpha/10^{11}$~cm$^{-2}$: density 
of recombination sites\\
$F(\tau)\equiv
[1+(\Omega\tau/2)^2]^2$~~~~~~~~~~~~~~~~~~~~~~~~~~~~~~~~~~~~~~~~~~~~~~~~~~$\tau_n$: nuclear
spin-spin relaxation rate\\
$\tau_{el}$: electron spin-spin relaxation rate~~~~~~~~~~~~~~~~~~~~~~~~~~~~~~~~$\mu_e\approx\mu_B$;
$\mu_B\equiv e\hbar/2m_ec$: Bohr magneton\\
$f_1(s)\equiv{s^{-6}(1+s^{2})^{2}}$~~~~~~~~~~~~~~~~~~~~~~~~~~~~~~~~~~~~~~~~~~~~~~~~~~~~~$f_2(s)\equiv(\frac{1+s^{2}}{s(1-s^{2})})^{1/2}$\\

}

\end{table}

\subsection{Paramagnetic Alignment}

The Davis-Greenstein (1951)
mechanism (henceforth D-G mechanism)
is based on the paramagnetic dissipation that is experienced
by a rotating grain. Paramagnetic materials contain unpaired
electrons that get oriented by the interstellar magnetic field ${\bf B}$.
The orientation of spins causes
grain magnetization and the latter
varies as the vector of magnetization rotates
 in the grain body coordinates. This causes paramagnetic loses
at the expense of the grain rotation energy.
Be mindful, that if the grain rotational velocity ${\Omegabold}$
is parallel to ${\bf B}$, the grain magnetization does not change with time
and therefore
no dissipation takes place. Thus the
paramagnetic dissipation  acts to decrease the component of ${\Omegabold}$
perpendicular to ${\bf B}$ and one may expect that eventually
grains will tend to rotate with ${\Omegabold}\| {\bf B}$
provided that the time of relaxation $t_{D-G}$ is much shorter than
the
time of randomization through chaotic gaseous bombardment, $t_{gas}$.
In practice, the last condition is difficult to satisfy. It is clear from 
Table~1 that for $10^{-5}$ cm
grains
in the diffuse interstellar medium,
$t_{D-G}$ is of the order of $10^{13}a_{(-5)}^2 s^2 B^{-2}_{(5)}$s ,
while  $t_{gas}$ is $5\times 10^{12}n_{(20)}T^{-1/2}_{(2)} a_{(-5)}$ s if
magnetic field is $10^{-5}$ G and
temperature and density of gas are $100$ K and $20$ cm$^{-3}$, respectively.

The first detailed analytical treatment of the problem of D-G
alignment was given by Jones \& Spitzer (1967) who described the alignment
of ${\bf J}$
using the Fokker-Planck equation. This
approach allowed them to account for magnetization fluctuations
within the grain material, and thus provided a more accurate picture of the
${\bf J}$ alignment.
The first numerical treatment of
D-G alignment was presented by Purcell (1969).
By that time, it became clear that the original D-G
mechanism is too weak to explain the observed grain alignment. However,
Jones \& Spitzer (1967) noticed that if interstellar grains
contain superparamagnetic, ferro- or ferrimagnetic
inclusions\footnote{The evidence for such inclusions was found much later
through the study of interstellar dust particles captured in
the atmosphere (Bradley 1994).}, the
$t_{D-G}$ may be reduced by orders of magnitude. Since $10\%$ of
atoms in interstellar dust are iron,
the formation of magnetic clusters in grains was not far fetched
(see Martin 1995).
However, detailed calculations in Roberge \& Lazarian
(1999) showed that the degree of alignment achievable cannot account for the
observed polarization coming from molecular clouds if grains rotate thermally.
 This is the consequence of the
thermal suppression of paramagnetic alignment first discussed
by Jones \& Spitzer (1967). These internal
magnetic fluctuations
randomize grain orientation with respect to the magnetic field if the
grain body temperature is close to the rotational one.

P79 pointed out that fast rotating grains are immune to
both gaseous and internal magnetic
randomization. Thermal trapping that we discussed in \S 3.2
limits the range of grain sizes
for which Purcell's torques can be efficient (LD99ab).
For grains that are less than the critical size, which can be $10^{-4}$~cm
and larger, rotation is essentially thermal (see section 3.2). The alignment of such grains
is expected to be in accordance with the DG mechanism predictions (see
Lazarian 1997,
Roberge \& Lazarian 1999), and seem to be sufficient to explain the residual alignment
of small grains that is seen in the Kim \& Martin (1995) inversion (see \S 6.5).

Lazarian \& Draine (2000) predicted
that PAH-type particles can be aligned paramagnetically due to the relaxation
that is faster than the DG predictions. In fact, they showed that the DG
alignment is not applicable to very swiftly rotating particles, for which
the Barnett magnetic field gets comparable to magnetic fields induced by uncompensated
spins in the paramagnetic material. For such grains, this relaxation is more efficient than the
one considered  by Davis \& Greenstein (1951). This effect, that is termed ``resonance relaxation''
in Lazarian \& Draine  (2000), allows the alignment of PAHs. These tiny ``spinning'' grains
are responsible for the anomalous foreground
microwave emission (Draine \& Lazarian 1998,
see also Lazarian \& Finkbeiner 2003 for a review).

\subsection{Mechanical Alignment}

The Gold (1951) mechanism is a process of mechanical alignment of grains.
Consider
a needle-like grain interacting with a stream of atoms. Assuming
that collisions are inelastic, it is easy to see that every
bombarding atom deposits with the grain an angular momentum $\delta {\bf J}=
m_{atom} {\bf r}\times {\bf v}_{atom}$,
which is directed perpendicular to both the
needle axis ${\bf r}$ and the
 velocity of atoms ${\bf v}_{atom}$. It is obvious
that the resulting
grain angular momenta will be in the plane perpendicular to the direction of
the stream. It is also easy to see that this type of alignment will
be efficient only if the flow is supersonic\footnote{Otherwise grains
 see atoms coming not from one direction, but from a wide cone of
directions (see Lazarian 1997a) and the efficiency of
alignment decreases.}.

Suprathermal rotation introduced in Purcell (1979) persuaded researchers
that mechanical alignment is marginal. Indeed, it seems natural to accept
that fast rotation makes
it difficult for gaseous bombardment to align grains. However, the
actual story is more interesting. First of all, it was proven that
 mechanical alignment of suprathermally rotating grains
is possible (Lazarian 1995). Two mechanisms that were termed
``crossover'' and ``cross section'' alignment were introduced
there. The mechanisms were further elaborated and quantified
in Lazarian \& Efroimsky (1996), Lazarian, Ozik \& Efroimsky (1996),
Efroimsky (2002b). Second, as we discussed in \S 3.3, the supersonic velocities
are available over substantial regions of interstellar medium,
both due to MHD turbulence and ambipolar diffusion.

In fact, the discovery of thermal trapping (\S 3.2) made the original
Gold (1951) mechanism more relevant. Therefore when grains are not thermally
trapped and rotate suprathermally
the crossover and cross section alignments should take place, while
for thermally trapped grains the original Gold mechanism remains in force. The quantitative
numerical study of the Gold alignment in Roberge et al. (1995) was
done under the assumption of  the perfect
coupling of ${\bf J}$ with the axis of maximal inertia (cf \S 3.1).
This study shows a good
correspondence with an analytical formulae for the alignment
of ${\bf J}$ vector in L94 when
the gas-grain velocities are transsonic. An analytical study in
Lazarian (1997) accounts for the incomplete internal alignment in
a more sophisticated way, compared to L94, and
predicts the Rayleigh reduction factors of $20\%$ and more for grains
interacting with the Alfven waves. A detailed numerical study would be in order
to test the predictions.

\subsection{Radiative Torque Alignment}

Anisotropic starlight radiation can both spin the grains and align them.
This was first realized by Dolginov \& Mytrophanov
(1976), that
radiative torques are bound to induce alignment.  In their paper they
considered a tilted oblate grain with the helicity axes coinciding with
the axis of maximal inertia, as well as a tilted prolate grain for which
the two axes were perpendicular. They concluded, that subjected to a radiation
flux, the tilted oblate grain will be aligned with longer axes perpendicular
to magnetic field, while the tilted prolate grain will be aligned with the
longer axes parallel to magnetic field. At that time
the internal relaxation was not yet a part of accepted grain dynamics.
The problem was revisited by Lazarian (1995), who took into account
the internal relaxation and concluded that both prolate  and oblate grains
will be aligned with longer axes perpendicular to the magnetic field. However,
Lazarian (1995) did not produce quantitative calculations and underestimated
the relative importance of radiative torque alignment compared to other
mechanisms.

\begin{figure}[h]
\includegraphics[width=3.3in]{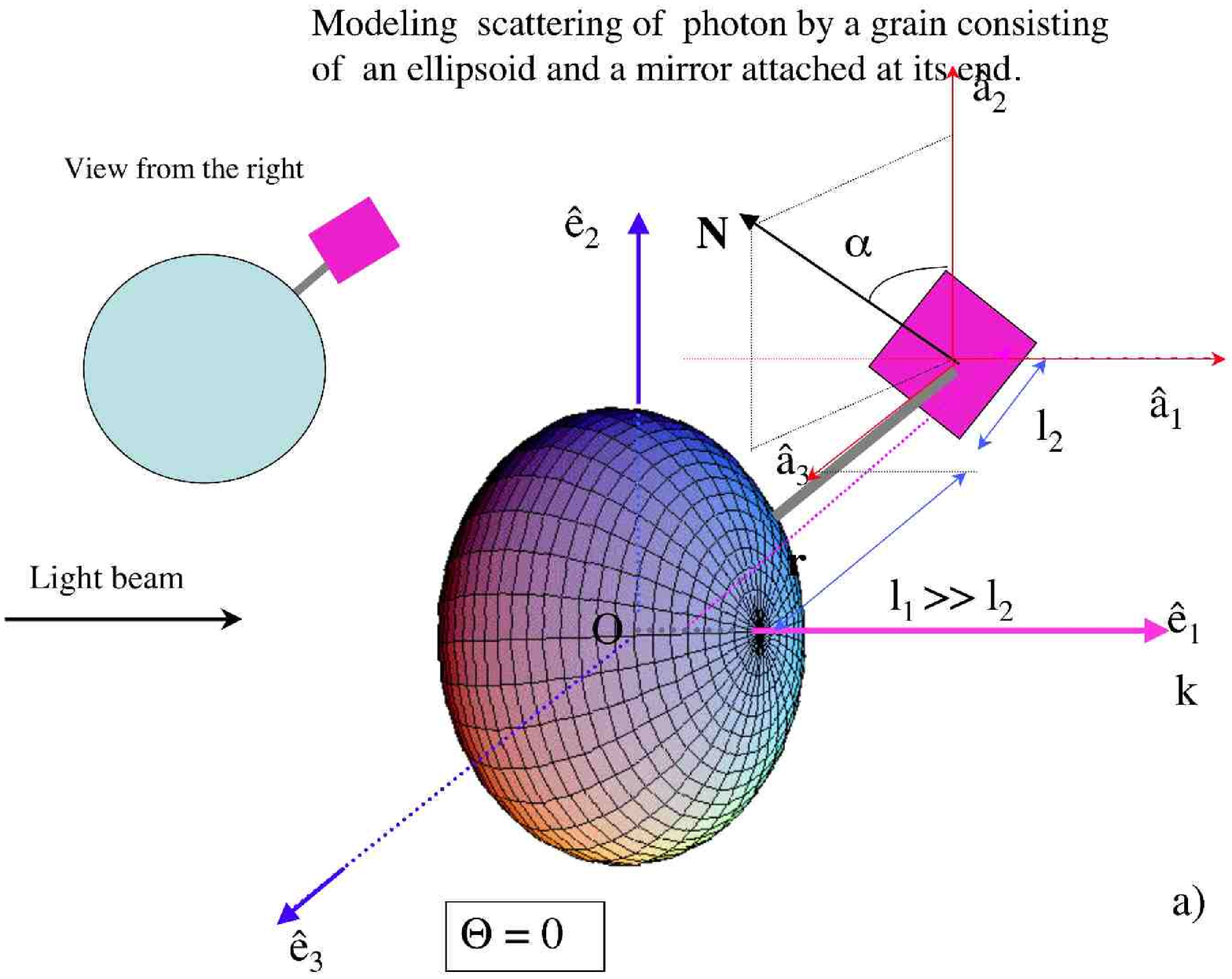}
\hfill
\includegraphics[width=2.5in]{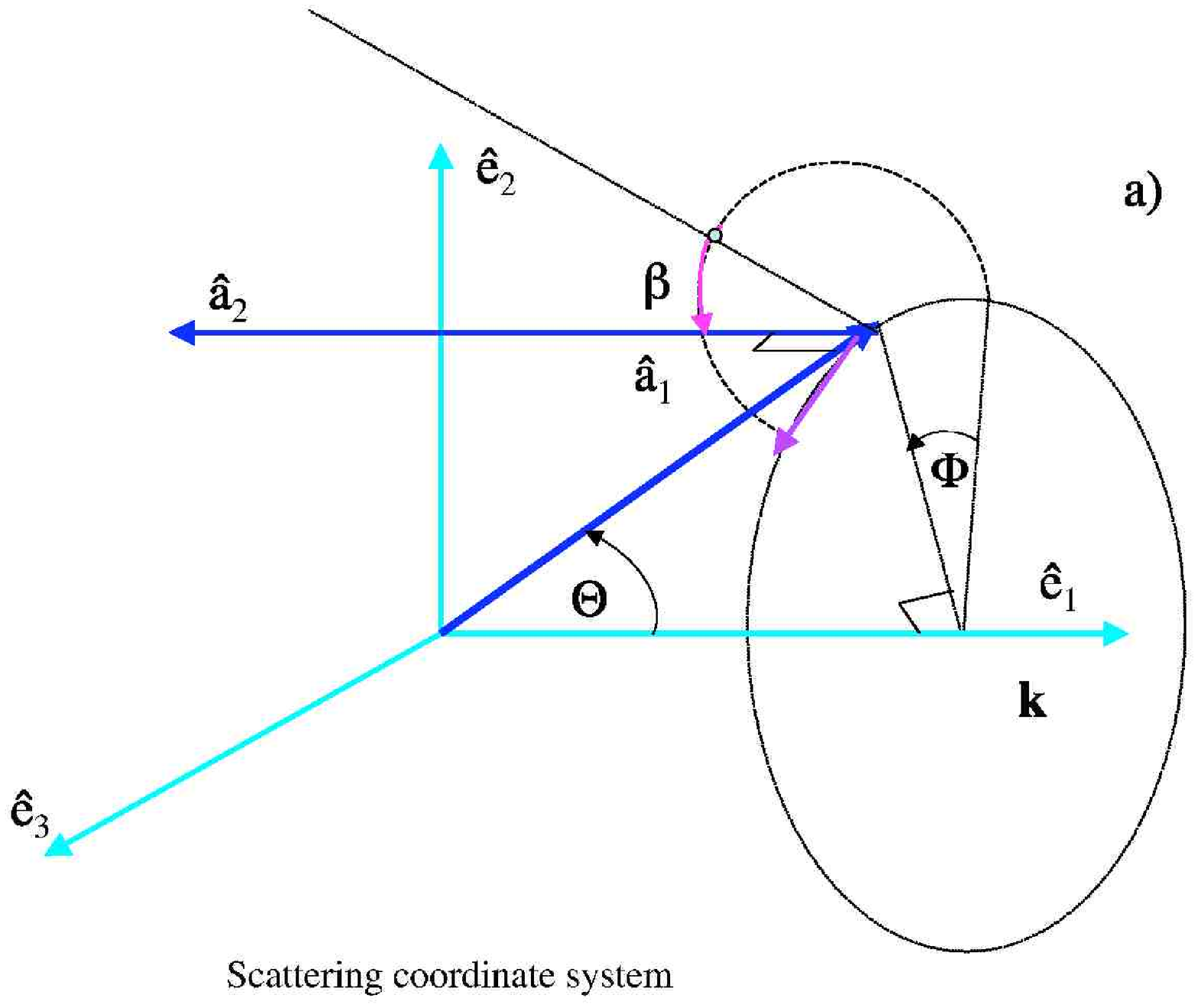}
\caption{
\small
{\it (a) Left panel}.-- A model of a ``helical'' grain,
that consists of a spheroidal grain with an inclined mirror attached to it,
reproduces well the radiative torques (from LH06).
 {\it (b) Right panel}.-- The ``scattering coordinate system'' which
illustrates the definition of torque components: ${\bf a_1}$ is directed
along the maximal inertia axis of the grain; ${\bf k}$ is the direction of radiation.
The projections of normalized radiative torques $Q_{e1}$,
$Q_{e2}$ and $Q_{e3}$ are calculated in this reference frame for $\Phi=0$.}
\label{fig:2Dspek}
\end{figure}

It happened that the Dolginov \& Mytrophanov (1976) study came before its time.
The researchers themselves did not have reliable tools
to study the dynamics of irregular grains and the impact of their work was initially low.
Curiously enough, Purcell studied the aforementioned work, appreciated the Barnett
magnetization described there, but did not recognized the importance to the
radiative torques. In fact, he had means to calculate them numerically
using the Discreet Dipole Approximation (DDA) code available to him.

The explosion of interest to the radiative torques we owe to
Bruce Draine, who realized that the torques
can be treated with the DDA
code by Draine \& Flatau (1994) and modified the code correspondingly.
The magnitude of torques were found to be substantial and present
for grains of all irregular shapes studied in Draine 1996, DW96 and DW97. After that it became impossible to ignore the radiative torque alignment. More recently, radiative
torques have been studied in laboratory conditions (Abbas et al. 2004).

Potentially, the isotropic radiative torques could ensure suprtathermal
rotation and provide the alignment in the spirit of P79
mechanism. Indeed, radiative torques are related to the volume of the
grain. Therefore a deposition of a monolayer of atoms over the grain surface, i.e.
resurfacing, that can reverse the direction of Purcell's torques, does
not affect the radiative torques. Long-lived suprathermal torques may
ensure efficient paramagnetic alignment. In this way the idea of radiative
torques is presented in a number of research papers. This
way of thinking about radiative torque alignment is {\it erroneous}, however.

In fact, isotropic torques are fixed in grain coordinates and in all
respect are similar to the Purcell's torques. Therefore,
typical interstellar grains driven only by isotropic radiative torques
cannot rotate suprathermally due to the thermal trapping effect that we
discussed in \S 3.2. 

Moreover, in most cases the radiation field that we deal with has
an appreciable anisotropic component. This component induces
 torques that can align grains. DW97 study confirmed that the torques tend to
align grains with long axes perpendicular to magnetic field.

Objectively, the DW96 and DW97 papers signified a qualitative change in
the landscape of grain alignment theory. These papers claimed that radiative
torques alignment may be the dominant alignment mechanism in the diffuse
interstellar medium. However, questions about the nature of the alignment
mechanism, the particular choice of grains studied,
as well as the efficiency of radiative torques
in different environments remained. In addition, the DW97 treatment ignored
the physics of crossovers (see \S 3.2).
In view of that, I recall my
 conversations with Lyman Spitzer, who was excited about
the efficiency of radiative torque, but complained that he was lacking
a clear physical picture of the alignment mechanism.

\begin{figure}[h]
\includegraphics[width=1.7in]{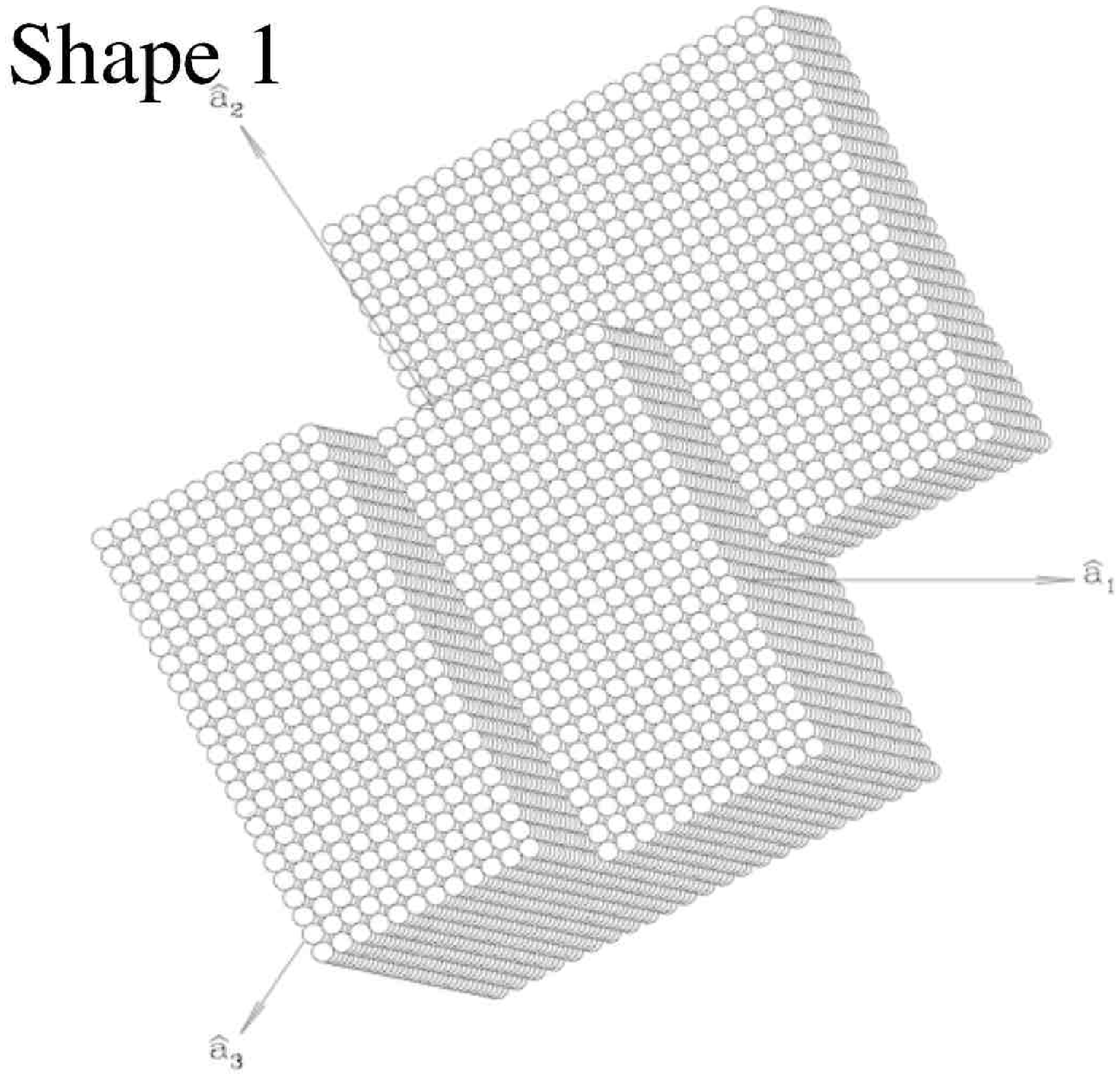}
\includegraphics[width=1.7in]{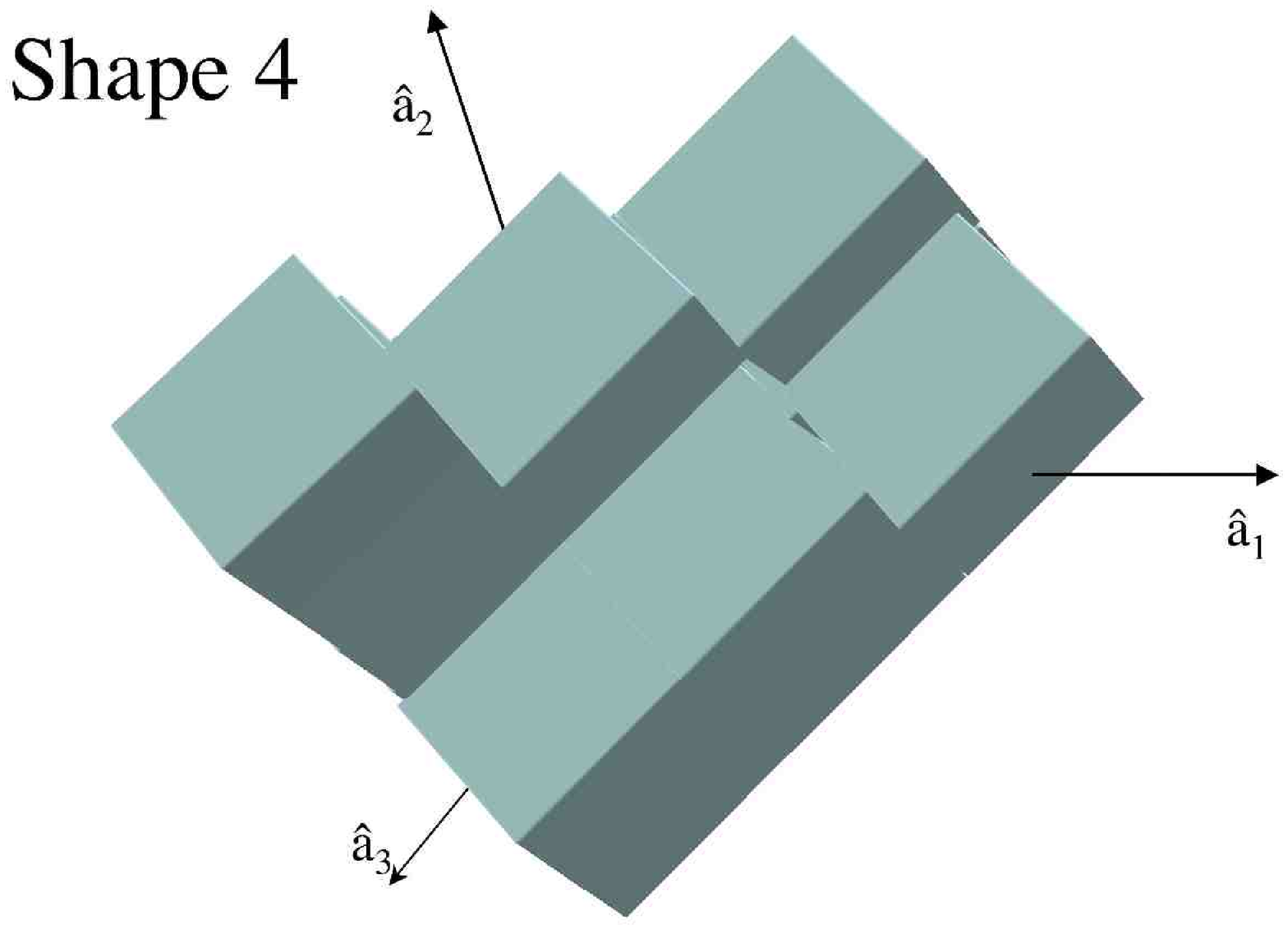}
\includegraphics[width=1.7in]{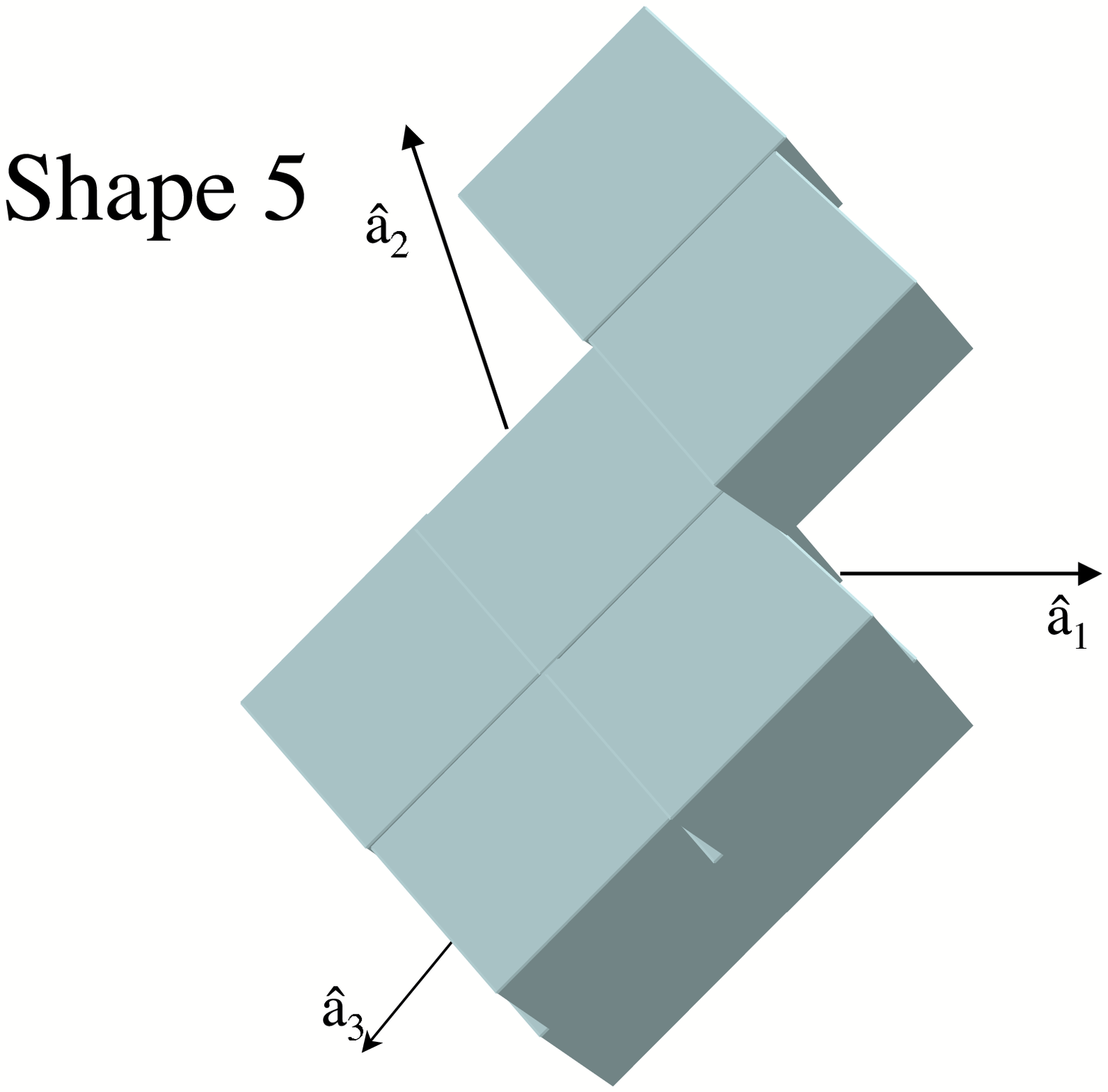}
\caption{\small Examples of irregular shapes studied in LH07.}
\end{figure}
\begin{figure}[h]
\includegraphics[width=3.in]{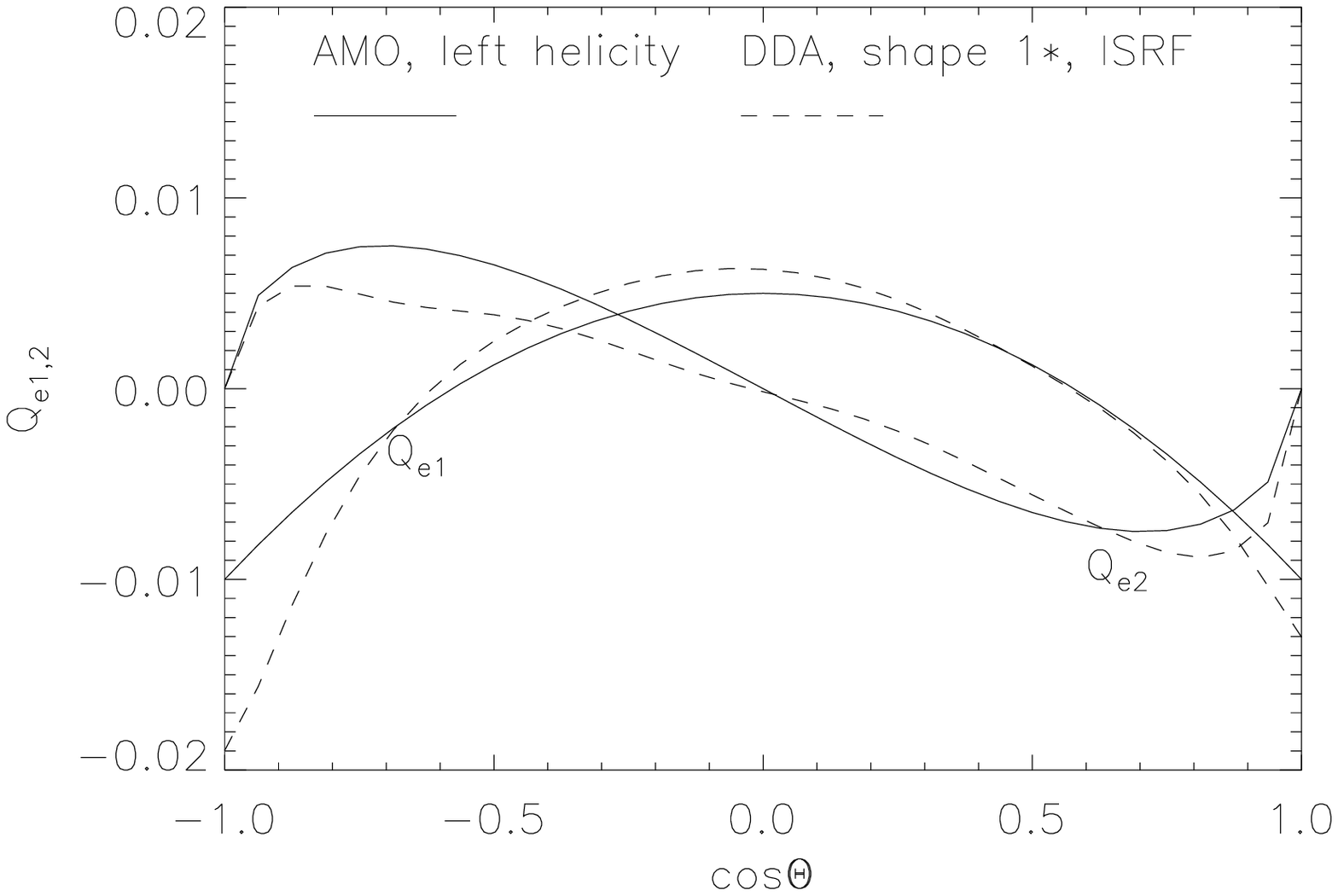}
\hfill
\includegraphics[width=3.in]{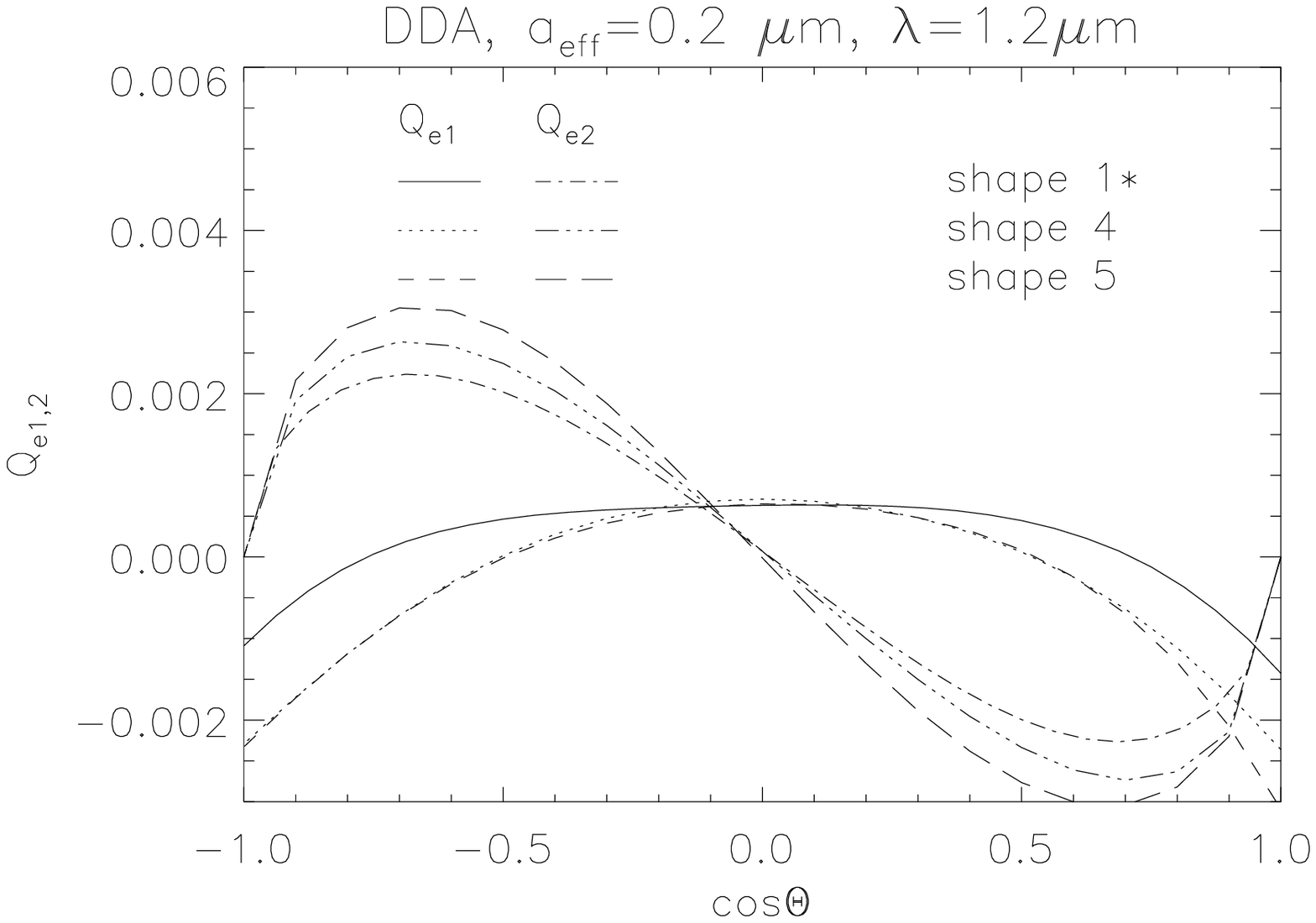}
\caption{
\small
{\it (a) Left panel}.-- Two components of the radiative torques
are shown for our
 analytical model (solid lines) in Fig.~7a and for an irregular
grain in Fig.~8 (dashed lines).
 {\it (b) Right panel}.-- Radiative torques for different grain shapes. From
Lazarian \& Hoang (2006). }
\end{figure}

To address this concern LH07
proposed a simple
model that reproduces well the essential basic properties of radiative
torques. The model consists of an oblate grain with a mirror attached to
its side (see Fig.~7a). This model allows an analytical treatment and provides
an physical insight why irregular grains get aligned. In fact, it shows that
for a range of angles between the radiation and the magnetic
field the alignment gets ``wrong'', i.e. with the
long axes parallel to magnetic field. However, this range is rather narrow (limited
to radiation direction nearly perpendicular to magnetic field) and
tends to disappear in the presence of internal wobbling (see \S 3.1).

In LH07 we concluded that the alignment of grains with longer axes perpendicular to magnetic 
field lines is a generic property of radiative torques that stems
from the basic symmetry properties of the radiative torque components.
Our work showed that the entire description of alignment may be obtained with
the two components of the radiative torques $Q_{e1}$ and $Q_{e2}$ as they
are defined in the caption of Fig.~7. The third component $Q_{e3}$ is
responsible for grain precession only.
The functional dependences of the torque components
 that are experienced by our model grain are similar to those
experienced by irregular grains shown in Fig.~8. It is really remarkable that
our model and grains of very different shapes have very similar
functional dependences of their torque components (see Fig.~9)!
 Note that the particular set of
grains is ``left-handed''. For ``right handed'' grains
both $Q_{e1}$ and $Q_{e2}$ change simultaneously in a well defined manner.
For our grain model to become ``right handed''
the mirror should be turned  by 90 degrees.

The phase trajectories in Fig.~10 show that only a small fraction of
grains get to attractor points with high angular momentum. It is most
probable for a arbitrary chosen grain to end up at the attractor point
that, in the absence of grain thermal wobbling and
gaseous bombardment, corresponds to $J\rightarrow 0$. Within this model it
is only natural to get grain aligned with $J\sim J_d$ when
thermal wobbling is included, as this is observed in WD03 (see \S 3.2).

\begin{figure}[h]
\includegraphics[width=3.in]{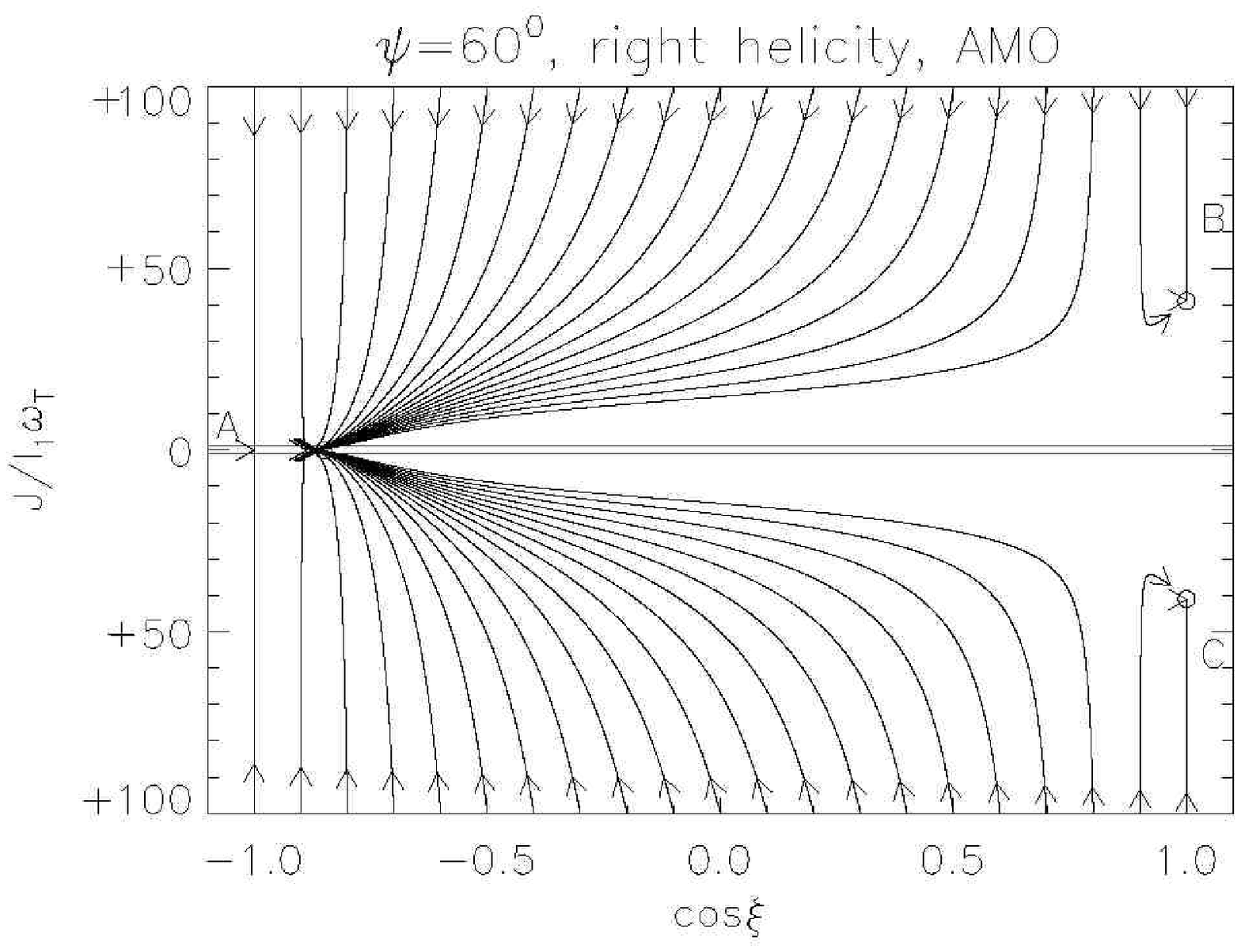}
\hfill
\includegraphics[width=3.in]{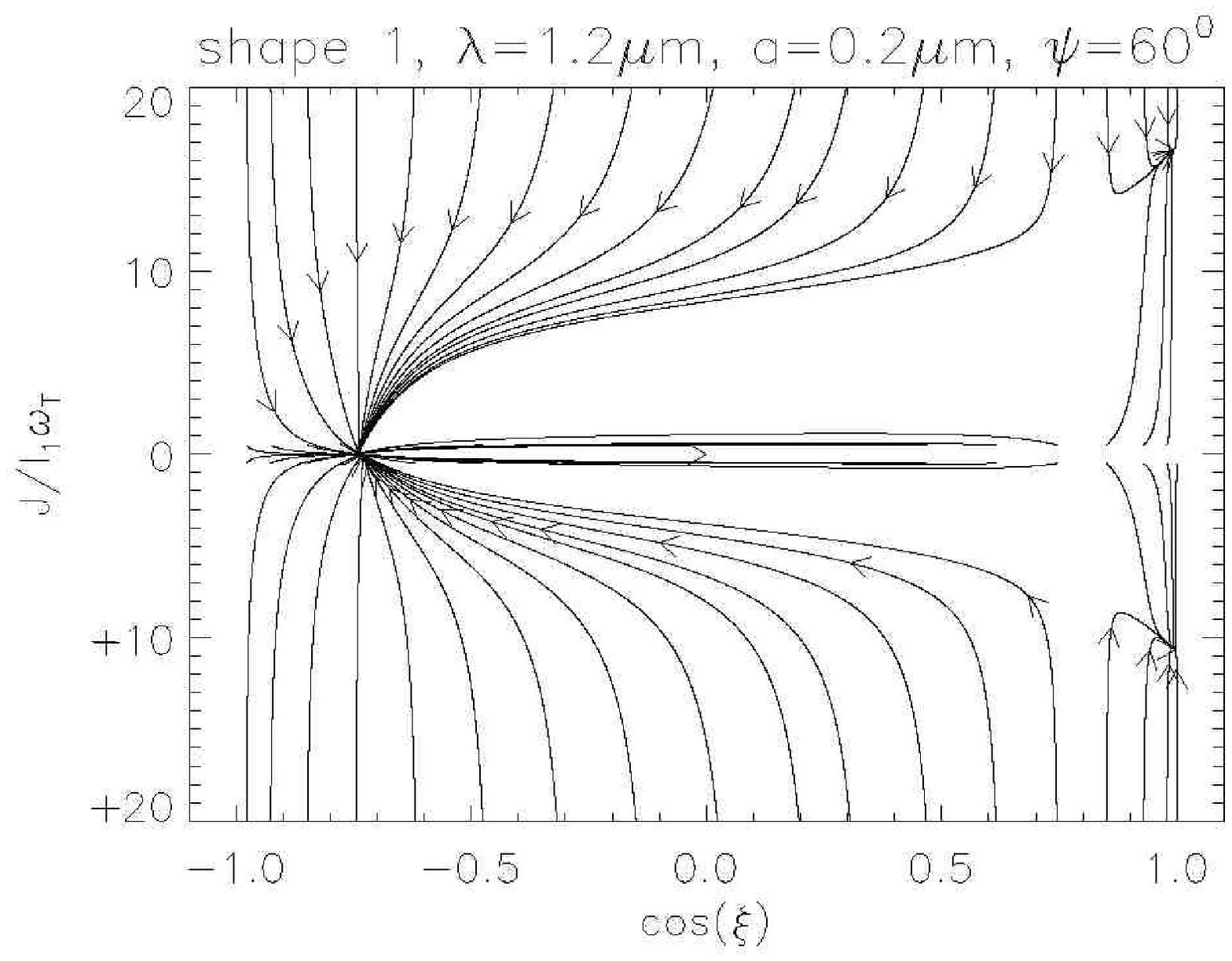}
\caption{
\small
{\it (a) Left panel}.-- Phase trajectory map obtained for the
model grain given shown in Fig.~7.
 {\it (b) Right panel}.-- The same for an irregular grain in Fig.~8 (shape 1).
 From Lazarian
\& Hoang (2006).}
\label{fig:2Dspek}
\end{figure}

What does make grains helical? Both rotation about a
well defined axis and grain irregularity do this. For instance, 
if we attach the weight-less mirror to
a sphere rather than an oblate body, this would 
average out the radiative torques as the mirror will be constantly changing 
its orientation in
respect to the rotational axis.

\subsection{Sub-sonic Mechanical Alignment as Next Challenger}

As we mentioned earlier, the requirement of the supersonic drift
limits the applicability of mechanical alignment. Such drift is, however,
not necessary for helical grains. The model grain in Fig.~7a is helical not
only in respect to radiation, but also to mechanical flows (see also LH07).
 In fact, the
functional dependence of the torques that we obtain for our model grain does
not depend on whether photons or atoms are reflected from the mirror. Therefore
we may predict that, if atoms bounce from the grain surface elastically, the helical
grains\footnote{The mechanical alignment of helical grains was briefly
discussed in Lazarian (1995) and
 Lazarian, Goodman \& Myers (1997), but was not elaborated there.} will
align with long grain axes perpendicular to the flow
in the absence of magnetic field.  If the
dynamically important magnetic field is present, the alignment is
expected with long axes perpendicular to ${\bf B}$.
 If atoms stick to the
grain surface and then are ejected from the place of their impact, this changes the values
of torques by a factor of order unity.

It is easy to understand why supersonic drift is not required for helical
grains. For such grains 
the momentum deposited by regular torques scales in proportion to the number of collisions,
while the randomization adds up only as a random walk. In fact, the difference
between the mechanical alignment of spheroidal and helical grains is similar
to the difference between the Harwit (1971) alignment by stochastic absorption
of photons and the radiative torque alignment. While the
Harwit alignment requires very special
conditions to overpower
 randomization, the radiative torques acting on a helical
grain easily beat randomization.

Similarly, as in the case of the radiative torques, it is
possible to disregard the Harwit process, it may be
possible to disregard
the Gold-type processes (see Table~2 and \S 4.3) for irregular grains.
As the grain
helicity does not change sign during grain flipping, the thermal trapping effects described in LD99a
are absent for the mechanical spin-up of helical grains.

The properties of helical grains require detailed studies.
 For instance, in the presence of Purcell's thrusters and no flipping (see Fig.~4a), the helical grain
may induce
its own translational motion as it rotates.

What would it take to make a grain helical for mechanical interactions? This
is a question similar to
one that worried researchers with the radiative torques before
Bruce Draine made his simulations. We do not have the simulations of mechanical
torques on irregular grains, but in analogy with the radiative torques, I would
claim that such torques should be generic for an irregular grain, provided that
there is a correlation of the place where an atom hits the grain and where it
evaporates from the grain. It is intuitively clear that
the effects of helicity should be more important for larger grains. As the
relative gas-grain drift induced by gyroresonance (Yan \& Lazarian 2003) is faster for larger grains
this can be used as another argument for relatively better alignment of
large helical grains.

As the physics of helical grain alignment and those previously known mechanical
alignment
mechanisms is different, we can talk of a completely new process of alignment
that can be tentatively termed  ``sub-sonic mechanical alignment''
to stress its independence of supersonic drift. The
traditional {\it supersonic} mechanical alignment mechanisms we discussed in \S 4.3
tend to minimize grain cross section. This
means, for instance, that for grains streaming along magnetic fields,
the stochastic torques tends to align grains with longer axes
parallel to magnetic field. On the contrary, our study in LH07 showed that the
mechanical torques on helical grains tend to align grains
in the same way as the radiative torques do, i.e., 
the helical grains will tend to be aligned perpendicular to magnetic field
irrespectively of the direction of the drift. Further work should
show in what situations the ``sub-sonic
mechanical alignment'' can reveal magnetic fields when
radiative torques fail to do this.

All in all, our considerations above suggest that the helicity is an
intrinsic property of rotating irregular grains and therefore the
mechanical alignment of helical grains
 should  overwhelm any mechanical alignment
process discussed in \S 4.3 when the two mechanism tend to align grains
in opposite directions. This raises questions of
whether we can ever expect to have alignment with grain long axes parallel
to magnetic field (cf. Rao et al. 1998, Cortes et al. 2006).
Can the alignment of helical grains fail? This can happen, for instance,
in the absence of correlation of the impact and evaporation sites of impinging
atoms. This issue can be clarified by laboratory studies.

\section{Dominant Mechanism: Progress and Problems}

\subsection{Niches for Mechanisms and Quantitative Theory}

It is clear that the major alignment mechanisms discussed in \S 4 have their own niches.
For instance, Davis-Greenstein mechanism should be important for
small paramagnetic grains as the ratio of the paramagnetic alignment
rate to the gaseous randomization rate scales inversely proportional to
grain size (see Lazarian \& Draine 2000). At the same time,
the most promising mechanism, the radiative torque one,
is not efficient for
sufficiently small grains (i.e. $\lambda \gg a$). We summarize the current
situation with the known alignment mechanisms by Table~2. Conservatively,
we did not include in the table the mechanical
alignment of helical grains, an interesting mechanism that have not been properly studied yet.

If grains are superparamagnetic (Jones \& Spitzer 1967, Mathis 1986),
 they can be aligned,
provided that their rotational temperature is larger than the grain
temperature. As the rate of paramagnetic relaxation for ``super''
grains is larger than the rate of collisional damping, it is this
faster rate that should strongly affect the phase trajectory of grains subjected
to radiative torques.

We showed that the
gyroresonance acceleration of grains discussed in \S 3.3  allowed
an efficient acceleration of grains to supersonic velocities. Note, that
the processes
 enabling a supersonic drift have been the stumbling block for the
mechanism. In this sense the Gold-type mechanisms for thermally rotating grains
and crossover and cross section mechanisms for suprathermally rotating grains
might look currently promising. However, the competition with the mechanical
alignment of helical grains and the radiative torque alignment
limits the range of circumstances where the process dominates.

The radiative torque alignment looks the most promising at the moment. As we have
discussed in \S 4.3 it allows predictions that correspond well to observational
data. Nevertheless, both the
observational
testing of the theory and the improvement of
the radiative torque ``cookbook'' are essential.
Some of the required improvements are obvious. For instance, the
position of the low-$J$ attractor points (see Fig.~10), at which most of the aligned grains reside,
show variations with the grain shape. Therefore to
predict the expected alignment measure, i.e., $R$ (see eq.~\ref{R}), more precisely,
 one may need to consider a variety of grain shapes. The calculation of radiative torques for
a given radiation spectrum, a given distribution of grain sizes and
a variety of shapes
 is a challenging computational task. Fortunately,
LH07 showed that, with satisfactory
accuracy the radiative torques demonstrate self-similarity, i.e.
can be presented as a function of
$\lambda/a$ only (see also Cho \& Lazarian 2006).

\begin{figure}[h]
\includegraphics[width=6.in]{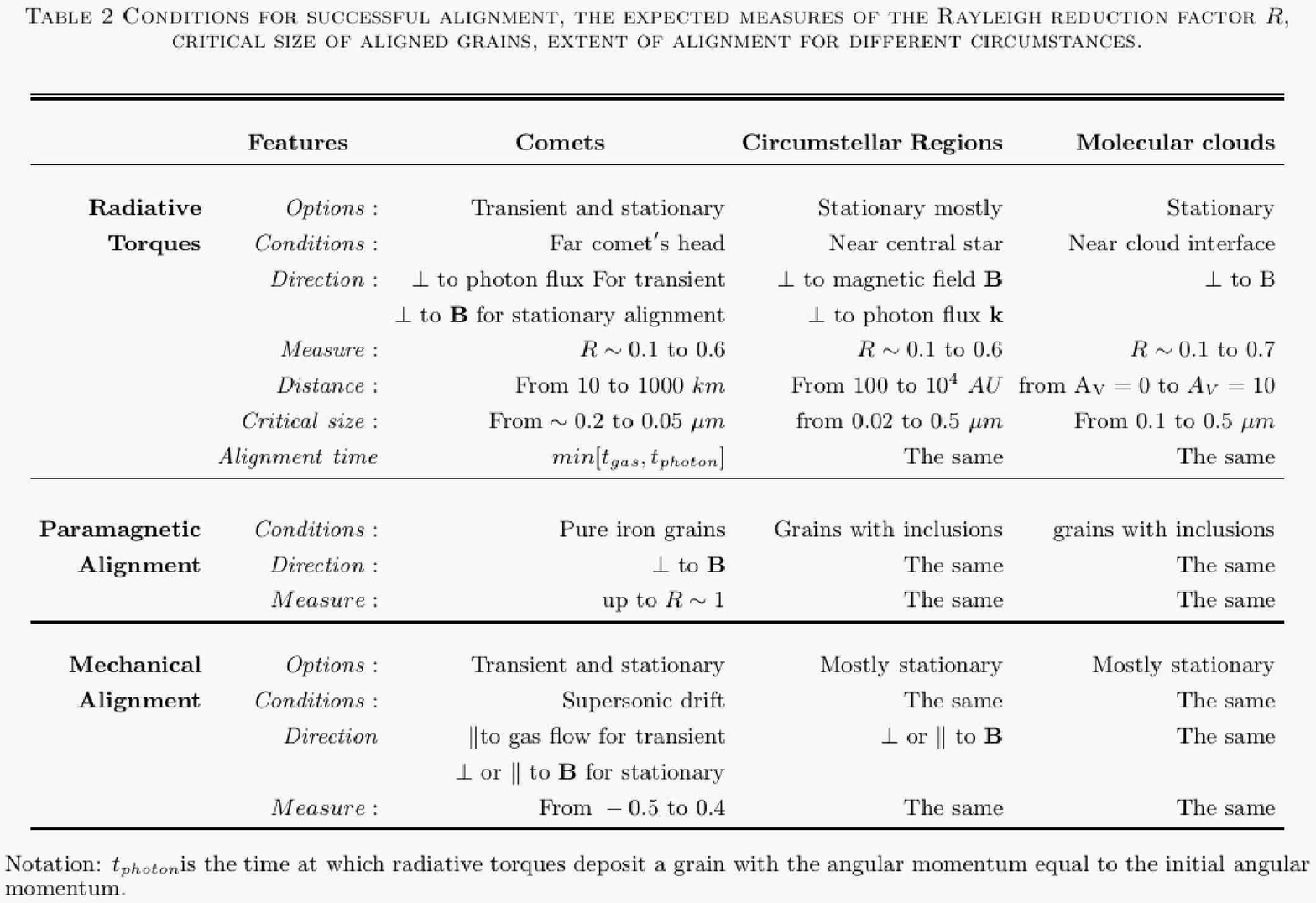}
\end{figure}

The quantitative theory for different mechanisms is at different stages of development.
For some mechanisms, e.g. for the Davis-Greenstein alignment, the
theory is detailed and well-developed for spheroidal grains (see Roberge \& Lazarian 1999 and
references therein).
There are reasons to believe that these results should be applicable also
to realisticly irregular grains. However, at the moment the mechanism does not
looks promising for alignment of grains larger than $2\times 10^{-6}$~cm. The paramagnetic alignment theory for suprathermally rotating grains had
been developed before the discovery of thermal flipping and thermal trapping effects (Lazarian
\& Draine 1999ab). Therefore
the model of alignment
 in Lazarian \& Draine (1997) is applicable to grains larger than a critical size $a_c$
which is approximately $10^{-4}$~cm. The
relative role of the Purcell suprathermal torques and the radiative torques requires further
studies for various astrophysical environments. Cho \& Lazarian (2005) claimed that 
the radiative torques are dominant for the molecular cloud interiors where
large grains are present. Similarly, the domain of applicability of the suprathermal mechanical
alignment (Lazarian 1995, Lazarian \& Efroimsky 1996, Lazarian et al. 1996, Efroimsky 2002a)
is also limited by the grains larger than $a_c$.

The radiative torque alignment mechanism has undergone dramatic changes in the last 
10 years. From the mostly forgotten one it has risen to the dominant one. The alignment has
been studied for grains assuming perfect alignment (DW97, LH07), as well as taking into
account thermal fluctuations (WD03, Hoang \& Lazarian 2007). Moreover, the process of alignment
is not any more a result of numerical experimentation. A simple analytical model in LH07 does
reproduce the essential features of the alignment. However, a more rigorous studies of the
effects of the incomplete internal alignment on radiative torques are necessary. An approach
based on the elimination of the fast variable presented in Roberge (1997) seems promising
if we want to get precise measures for the grain alignment (see eq.~{\ref{R})). 

Obtaining alignment measures when several alignment processes act simultaneously is another
challenge for quantitative studies. It has been addressed in Roberge et al. (1995) numerically
and in Lazarian (1997) analytically for the situation when the 
mechanical and paramagnetic alignment mechanisms
act simultaneously. In reality, a number of possible combinations is higher and the 
interaction of different mechanisms may be very non-linear. For instance, radiative torques 
can prevent some grains from thermal flipping thus changing the conditions for other mechanisms
to act. The studies in WD03, LH07 and Hoang \& Lazarian (2007)
show that the fraction of such suprathermal grains is not large, however.

\subsection{New Situations, New Challenges}

As the grain alignment theory matures, it starts to deal with a wider variety
of astrophysical situation, rather than just interstellar grains. This
opens new opportunities for astrophysical magnetic fields studies, but also poses new
challenges.

Consider, for instance, the alignment of grains in accretion disks. The
grains their may be very large, up to ``pebble'' size. As grains get larger
the physics of their alignment changes\footnote{The observations of very
large aligned grains is a separate issue that we do not dwell upon.
If aligned grains are much larger than the wavelength of observations, they do
not produce polarized signal. This means that to study the alignment of
large grains one should increase the wavelength of observations. The magnetic
field mapping with aligned large grains may
require taking into account polarized synchrotron foreground.}. For instance, for grains larger
than $10^{-3}$cm the mechanical alignment arising from the difference in
the positions of the center of pressure and the center of gravity, the
so-called ``weathercock mechanism'' (Lazarian 1994b), gets important.
In addition, for larger grains, the internal alignment through nuclear
relaxation gets subdominant compared to inelastic relaxation (P79,
Lazarian \& Efroimsky 1999). Eventually, all internal alignment mechanisms
get inefficient.
This is a regime that earlier researchers, who were unaware of internal
relaxation processes, dealt with (see Dolginov
\& Mytrophanov 1976).

Interestingly enough, some earlier abandoned mechanisms may get important
in new situations. Take, for instance, the ``iron fillings'' mechanism,
that considers alignment of iron needles along magnetic fields. This
mechanisms proposed
by Spitzer \& Tukey (1951) at the very beginning of the grain alignment
studies, may still
be important if grains are sufficiently large and magnetic fields are strong.

Environments for alignment may be quite exotic. For instance, it is a good bet to disregards electric fields in interstellar gas. However,
it may not always be true. According to private communication
from Jim Hough electric fields could align dust grains in the Earth atmosphere.
Serezhkin (2000) estimated electric fields that may be present in
comet comas. This opens a completely new avenue for research. Indeed,
first of all, electric fields can serve as  the ``axis of alignment'' provided that
grains have dipole moments\footnote{Even in the absence of electric field
grain dipole moments can affect grain dynamics (see Draine \& Lazarian 1998b, Yan et al. 2004,
Weingartner 2006).}  (see a discussion of the latter point in
Draine \& Lazarian 1998). Thus, the radiative torque, subsonic and supersonic
mechanical alignment processes can happen in respect to the electric field.
Then, an analog of the ``iron fillings'' alignment
is possible, especially, if grains have properties of electrets (Hilczer \& Malecki 1986). Moreover, an
electric analog of paramagnetic relaxation
 is possible as grains rotate in electric field. Some materials, e.g. segnetoelectrics (see Mantese and Alphay 2005), are
particularly dissipative and can act the same way as superparamagnetic inclusions
act to enhance the efficiency of the D-G relaxation.

The issue of the direction of alignment requires care when the parameters
of the environment changes. For instance, it was discussed in
Lazarian (2003) that the alignment in typical interstellar medium conditions
would happen in respect to magnetic field, irrespectively of
the mechanism of alignment. This is the consequence of the fast Larmor
precession.
Even if magnetic field changes its direction
over the time scales longer compared to the Larmor period, the angle
between ${\bf J}$ and local ${\bf B}$ is preserved as the consequence
of the preservation of the adiabatic invariant. Note, that depending on the mechanism,
the  grains
 may align with their longer axes either {\it perpendicular} or {\it parallel}
to magnetic field, however.

Other situations when magnetic field is not the axis of alignment are also possible.
Consider, for instance, radiative torques. Whether the radiation direction
or magnetic field is the axis of alignment depends on the
precession rate around these axes (see Table~1).
For instance, in the vicinity of stars the grains can to get aligned in
respect to the radiation flux, however. For a star the radius
at which the light acts as the axis of alignment changes from
of $1$ AU for magnetic field of $10^{-3}$~G to $10^3$ AU for
the field of $10^{-6}$~G (LH07). Light flashes from supernovae explosions
may impose the direction of the photon flux as the alignment axis over larger scales.
At the same time, one can check that for typical diffuse ISM (see Table~1) is
 $t_{L}/t_{RT} \sim 10^{-3}$, i.e. the Larmor precession is much faster than
precession induced by radiation. Therefore magnetic field stays
 the alignment axis as it was assumed in the earlier work.

Similarly, gas streaming can induce its own alignment direction.
Dolginov \& Mytrophanov (1976) assumed that
whether magnetic field or a gaseous flow defines the axis of
alignment depended on the ratio of Larmor precession time to
that of mechanical
alignment. LH07, however, concluded that the precession time of
a grain in a gaseous flow (an analog of $t_{RT}$ in Table~1) should be taken instead. The 
latter time is orders of magnitude less than the time assumed in
Dolginov \& Mytrophanov (1976). As the result, high density
molecular outflows can overwhelm the magnetic field
and impose its direction as the direction
of alignment, provided that the directions of the
outflow and magnetic field do not coincide (LH07).   Interestingly enough,
the mechanical flows can define the axis of alignment even for
subsonic flow velocities, i.e. at those velocities for which the
process considered by Dolginov \& Mytrophanov (1976) is not efficient.

Other processes may also be important in more restricted situations. Consider,
 for instance, the grain spun-up by cosmic rays (Sorrell 1995ab).
The calculations by Lazarian \& Roberge (1997b) show, that for the cosmic-ray-induced
torques to be important, the enhancement of the low energy 
cosmic ray flux over its typical interstellar value by a factor of 
more than $10^3(10^{-5}~{\rm cm}/a)$ is necessary. Therefore this process could only be 
important over localized regions near cosmic ray sources.

\subsection{Avenues for Theory Advancement}

It is easy to notice that both
studies of irregular grains and subtle physical
effects have provided the major boost for the grain alignment theory. Indeed,
the theory started with the favorite with physicists ``spherical cow'' model,
which literally corresponded to the assumption of spherical grains in
D-G model. Later, the studies of the alignment of oblate and prolate grains have been undertaken.
However, completely new effects were revealed when irregular grains were
considered. Indeed, the helicity, which is the key ingredient for both the
operation of the radiative torques (see \S 4.4) and the subsonic mechanical alignment (see \S 4.5), is zero for spheroidal grains.

Similarly, an adequate treatment is necessary for grain properties.
Originally grains were considered as
solid absolutely rigid passive bricks without internal structure. 
It is only later, that effects of elasticity
as well as magneto-mechanical effects were considered. The back-reaction of
thermal fluctuations on grain dynamics through these effects changed
drasticly our understanding of both grain dynamics and alignment.
Improvements in this direction can be obtained by accounting for the triaxial
 ellipsoids of inertia corresponding to irregular grains. Some work in this
direction has been already done for the inelastic relaxation (see
Efroimsky 2000).

We believe that more effects will be considered as grain alignment theory
matures and is being applied to new astrophysical environments.
For instance, we have discussed above, that potentially grain
surface physics may be essential
for the mechanical alignment of helical grains. Plasma-grain interactions
seem to be another promising direction, which has been marginally developed
so far (see Draine \& Lazarian 1998b, Yan et al. 2004, Shukla \& Stenflo 2005).

\section{Polarimetry and Grain alignment}

\subsection{Grain Alignment in Molecular Clouds}

Polarization arising from aligned grains
provides a unique source of information about magnetic
fields in molecular clouds. For many years this has been the
most important practical motivation for developing the grain
alignment theory.

The data on grain alignment in molecular clouds looked at some point
very confusing. On one hand, optical and near-infrared polarimetry
of background stars did not show an increase of polarization degree
with the optical depth starting with a threshold of the order of a few (Goodman et al. 1995,
Arce et al. 1998). This increase
would be expected if absorbing grains were aligned by magnetic field
within molecular clouds.
On the other hand, far-infrared measurements (see Hildebrand 2000, henceforth
H00)
showed strong polarization that was consistent with emission from
aligned grains. A quite general explanation to those facts was given
in Lazarian, Goodman \& Myers (1997, henceforth LGM97), where it was argued that
all the suspected alignment mechanisms are based on non-equilibrium processes
that require free energy to operate. Within the bulk of molecular clouds
the conditions are close to equilibrium, e.g. the temperature difference
of dust and gas drops, the content of atomic hydrogen is substantially
reduced, and the starlight is substantially attenuated. As the result
the major mechanisms fail in the bulk part of molecular clouds apart
from regions close to the newly formed stars as well as the cloud
exteriors that can be revealed by far-infrared polarimetry.

The alternative explanations look less appealing. For
instance, Wiebe \& Watson (2001) noted that small scale turbulence
in molecular clouds can reduce considerably the polarization degree even if
grain alignment stays efficient. This, however, is inconsistent with
the results of the
far-infrared polarimetry that
revealed quite regular pattern of magnetic
field in molecular clouds (see H00).

An extremely important study of alignment efficiency
has been undertaken by Hildebrand and his
coworkers (Hildebrand et al. 1999, Hildebrand 2000, 2002). They pointed out that
for a uniform
sample of aligned grains, made of dielectric material consistent
with the rest of observational data, polarization degree, $P(\lambda)$, should stay constant
for $\lambda$ within the far-infrared range. The data
at 60 $\mu$m, 100 $\mu$m from Stockes on the Kuiper Airborne Observatory,
350 $\mu$m from Hertz on Caltech Submillimeter Observatory, and
850 $\mu$m from SCUBA on the JCMT revealed a very different picture.
This was explained (see Hildebrand 2002) as the evidence for the
existence of the populations of dust grains with different temperature
and different degree of alignment. The data is consistent with cold
(T=10~K) and hot (T=40~K) dust being aligned, while warm (T=20~K) grains
being randomly oriented (H00).
If cold grains are identified
with the outer regions of molecular clouds, hot grains with regions
near the stars and warm with the grains in the bulk of molecular clouds
the picture gets similar to that in LGM97.

\begin{figure}[h!t]

\includegraphics[width=.46\textwidth]{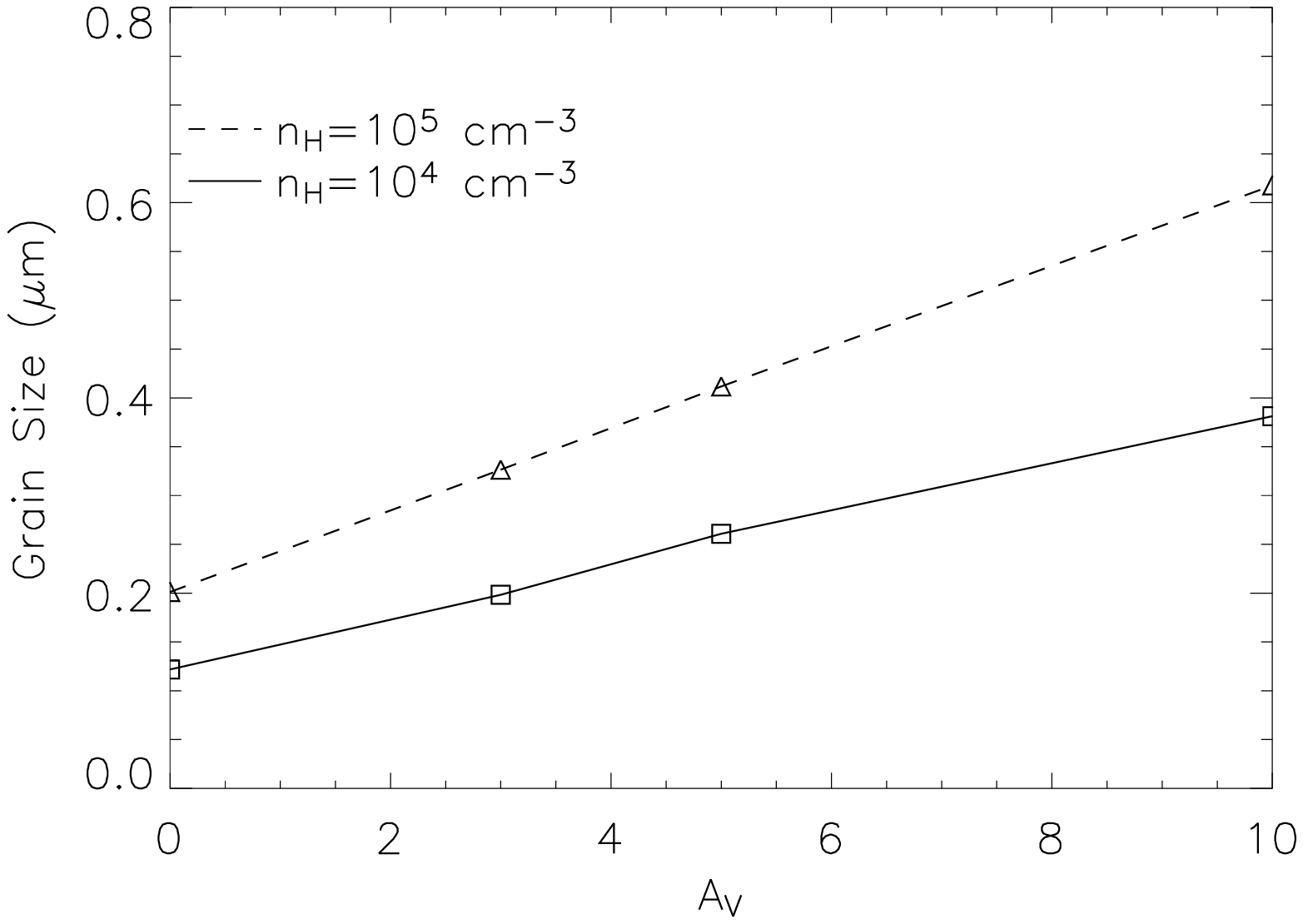}
\includegraphics[width=.48\textwidth]{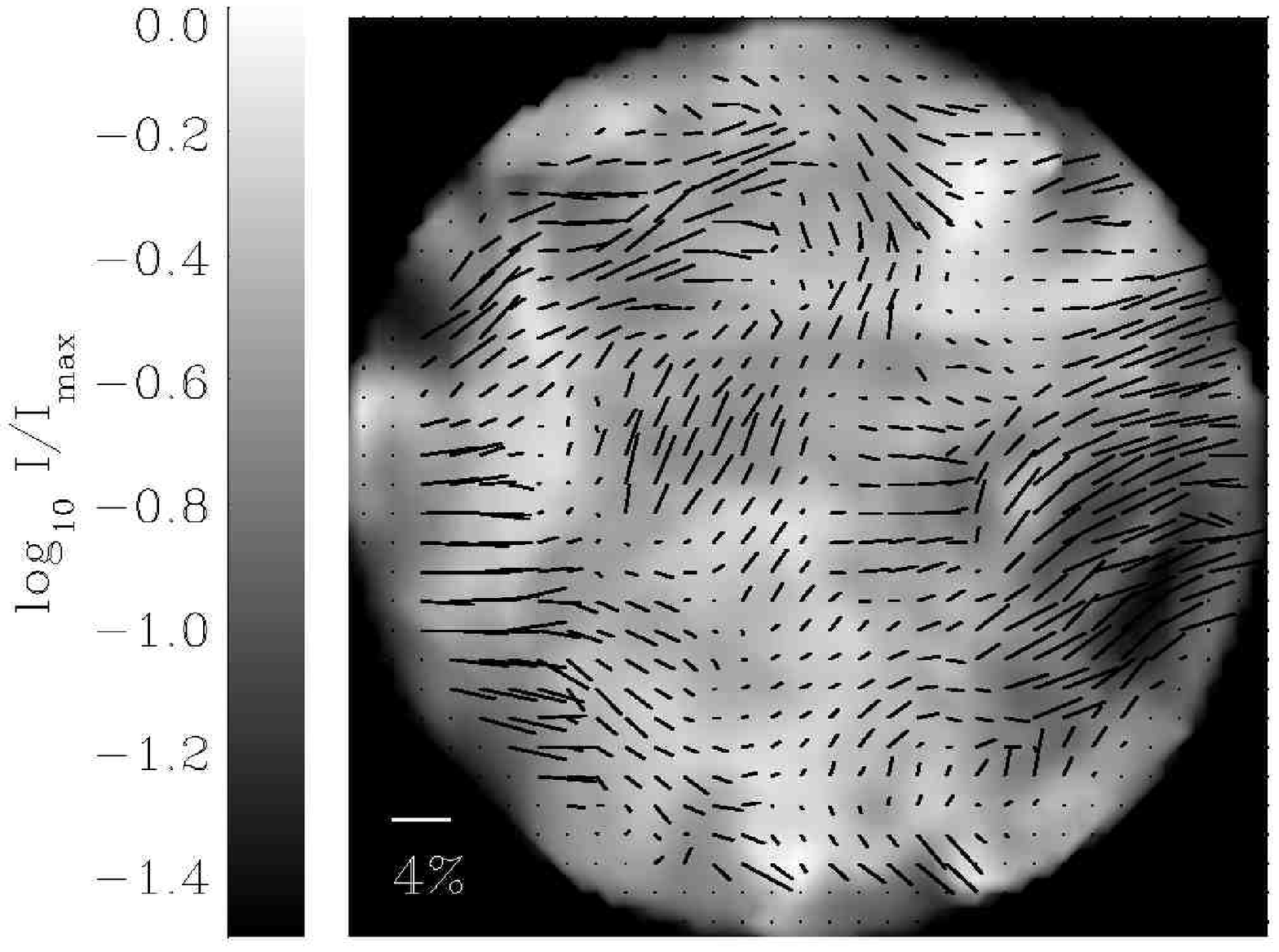}
\caption{
\small
{\it (a) Left Panel}: Aligned grain size vs. visual extinction $A_V$. For the threshold
suprathermal angular velocity 5 times larger than the thermal angular
velocity was chosen. It is clear that increase of grain size can compensate
for the extinction of light in cloud cores.
 Solid line: $n_H=10^4 cm^{-3}$; Dotted line: $n_H=10^5 cm^{-3}$ in the cloud.
(from Cho \& Lazarian 2005)
{\it (b) Right Panel}: The 850 $\mu$m emission map of the model cloud.
Superimposed are the projected polarization vectors (from Bethell et al. 2006).
}
\end{figure}

However, the data obtained for pre-stellar cores in Ward-Thompson et al.
(2000) seem to be at odds with the LGM97 predictions.
Indeed, the properties of these cores summarized in Ward-Thompson et al.
(2002) and Crutcher et al. (2004) fit into the category of zones 
that, according to LGM97, should not contain aligned grains. 

What could be wrong with the LGM97 arguments? The latter paper treats grains
of $10^{-5}$~cm size. The
grains in prestellar cores can be substantially larger. Grain alignment 
efficiency depends on grain size. Therefore the estimates in LGM97 had to 
be reevaluated.

Cho \& Lazarian (2005, henceforth CL05) revealed a steep dependence of
radiative torque
efficiency on grain size. While an earlier study by Draine \& Weingartner
(1996) was limited by grains with size $a \leq 2 \times 10^{-5}$~cm,
CL05 studied grains up
to $3\times 10^{-4}$~cm size subjected to the attenuated radiative field calculated
in accordance with the prescriptions in Mathis, Mezger \& Panagia (1983).
Fig.~11a shows that large grains can be efficiently span up by radiative
torques even at the extinction of $A_v$ of 10 and higher.  A numerical
treatment of the radiative transfer was used in the papers
that followed, e.g. Pelkonen et al. (2006), Bethell et al. (2006) (see
Fig.~11b and 12).

\begin{figure}[htb]

\includegraphics[width=.58\textwidth]{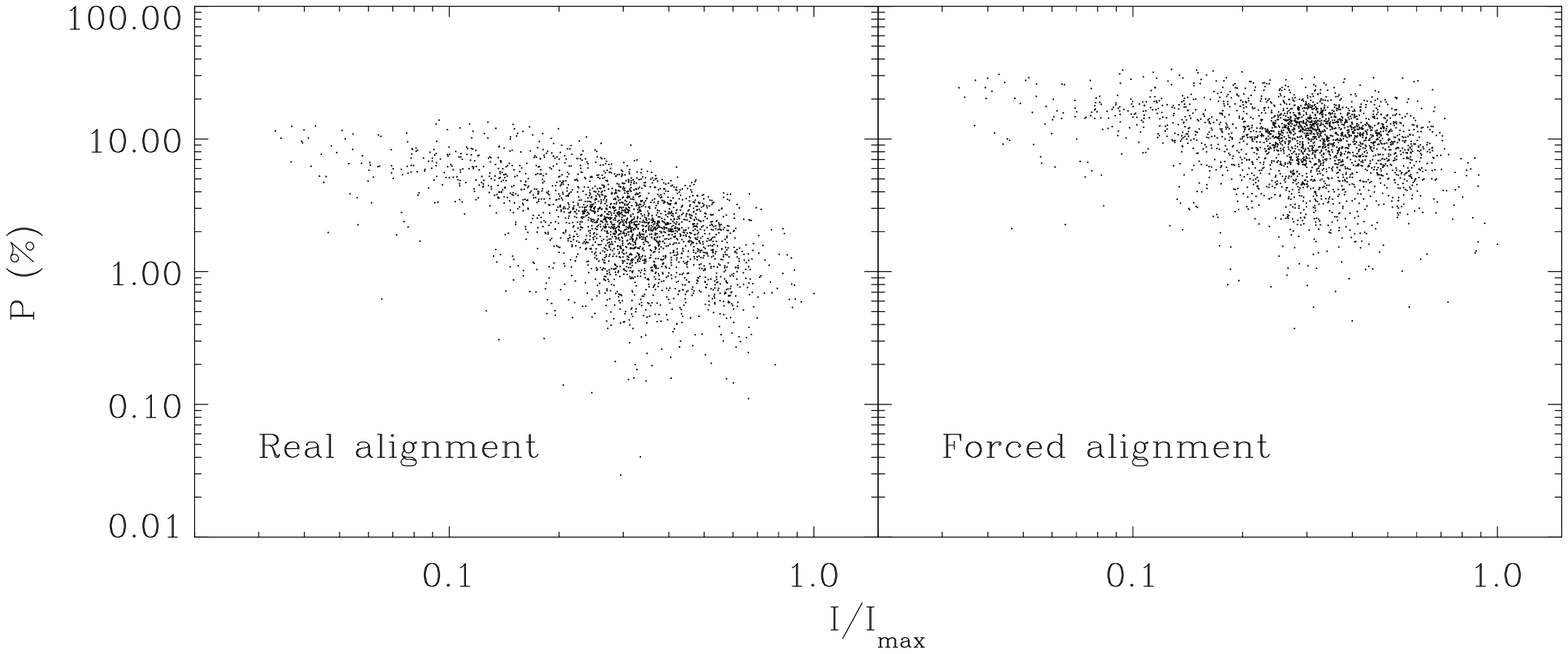}
\includegraphics[width=.37\textwidth]{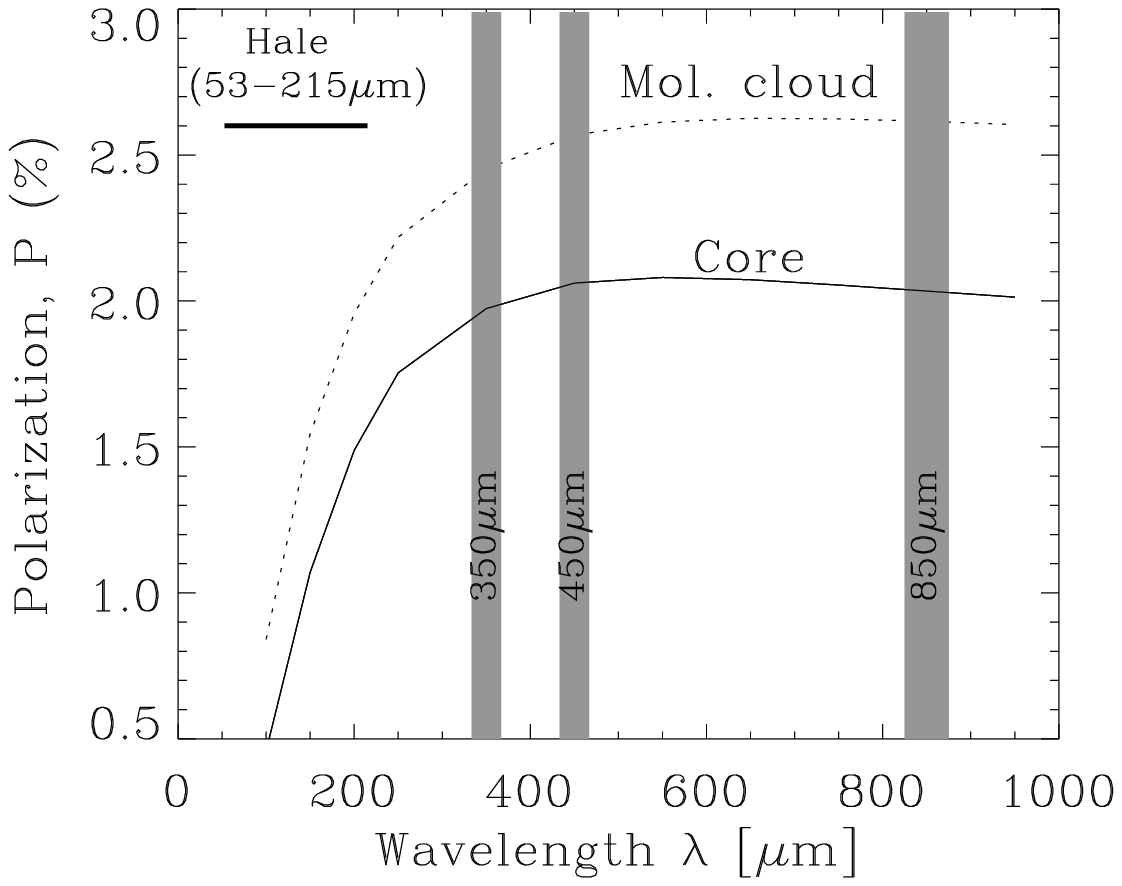}
\caption
{
\small
{\it (a) Left Panel}.-- The polarization degree  for 850 $\mu$m emission
from a cloud 
as a function of normalized emission intensity
for
 the actual calculated degrees of alignment (real alignment) and assuming that all grains are
perfectly aligned.
{\it (b) Right Panel}.-- The polarization spectra of a model core and
a ``starless'' molecular  cloud. The projected Hale polarimeter wave-band coverage is
also shown. From Bethell et al. (2006).
}
\label{fig1}
\end{figure}

Fig.~12a illustrates that a naive assumption of the perfect alignment results
in a substantial overestimation of the polarization degree. While the
polarization spectra in Fig.~12b is obtained for a starless core/cloud, a more
non-trivial behavior is expected for a cloud with active star formation. This
calls for multi-frequency observations (see also H00).

We note, that in CL05 and the subsequent papers the efficiencies of radiative
torques in terms of alignment were parameterized in terms of maximal rotational
velocities $\Omega_{max}$ achievable by  the torques. As we discussed in \S 4.4, most of
the interstellar grains do rotate thermally in the presence of
radiative torques. Nevertheless, the above parameterization does characterize
the relative role of the randomizing atomic collisions and aligning effects
of the radiative torques. Our tests that include simulated
gaseous bombardment in Hoang \& Lazarian (2007)
 show that grains are being aligned by radiative torques when $\Omega_{max}>
3 \Omega_{thermal, gas}$.

\subsection{Testing Alignment at the Diffuse/Dense Cloud Interface}

The grain alignment theory can be directly tested at the cloud interface.
 Mathis (1986) explained the dependence of the polarization
degree versus wavelength , namely the Serkowski law (Serkowski 1973) (see
also Fig.~13a)
\begin{equation}
P(\lambda)/P_{max}=exp \left(-K ln^2(\lambda_{max}/\lambda) \right)~,
\end{equation}
(where $\lambda_{max}$ corresponds to the peak percentage polarization
$P_{max}$ and $K$ is a free parameter),
assuming that it is only the grains larger than
the critical size that are aligned. Those grains were identified in
Mathis (1986) as having superparamagnetic inclusions and therefore subjected
to more efficient paramagnetic dissipation. 

The ratio of the total to
selective extinction $R_v\equiv A_v/E_{B-V}$ reflects the mean size of
grains present in the studied volume. It
spans from $\sim 3.0$ in diffuse ISM to $\sim 5.5$ in dark clouds
(see Whittet 1992 and references therein) as
the mean size of grain increases due to coagulation or/and mantle
growth. The earlier studies were consistent with the assumption that the
growth of $R_v$ was accompanied by the corresponding growth of
$\lambda_{max}$ (see Whittet \& van Brenda 1978). The standard interpretation
for this fact was that as grains get bigger, the larger is the critical
size starting with which grains get aligned. This interpretation was
in good agreement with Mathis' (1986) hypothesis of larger grains having superparamagnetic inclusions.
However, a more recent study by Whittet et al. (2001) showed that grains
at the interface of the Taurus dark cloud do not exhibit the
correlation of $R_v$ and $\lambda_{max}$. This surprising behavior
was interpreted in Whittet et al (2001) as the result of {\it
size-dependent variations in grain alignment} with small grains losing
their alignment first as deeper layers of the cloud are sampled.
Whittet et al (2001) did not specify the alignment mechanism,
but their results posed big problems to the
superparamagnetic mechanism (see \S 4.2). Indeed, the data is suggestive that
$R_v$ and therefore the mean grain size may not grow  with extinction
while the critical size for grain alignment grows.

\begin{figure}[htb]

\includegraphics[width=.46\textwidth]{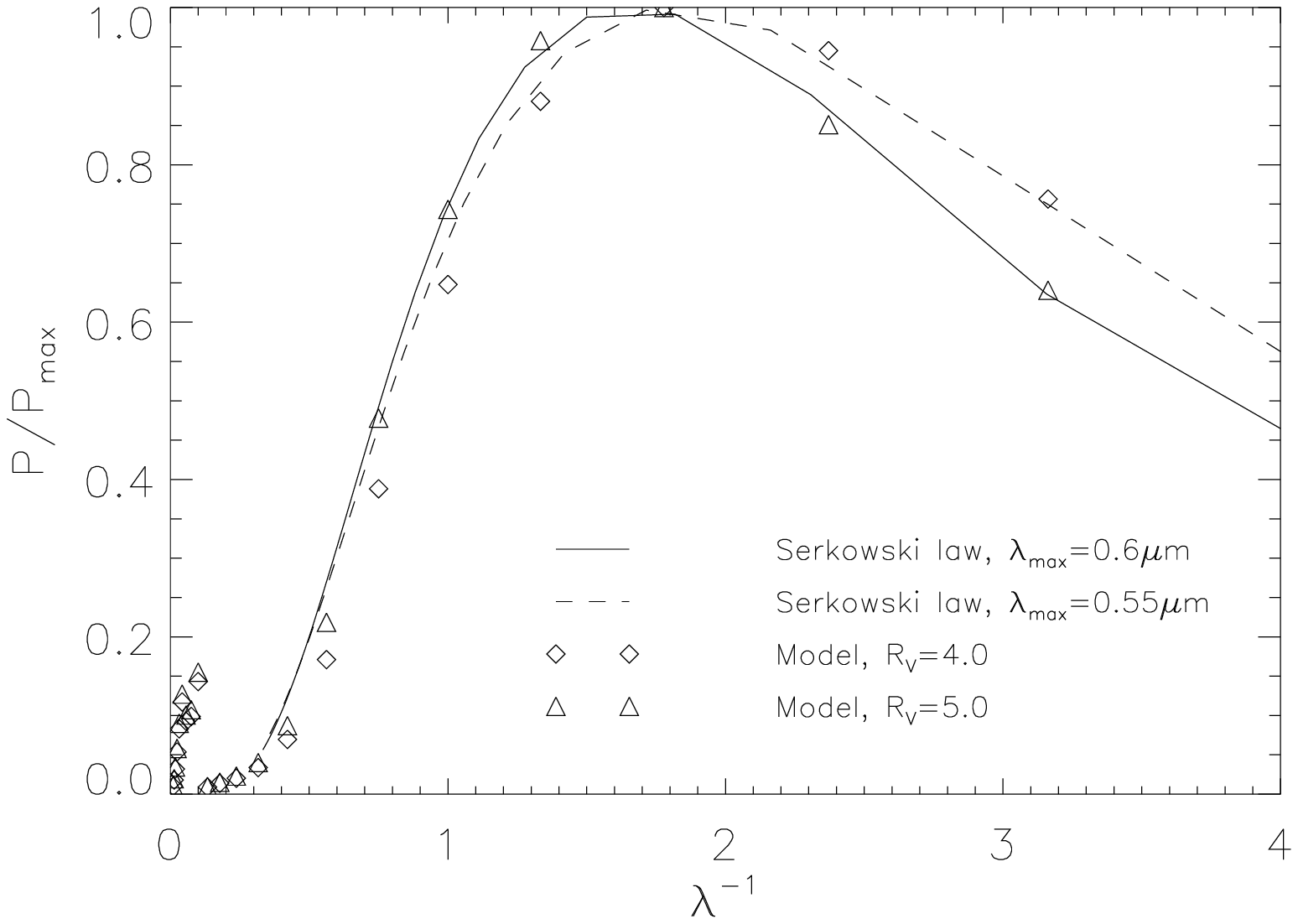}
\includegraphics[width=.48\textwidth]{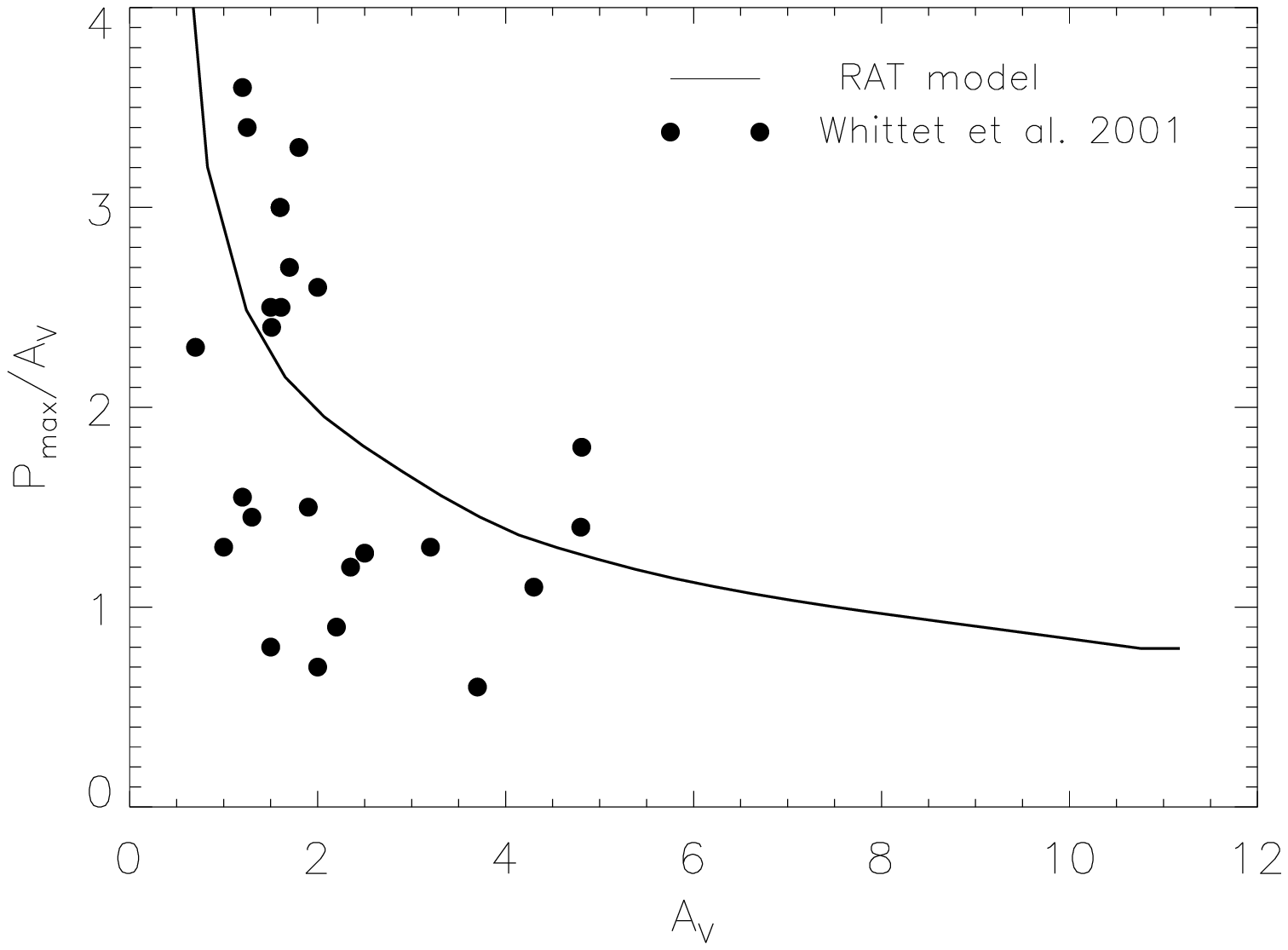}
\caption
{
\small
{\it (a) Left Panel}: Serkowski curves and fits by radiative
torque models. {\it (b) Right Panel}:
$p_{max}/A_{V} $ as function of $A_{V}$ from our calculations
with radiative torques (solid line) and the
observation data by Whittet et al. (2001). The interface is simulated as a
 homogeneous slab. The MRN distribution of dust with
$a_{max}=0.35 \mu m$ was used. From Hoang \& Lazarian (in preparation).
}
\end{figure}

Lazarian (2003) noticed that the Whittet data agrees well with the
expectations of the radiative torque mechanism. We present in Fig.~13b
our recent fit for the data using the radiative torques that arise
from the attenuated interstellar radiation field.

\subsection{Alignment in Magnetized Disks around stars}

Magnetic field plays important roles in the evolution of protostellar disks.
Magnetic pressure can provide extra support to the disks and
magnetic field can promote removal of angular momentum from disks (see
Velikov 1959; Chandrasekhar 1961; Balbus \& Hawley 1991).
However, there are many uncertainties in the structure and the effects
of the magnetic field in protostellar disks. Quantitative studies of magnetic fields
in the disks are essential therefore.

Consider T Tauri stars first.
Tamura et al. (1999) detected polarized emission from T Tauri stars, which are
low mass protostars. Aitken et al. (2002) studied polarization that can
arise from magnetized accretion disks.
They considered a single grain component consisting of
the 0.6$\mu m$ silicate and used an ad hoc assumption
that all grains at all optical depths are  aligned
with $R=0.25$.

Cho \& Lazarian (2006) used a more sophisticated model for grain alignment.
They calculated the radiative torques acting on grains, assuming the
model of the disk in Fig.~14a. The results of their calculations are shown
in Fig.~14b. It is clear that with multiwavelengths observations it should
be possible to separate the  contributions arising from the disk surface and interior.

\begin{figure*}[h!t]
\includegraphics[width=.38\textwidth]{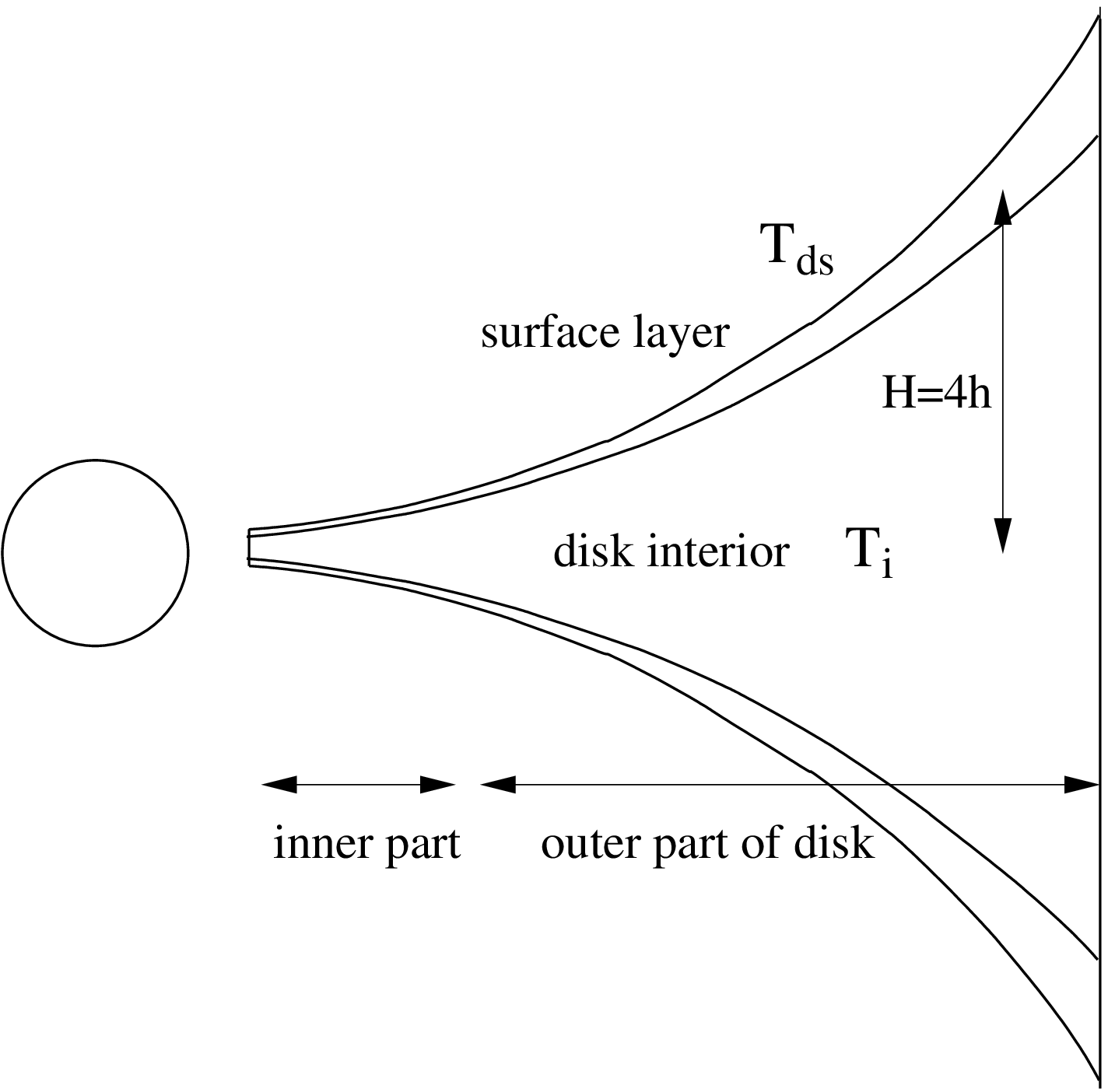}
\includegraphics[width=.50\textwidth]{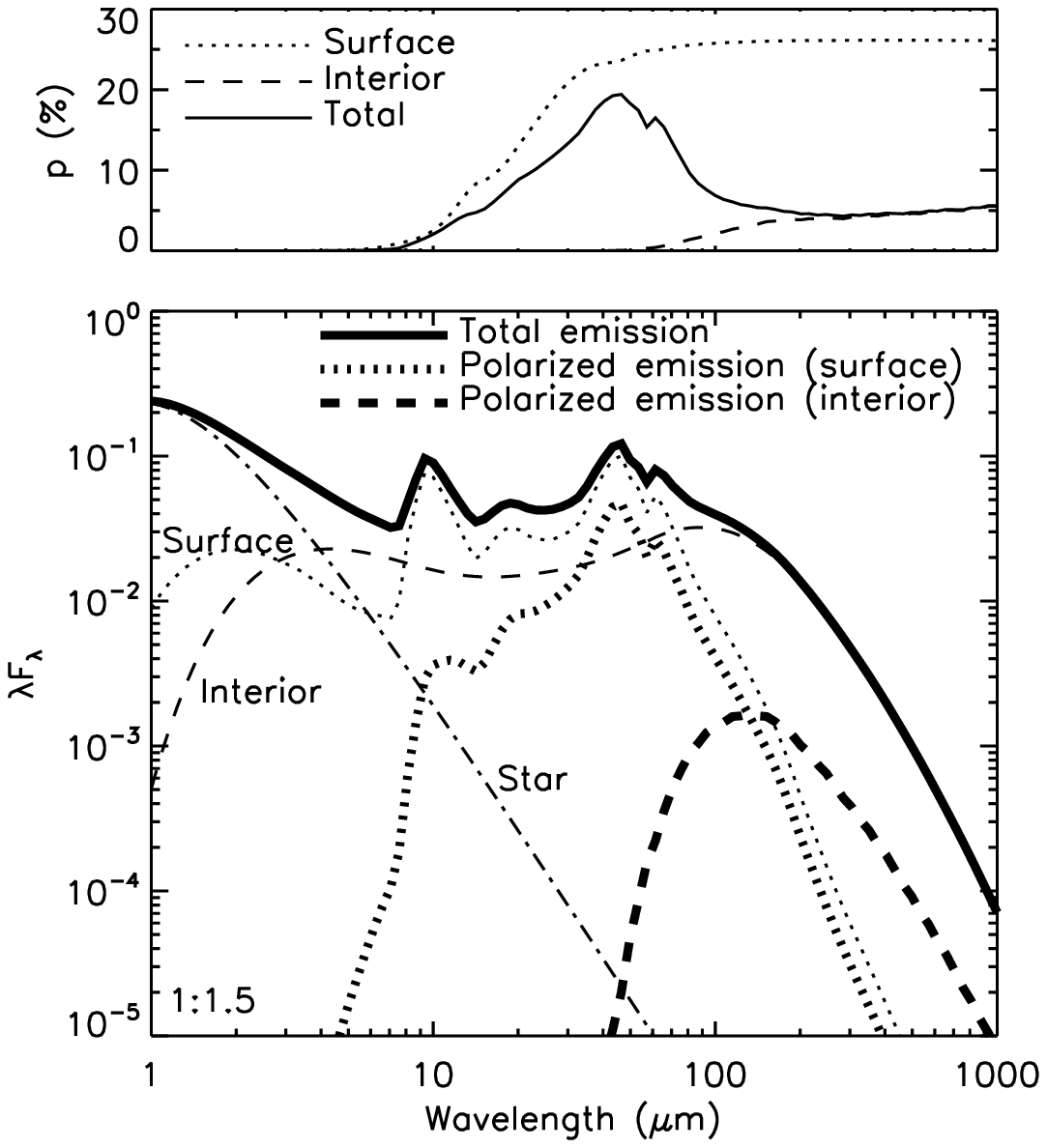}
\caption{
\small
{\it (a) Left panel}.-- A schematic view of the disk model.
The surface layer is hotter and heated by the star light.
The disk interior is heated by re-processed light from the
surface layers. We assume that the disk height, $H$, is 4 time the
disk scale height, $h$. {\it (b) Right panel}.-- Spectral energy distribution.
The vertical axis (i.e. $\lambda F_{\lambda}$)
is in arbitrary unit.
Results are for oblate spheroid grains with axis ratio of 1.5:1. From
Cho \& Lazarian 2006.
}
\label{fig:model}
\end{figure*}

This is the first attempt to simulate
polarization from a disk on the basis of grain alignment theory.
More attempts should follow. In fact, it has been known for decades that
various stars, both young and evolved, exhibit linear polarization
(see a list of polarization maps in Bastien \& Menard 1988). While
multiple scattering has been usually quoted as the cause of the polarization,
recent observations indicate the existence of aligned dust around
eta Carinae (Aitken et al. 1995) and  evolved stars
(Kahane et al. 1997).
This suggests that for other stars the dust should be also aligned (Chrisostomou et al 2000).
In fact, some of the arguments that
were used against aligned grains are, in fact, favor them. For
instance, Bastien \& Menard (1988) point out that if polarization
measurements of young stellar object were interpreted in terms of grain
alignment with longer grain axes perpendicular to magnetic field,
the magnetic field of accretion disks were in the disk plane. This is exactly what the
present day models of accretion disks envisage.

Interestingly enough,  alignment of dust in environments
different of diffuse ISM and molecular clouds was professed by
a number of pioneers of the grain alignment research. For instance, Greenberg (1970) claimed that
interplanetary dust should be mechanically aligned. Dolginov \& Mytrophanov
(1976) conjectured that comet dust and dust in circumstellar regions
was aligned.
However, both the problems in understanding of grain alignment
and the inadequacy of  polarimetric data did not allow those
views to become prevalent (although see  Wolstencroft 1985, Briggs
\& Aitken 1986 where alignment was supported). I feel that now we
have a much better case to include alignment while dealing
with polarization from dust in various environments. Quantitative modeling
should both test grain alignment theory and environments under study.

\subsection{Grain Alignment in Comets}

The
``anomalies'' of polarization from comets\footnote{When light is scattered
by the randomly oriented particles with sizes much
less than the wavelength, the scattered light is
 polarized perpendicular to the scattering plane, which is the
plane passing through the Sun, the comet and the observer.
Linear polarization from comets has been long known to exhibit
polarization that is not perpendicular to the scattering plane.}
(see  Martel 1960,
Beskrovnaja et al 1987, Ganesh et al 1998) as well as circular
polarization from comets (
Metz \& Haefner 1987, Dollfus \& Suchail 1987, Morozhenko et al 1987)
are indicative of grain alignment.

However, conclusive arguments in favor of grain alignment 
were produced for the Levi (1990 20) comet through direct
measurements of starlight polarization, as the starlight was passing
through comet coma (Rosenbush et al 1994). The data conclusively
proved the existence of aligned grains in comets.

Note, that the issue of circular polarization was controversial for a while.
When both left and right handed polarization is present in different
parts of coma the average over entire coma may get the 
polarization degree close to zero.
This probably
explains why earlier researchers were unsuccessful attempting
to measure circular polarization while using large
apertures. Recent measurements by Rosenbush et al. (1999), Manset et al.
(2000) of circular polarization from Hale-Bopp Comet support the notion
that circular polarization is a rule rather than an exception.

\begin{figure}[htb]

\includegraphics[width=.95\textwidth]{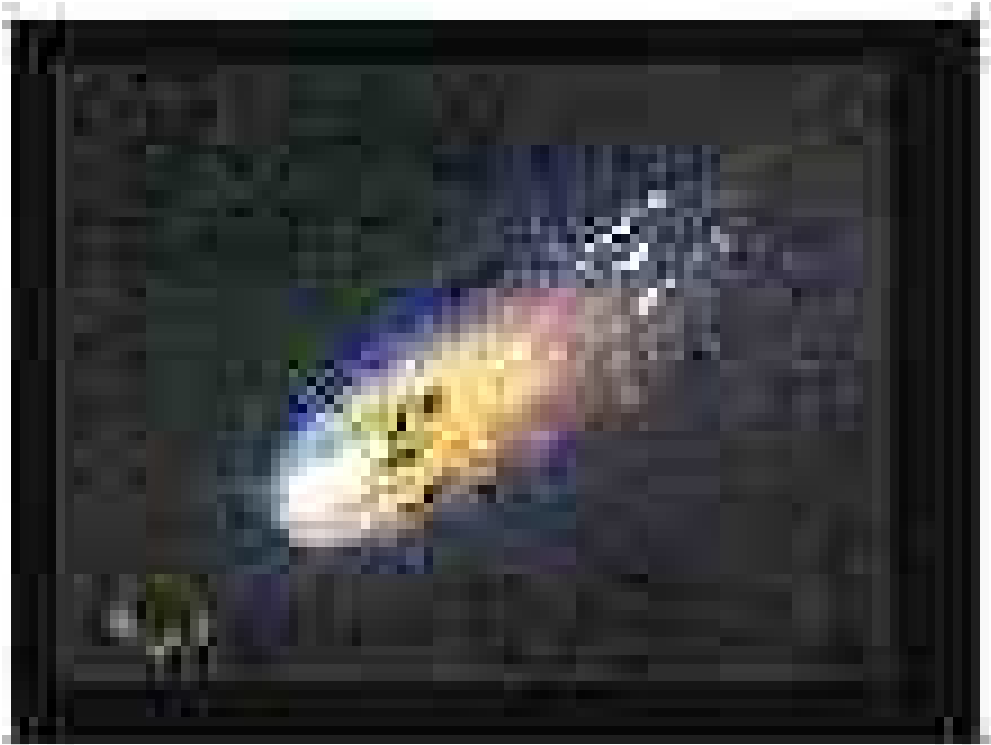}
\caption
{
\small
Zones of grain alignment in respect to magnetic field and
in respect to radiation/electric field for a comet at 1AU from the
Sun. Radiative torque alignment. From Hoang \& Lazarian (2007).
}
\label{fig1}
\end{figure}

A more recent paper by Rosenbush et al. (2006) reports circular polarization
from a comet C/1999 S4(LINEAR). The data indicates that the polarization arises
from aligned grains. The mechanism of alignment requires further studies, however.
 If magnetic fields do not penetrate into coma,
the alignment happens in respect to direction of radiation (see \S 5.2)
no circular polarization is possible (see Eq.~(\ref{circular})).  Outflow
velocities are not vividly supersonic to allow efficient Gold alignment.
What is the mechanism that produces the circular polarization?
 Several explanations are possible on the basis of
our earlier discussion. First of all, the alignment observed may be the
sub-sonic mechanical
alignment of irregular grains (see \S 4.5).
 Second, the alignment may be due to
radiative torques, but the outflow could alter the direction of the axis of
alignment. The structure of the ``precessing'' radiative torque component is
such that the precession rate goes to zero as the grain gets aligned in
respect to the radiation. Therefore it is easy to perturb the alignment axis
for radiative torques. Third, as we discussed in \S 5.3 electric field could
cause grain precession and even grain alignment. The choice between these possibilities
should be made on the basis of comparing the results of modeling with observations. We illustrate the model of alignment in a comet in Fig.~15.

\subsection{Alignment of Small Grains}

For particles much less than the wavelength the efficiency of radiative
torques drops as $(a/\lambda)^4$ (see L95).
 Within circumstellar regions, where UV flux is enhanced
smaller
grains can be aligned by radiative torques. This could present a possible solution
for the reported anomalies of polarization in the 2175 {\,{\rm \AA}}
 ~~extinction feature (see Anderson et al 1996) which have been interpreted
as evidence of graphite grain alignment (Wolff et al 1997). If
this alignment happens in the vicinity of particular
stars with enhanced UV flux
and having graphite grains in their circumstellar regions, this may
explain why no similar effect is observed along other lines of sight.

The maximum entropy inversion technique in Kim \& Martin (1995) indicates that
 grains larger than a particular critical size are aligned. This is
consistent with our earlier discussion of radiative torques and the Serkowski law (see \S 6.2).
 However, an interesting feature of
the inversion is that it is suggestive of smaller grains being partially
aligned. Initially, this effect was attributed to the problems with the
assumed dielectric constants employed in the inversion, but a further
analysis that we undertook with Peter Martin indicated that the alignment
of small grains is real. Indeed, paramagnetic (DG) alignment must act on
the small grains\footnote{To avoid a confusion we should specify that we
are talking about grains of $10^{-6}$~cm. For those grains the results
of DG relaxation coincide with those through resonance relation in
Lazarian \& Draine (2000). It is for grains of the size less than
$10^{-7}$~cm that the resonance relaxation is dominant.}.
An important feature of this weak alignment is that it is proportional to
the energy density of magnetic field. This opens a way for
a new type of magnetic field diagnostics.
As very small grains may emit polarized radiation as they rotate (see \S 4.2) both
UV and microwave polarimetry may be used to estimate the intensities of magnetic field.

\section{Concluding remarks}

\subsection{Present situation}

Historically the goal of the grain alignment theory was to account for puzzling
polarimetric observations. The situation has changed, however, as
grain alignment became a predictive theory. This calls for
more quantitative modeling and for more further polarimetry data acquisition, to test the models.

It was not  possible in the present review to
discuss all the interesting cases where grain alignment may
be important. Theoretical  considerations suggest that
grain alignment should take place within various astrophysical environments.
Polarized radiation from neighboring galaxies (Jones 2000),
galactic nuclei (see Tadhunter et al 2000),
AGNs, Seyfet galaxies (see Lumsden et al 2001)  can be partially due to
aligned particles. Revealing this contribution would allow us to study
magnetic fields in those and other interesting settings.

Polarization from aligned grains can benchmark other techniques for magnetic
field studies. For instance, anisotropies of the magnetohydrodynamic (MHD)
turbulence reveal the local direction of the magnetic fields; those can be measured
with observations of the Doppler-shifted spectral lines (Lazarian
et al. 2001, Esquivel \& Lazarian 2005 and references therein). Polarimetry of
aligned grains provides a way of testing the accuracy of
this new technique. Similarly, aligned grains can remove the 90 degrees uncertainty arising
 in the magnetic field studies based upon the Goldreich-Kylafis effect or
alignment of atoms/ion with fine or hyperfine structure\footnote{For the Goldreich-Kylafis (1982)
effect this uncertainty is intrinsic, while for the technique proposed in Yan \& Lazarian (2006, 2007)
the uncertainty can be removed by using several aligned species.}, as proposed by 
Yan \& Lazarian (2006, 2007).

Recently, promising attempts have been made to test the predictions 
of the grain alignment theory (see Hildebrand 2003, Andersson \& Potter 2005, 2006, 2007),
 or to use the grain 
alignment theory to explain observations (see Frisch 2006, Cortes \& Crutcher 2006). It is 
significant that the numerical
simulations that include theory-motivated prescriptions for grain alignment (see \S 6) allow easy
comparisons with observations. If combined with new polarimetric instruments, that have been built
recently or are to be built in the near future, this ensures progress in reliable tracing of
magnetic fields using aligned grains.

\subsection{Important questions}

In regards to practical studies of magnetic fields
a few questions will be in order.

$\bullet$
What is the advantage of the far-infrared polarimetry for studies of magnetic
fields in molecular clouds compared to the optical and
near-infrared observations? An immediate answer would be that the far infrared
polarimetry reveals aligned grains near newly born stars, unaccessible
to optical or near-infrared photons.
An additional advantage of the far infrared
spectropolarimetry  stems from the fact that it allows us
to separate contributions from different parts of the cloud (see Hildebrand
2000). This will
enable us to carry out tomography of the magnetic field structure.

$\bullet$
What is the future of the optical and near-infrared polarimetry?
It would be wrong to think that with the advent of the far-infrared
polarimetry there is a bleak future for extinction polarimetry at
shorter wavelengths. In fact, its potential for studies
of magnetic fields in the Galaxy is enormous (see Fosalba et al.
2002, Cho \& Lazarian 2002a). The possibility of using
stars at different distances from the observer allows to get an insight
into the 3D distribution of magnetic fields. In general, however,
it is extremely advantageous to combine polarimetric measurements in optical/near-infrared and
far-infrared wavelengths. For instance, it may be pretty challenging
to trace the connection of Giant Molecular Clouds (GMCs) with the
ambient interstellar medium using just far-infrared measurement.
However, if extinction polarimetry of the nearby stars is included,
the task gets feasible (see Poidevin \& Bastien 2006).
 Similarly, testing modern concepts of MHD
turbulence (see Goldreich \& Shridhar 1995)
and turbulent cloud support (see
McKee 1999) would require data from both  diffuse and dense media.

$\bullet$
Is it possible to study magnetic fields using radiation {\it scattered}
by aligned grains? The studies of molecular cloud column densities with
the near infrared scattered light were presented in Padoan et al (2006)
and Juvela et al. (2006). Those have shown that large scale mapping of scattered
intensity is possible up to $A_v\sim 10$ even for clouds illuminated by the average interstellar
radiation field. The polarization of scattered light should be affected by
grain alignment. This opens interesting prospects of detailed mapping of magnetic
fields at sub-arcsecond resolution, which for the closest star forming
regions corresponds to the scale of $\sim 100$~AU. This can bring to a new level
both the studies of magnetic fields in star forming regions and observational
studies of magnetic turbulence.   

$\bullet$
What is the advantage of doing polarimetry for different wavelengths?
The list of advantages is rather long. It is clear that
aligned grains can be successfully used as pick up
devices for various physical and chemical processes, provided
that we understand the causes of alignment. Differences in alignment
of grains of different chemical composition (see Smith et al. 2000)
provides a unique source of the valuable information. Comets present
another case in support of simultaneous multifrequency
studies. There the
properties of dust evolve in a poorly understood
fashion and this makes an interpretation
of optical polarimetry rather difficult. Degrees and directions
of dust alignment, that can be obtained that can be obtained via
far infrared polarimetry, can be used to get a self-consistent
picture of the dust evolution and grain alignment.

$\bullet$
Do we need the grain alignment theory to deal with polarized CMB foregrounds?
Polarized emission spectra arising from aligned dust may be very complex
if grains of different temperatures exhibit different degrees of
alignment. In this situation, the use of the naive power-law templates
may result in huge errors unless we understand grain alignment properly.
Needless to say, a well developed grain alignment theory is required to predict the
spectra of polarized emission from PAHs in the range of 10-100~GHz.

$\bullet$
What is the optical depth $A_v$ at which aligned grains fail to trace magnetic fields?
The answer depends on the grain size and the grain environments. If
we consider a starless cloud/core illuminated by the interstellar radiation field,
for grains of $10^{-5}$~cm the radiative alignment fails at $A_v$ $\sim 1.4$ (see Arce et al. 1998). 
However, larger grains in cloud cores
can be aligned at $A_v>20$
as was shown by Cho \& Lazarian (2005, 2006), which is a great news for
polarimetric studies of star formation. In the vicinity of stars and in the presence
of grain-gas drift smaller grains can also be aligned.

$\bullet$
What is the niche for the magnetic field studies with aligned grains? If we
try to answer this question briefly, we can point out that the aligned grains
trace magnetic fields in molecular clouds and cold diffuse gas, where so far
they have little competition from other techniques. Both observations and theory show that
grain alignment is a robust process that can operate in the presence of very weak
magnetic fields. I would like to stress the synergy of the starlight/dust emission
polarimetry and other techniques of magnetic field studies. Indeed, the different techniques
provide us with the data on magnetic fields in
different environments, e.g. different phases of the interstellar
medium. We can obtain an adequate picture of magnetized astrophysical settings by
combining the techniques, e.g. dust polarimetry, synchrotron polarimetry, polarimetry
of aligned atoms/ions and molecules.  

\subsection{Brief Summary}

The principal points discussed above are as follows:
\begin{itemize}
\item Grain alignment results in linear and circular polarization.
The degree of polarization depends on the degree of grain alignment,
the latter being the subject of the grain alignment theory.
\item Substantial advances in understanding grain dynamics, subtle
magneto-mechanical effects, as well as the back-reaction of thermal
fluctuations on grain rotation have paved the way for the advances
in understanding of grain alignment.
\item The grain helicity has been established as an essential property
of irregular grains rotating about their axis of the maximal inertia.
This allowed for a better physical understanding of the radiative torque's role,
and allowed to introduce new alignment mechanisms, e.g. the sub-sonic mechanical
alignment.
\item The grain alignment theory has, at last, reached its mature state
when predictions are possible. In most cases grain alignment takes place with
 respect to magnetic field, thereby revealing the magnetic field direction,
even if the alignment mechanism is not magnetic.
\item The radiative torque alignment, after having been ignored for many years, has
become the most promising mechanism which predictions agree well with
interstellar observations. To create alignment, this mechanism does not
rely on paramagnetic relaxation.
\item It is clear that the importance of grain alignment is not limited
to interstellar medium and molecular clouds. Polarimetry can be used
to study magnetic fields in accretion disks,
AGN, circumstellar regions, comets etc.
\item As astrophysical environments exhibit a wide variety
of conditions, various alignment mechanisms have their own niches. The
importance of studying the alternative mechanisms increases as attempts
are made to trace magnetic fields with aligned grains in the environments
other than the interstellar one.
\end{itemize}

{\bf Acknowledgments}.
I thank Bruce Draine, Michael Efroimsky,
 Roger Hildebrand, and Giles Novak for illuminating
discussions. Help by Hoang Thiem was extremely valuable.
I acknowledge the support by the NSF grant  AST-0507164,
as well as by the NSF Center for Magnetic Self-Organization in Laboratory and Astrophysical
Plasmas.


\begin{thebibliography}{12}
\bibitem{} Abbas, M. M., Craven, P. D., Spann, J. F., Tankosic, D.,
 LeClair, A., Gallagher, D. L., West, E. A., Weingartner, J. C.,
 Witherow, W. K., Tielens, A. G. G. M. 2004, ApJ, 614, 781-795
\bibitem{} Aitken, D. K., Briggs, G. P., Roche, P. F., Bailey, J. A.,
Hough, J. H. 1986, MNRAS, 218, 363-384.
\bibitem{} Aitken, D.K., Efstathiou, A., McCall, A.,
Hough, J.H, 2002, MNRAS, 329, 647-669
\bibitem{} Aitken, D.K., Smith, C.H., Moore, T.J.T, Roche, P.F. 1995, MNRAS 273, 359
\bibitem{} Anderson, C.M. et al 1996, AJ, 112, 2726-2743
\bibitem{} Andersson, B.-G, \& Potter, S.B. 2005, BAAS, 37, 1359
\bibitem{} Andersson, B.-G, \& Potter, S.B. 2006, ApJL, L640-L643
\bibitem{} Andersson, B.-G, \& Potter, S.B. 2007, ApJ, in press
\bibitem{} Arce, H., Goodman, A., Bastien, P., Manset, N. \& Summer, M. 1998,
ApJL, 499, L93-L96 
\bibitem{} Balbus, S. A., \& Hawley, J. F. 1991, ApJ, 376, 214
\bibitem{} Bastien, P. 1988, in ``Polarized Radiation of Circumstellar
Origin'', eds. G.V Coyne, A.M. Magalhaes, A.F.J. Moffat, R.E. Schulte-Ladbeck,
S. Tapia, D.T. Wickramasinghe, Vatican Observatory, p. 541-577
\bibitem{} Bastien, P. 1996,  in Polarimetry of the Interstellar Medium,
eds Roberge W.G. and Whittet, D.C.B. p.~297-314
\bibitem{} Bastien, P, Jenness, T., Molnar, J., 2005, ASPC, 343, 69-70
\bibitem{} Bastien, P. \& Landstreet, J.D. 1979, ApJ, 299, L137-L140
\bibitem{} Bastien, P. \& Menard, F. 1988, ApJ, 326, 334-338
\bibitem{} Beskrovnaja, N.G., Silantev N.A., Kiselev, N.N., \&
Chrenova, G.P., 1987, in Diversity and Similarity of Comets, Burssels,
ESA SP-278, p. 681-683
\bibitem{} Bethell, T., Chepurnov, A., Lazarian, A., \& Kim, J. 2006,
ApJ, in press
\bibitem{} Bradley, J.P., 1994, Science, 265(5174), 925-929
\bibitem{} Briggs, G.P., Aitken, D.K. 1986, Proc. Astron.
Soc. Australia, 6, 145-149
\bibitem{} Chandrasekhar, C. 1961, Hydrodynamic and Hydromagnetic Stability, Oxford: Oxford University
Press
\bibitem{} Cho, J., \& Lazarian, A. 2002a, ApJ, 575, L63-L66
\bibitem{} \rule{1.2cm}{0.2mm} 2002b, Phys.Rev.Lett., 84(24), 245001-1-4
\bibitem{} \rule{1.2cm}{0.2mm} 2003, NewAR, 47, 1143-1149
\bibitem{} \rule{1.2cm}{0.2mm} 2005, ApJ, 631, 361 (CL05)
\bibitem{} \rule{1.2cm}{0.2mm} 2006, submitted
\bibitem{} Chrysostomou, A., Gledhill, T.M., Menard, F.,
Hough, J.H., Tamura,M., Bailey, J. 2000 MNRAS, 312, 103-115
\bibitem{}  Chrysostomou A.,
Hough, J.H., Whittet, D.C.B., Aitken, D.K., Roche, P.F.,
Lazarian, A. 1996, ApJ, 465, L61-L64
\bibitem{} Cortes, P., \& Crutcher, R. 2006, ApJ, 639, 965-968
\bibitem{} Cortes, P., Crutcher, R., Matthews, B. 2006, ApJ, 650, 246-251
\bibitem{} Crutcher, R., Nutter, D. Ward-Thompson, D., \& Kirk, J. 2004, ApJ, 600, 279-285
\bibitem{}  Davis, L. 1955, Vistas in Astronomy, ed. A.Beer, 1, 336-
\bibitem{} Davis, L. \& Greenstein, J.L., 1951, ApJ,  114, 206-240
\bibitem{} Dolginov A.Z. 1972, Ap\&SS, 16, 337-349
\bibitem{} Dolginov A.Z. \& Mytrophanov, I.G. 1978, A\&A, 69, 421-430
\bibitem{} Dolginov A.Z. \& Mytrophanov, I.G. 1976, Ap\&SS, 43, 291-317
\bibitem{} Dollfus, A. \& Suchail, J.-L. 1987, A\&A, 187, 669-688
\bibitem{} Draine, B.T. 1985, in Protostars \& Planets II, Tucson,: University of Arizona Press, p. 621-640
\bibitem{} \rule{1.2cm}{0.2mm}, in Polarimetry of the Interstellar Medium,
eds Roberge W.G. and Whittet, D.C.B., A.S.P. 97. 16-25
\bibitem{}\rule{1.2cm}{0.2mm}, 2003, ARA\&A, 41, 241-289
\bibitem{} Draine, B.T., \& Flatau, P.J. 1994, J.Opt.Soc.Am.A., 11,
1491
\bibitem{} Draine, B.T. \& Lazarian A. 1998a, ApJ, 494, L19-L22
\bibitem{}\rule{1.2cm}{0.2mm} 1998b, ApJ, 508, 157-179
\bibitem{}\rule{1.2cm}{0.2mm} 1999, ApJ, 512, 740-754
\bibitem{} Draine, B.T. \& Salpeter, E.E. 1979, ApJ, 231, 77-94
\bibitem{} Draine, B.T. \& Weingartner, J.C. 1996, ApJ, 470, 551-565 (DW96)
\bibitem{}\rule{1.2cm}{0.2mm}  1997, ApJ, 480, 633-646 (DW97)
\bibitem{} Efroimsky, M. 2000, JMP, 41, issue 4, 1854-1888
\bibitem{} Efroimsky, M. 2002a, ApJ, 575, 886-899
\bibitem{} \rule{1.2cm}{0.2mm} 2002b, P\&SS, 49, 937-955
\bibitem{} Efroimsky, M. \& Lazarian, A. 2000, MNRAS, 311, 269-278
\bibitem{} Elmegreen, B., \& Scalo, J. 2004, ARA\&A, 42, 211-273
\bibitem{} Epstein, R. I. 1980, MNRAS, 193, 723
\bibitem{} Esquivel, A., \& Lazarian, A. 2005, ApJ, 631, 320-350
\bibitem{} Fosalba, P., Lazarian, A., Prunet, S., Tauber, J.A. 2002, ApJ, 564, 762-772
\bibitem{} Frisch, P. 2006, ApJ, submitted, astro-ph/0603745
\bibitem{} Ganesh, S., Joshi, U.C., Baliyan, K.S., Deshpande,
M.R. 1998, A \& A Suppl. 129, 489-493
\bibitem{} Girart, J., Crutcher, R., \& Rao, R. 1999, ApJ, L109-L113.
\bibitem{}  Gold, T. 1951,  Nature, 169, 322-323
\bibitem{} Goldreich, P., \& Kylafis, N. 1982, ApJ, 253, 606
\bibitem{} Goldreich, P., \& Shridhar, H. 1995, ApJ, 438, 763-775
\bibitem{} Goodman, A.A., Jones, T.J., Lada, E.A.; Myers, P.C. 1995,
ApJ, 448, 748-765
\bibitem{}  Goodman, A.A., \& Whittet, D.C.B. 1995, ApJ, 455, L181-L184
\bibitem{} Greenberg, J.M. 1968, in Nebulae and Interstellar Matter,
eds, B.M. Middlehurst \& L.H. Aller (University of Chicago Press), 221-230
\bibitem{} Greenberg, J.M., \& Habing, H.J., 1970, in Interstellar Gas Dynamics: Proceedings of the 6th
Symposium on Cosmical Gas Dynamics, IAU Symposium, vol. 39, 306-312
\bibitem{} Jones, T.J. 2000a, AJ, 120, 2920-2927
\bibitem{} Jones, T.J. \& Gehrz, R.D. 2000, Icarus, 143, 338-346
\bibitem{} Jones, R.V. \& Spitzer, L. 1967, ApJ, 147, 943-964
\bibitem{} Juvela, M., Pelkonen, V., Padoan, P. \& Mattila, K. 2006, A\&A, 457, 877-889
\bibitem{} Hall, J.S. 1949, Science, 109, 166-168
\bibitem{} Harwit, M. 1970, Nature, 226, 61-63
\bibitem{} Hilczer, B., \& Malecki, J., 1986, Electrets sutdies in electrical and electronic enginering,
Amsterdam: Elsevier Science Ltd.
\bibitem{}  Hildebrand, R.H. 1988, {\it QJRAS},  29, 327-351
\bibitem{} \rule{1.2cm}{0.2mm}  2002, in Astrophysical spectropolarimetry,
edited by by J. Trujillo-Bueno, F. Moreno-Insertis, and F. Sanchez, CUP,
p. 265-302
\bibitem{} \rule{1.2cm}{0.2mm}  2003, NewAR, 47, 1009-1015
\bibitem{}  Hildebrand et al. 2000, PASP, 112, 1215-1235 (H00)
\bibitem{} Hildebrand, R.H., \& Dragovan, M. \& Novak, G. 1984, ApJ,
284, L51-L54
\bibitem{} Hildebrand, R.H., Gonatas, D.P., Platt, S.R., Wu, X.D.,
Davidson,
J.A., \& Werner, M.W. 1990, ApJ, 362, 114-119
\bibitem{} Hildebrand, R. H., Dotson, J. L., Dowell, C. D., Schleuning, D. A.,
 Vaillancourt, J. E. 1999, ApJ, 516, 834-842
\bibitem{} Hildebrand, R. H., \& Dragovan, M. 1995, ApJ, 450, 663-666
\bibitem{} Hiltner, W.A. 1949, ApJ, 109, 471-480
\bibitem{} Hoang, T. \& Lazarian, A. 2007, MNRAS, accepted
\bibitem{} Jones, R.V., \& Spitzer, L.,Jr, 1967, ApJ,  147, 943-964
\bibitem{} Kahane, C., Viard, E., Menard, F., Bastien, P., Manset, N., 1997, Ap\&SS, 231, 223-226 
\bibitem{}  Kim, S.-H., \& Martin, P., G. 1995, ApJ, 444, 293-305
\bibitem{} Lazarian, A.  1994, MNRAS,  268, 713-723, (L94)
\bibitem{} \rule{1.2cm}{0.2mm} 1994b, Ap\&SS, 216, 235-237
\bibitem{} \rule{1.2cm}{0.2mm}   1995a, ApJ, , 453, 229-237
\bibitem{} \rule{1.2cm}{0.2mm}   1995b, MNRAS, 277, 1235-1242 (L95)
\bibitem{} \rule{1.2cm}{0.2mm}   1995c,  MNRAS, 274, 679-688
\bibitem{} \rule{1.2cm}{0.2mm} 1997a, ApJ, 483, 296-308
\bibitem{} \rule{1.2cm}{0.2mm}  1997b, MNRAS, 288, 609-617
\bibitem{} \rule{1.2cm}{0.2mm}  2003, Journ. of Quant. Spectroscopy \&
Radiative Transfer, 79-80, 881-902
\bibitem{} Lazarian, A. \& Cho, J. 2005, ASPS, 434, 333-345
\bibitem{} Lazarian, A., Goodman, A.A. \& Myers, P.C. 1997, ApJ, 490,
273-280 (LGM97)
\bibitem{} Lazarian, A., \& Efroimsky, M. 1996, ApJ, 466, 274-281
\bibitem{} \rule{1.2cm}{0.2mm} 1999, MNRAS, 303, 673-684
\bibitem{} Lazarian, A., \& Efroimsky, M., Ozik, J, 1996, ApJ, 472, 240-244
\bibitem{} Lazarian, A., \& Draine, B.T., 1997, ApJ, 487, 248-258
\bibitem{} \rule{1.2cm}{0.2mm} 1999a, ApJ, 516, L37-L40
\bibitem{} \rule{1.2cm}{0.2mm} 1999b, ApJ, 520, L67-L70
\bibitem{} \rule{1.2cm}{0.2mm} 2000, ApJ, 536, L15-L18
\bibitem{} Lazarian, A. \& Finkbeiner, D. 2003, NewAR, 47,
issue 11-12, 1107-1116
\bibitem{} Lazarian, A. \& Hoang, T. 2007, MNRAS, 378, 910 (LH07)
\bibitem{} Lazarian, A., Pogosyan, D., Esquivel, A. 2002, in Seeing through the Dust: the Detection of
HI and the exploration of the ISM in Galaxies, eds, Taylor, A.R., Landecker, T.L, Willis, A.G., ASP
conference Series, Vol. 276, San Francisco
\bibitem{} Lazarian, A., Pogosyan, D., Esquivel, A. 2002, in ``Seeing Through Dust: The Detection of
HI and Exploration of the ISM in Galaxies, eds. A.R. Taylor, T.L. Landecker, and A.G. Willis,
APS, vol. 276, 182-192
\bibitem{} Lazarian, A., \& Roberge, W.G., 1997a ApJ, 484, 230-237, (LR97)
\bibitem{}\rule{1.2cm}{0.2mm} 1997b, MNRAS, 287, 941-946
\bibitem{} Lazarian, A., \& Yan, H. 2002, ApJ, 566, L105-L108
\bibitem{} Lazarian, A., \& Yan, H. 2004, in ``Astrophysics of Dust'', eds. A.N. Witt, G.C. Clayton,
B.T. Draine, ASP, Vol. 309, 479-499
\bibitem{} Lazarian, A., \& Yan, H. 2005, in ``Magnetic Fields in the Universe'', eds. 
E. M. de Gouveia Dal Pino, G. Lugones, A. Lazarian p. 495-506 
\bibitem{} Lee, H.M., \& Draine, B.T. 1985, ApJ, 290, 221-228
\bibitem{} Lumsden, S. L., Heisler, C. A., Bailey, J. A.,
Hough, J. H., Young, S. 2001, MNRAS, 327, 459-474
\bibitem{} Manset, N., Bastien, P. 2000, Icarus, 145, 203-219
\bibitem{} MNRWAW, J.V., \& ALPAY, S.P., 2005, Graded Ferroelectrics, transpacitors and trasponents, Berlin: Springer
\bibitem{} Martin, P.G. 1971, MNRAS, 153, 279-286
\bibitem{}\rule{1.2cm}{0.2mm} 1995, ApJ, 445, L63-L66
\bibitem{} Menard, F., Bastien, P., \& Robert, C. 1988, ApJ, 335, 290-294
\bibitem{} Metz, K., Haefner, R. 1987, A\&A, 187, 539-542.
\bibitem{} Morish, A.H. 1980, The Physical Principles of Magnetism, New York: W.A. Benjamin
\bibitem{} Morozhenko, A. V., Kiselev, N. N., Gural'Chuk, A. L. 1987,
Kinematika i Fizika Nebesnykh Tel (ISSN 0233-7665), 3, 89-95
\bibitem{} Martel, M.-T. 1960, Annales d'Astrophysique, 23, 498-502
\bibitem{} Mathis, J.S.  1986,  ApJ,  308, 281-287
\bibitem{} Mathis, J., Mezger, P., \& Panagia, N. 1983, A\&A, 128, 212-229
\bibitem{} McKee, C.F. 1999, in The Origin of Stars and Planetary Systems,
Eds. Charles J. Lada and Nikolaos D. Kylafis, Kluwer, p. 29-45
\bibitem{} Netzer, N., \& Elitzur, M. 1993, ApJ, 410, 701-713
\bibitem{} Novak, G. et al. 2004, in {\it Millimeter and
           Submillimeter Detectors for Astronomy II}, eds.
           J. Antebi \& D. Lemke, Proceedings of the SPIE, Vol. 5498, p. 278
\bibitem{} Padoan, P., Goodman, A., Draine, B. T., Juvela, M.,
Nordlund, A., \& Rognvaldson, O.E. 2001, 559, 1005-1018
\bibitem{} Padoan, P., Juvela, M., \& Pelkonen, V. 2006, ApJL, 636, L101-L103
\bibitem{} Pelkonen, V., Juvela, M., Padoan, P. 2007, A\&A, 461, 551-564
\bibitem{} Poidevin, F. \& Bastien, P. 2006, ApJ, 650, 945-955
\bibitem{} Purcell, E.M. 1969, On the Alignment of Interstellar Dust,
 Physica,  41, 100-
\bibitem{}\rule{1.2cm}{0.2mm}  1975, in Dusty Universe, eds. G.B. Field
\& A.G.W. Cameron, New York, Neal Watson, p. 155-165
\bibitem{}\rule{1.2cm}{0.2mm} 1979, ApJ, 231, 404-416
\bibitem{}  Purcell, E.M., \& Spitzer, L., Jr  1971, ApJ,  167, 31-62
\bibitem{} Rao, R, Crutcher, R.M., Plambeck, R.L., Wright, M.C.H. 1998,
ApJ, 502, L75-L79
\bibitem{} Roberge, W.G. 1996 in Polarimetry of the Interstellar Medium,
eds, Roberge W.G. and Whittet, D.C.B., A.S.P. Vol. 97, p. 401-416
\bibitem{} \rule{1.2cm}{0.2mm} 1997, MNRAS, 291, 345-352
\bibitem{} \rule{1.2cm}{0.2mm} 2004, in ``Astrophysics of Dust'', eds. A.N. Witt, G.C. Clayton,
B.T. Draine, ASP, Vol. 309, 467-479
\bibitem{} Roberge, W.G., \& Hanany, S. 1990, B.A.A.S., 22, 862
\bibitem{} Roberge, W.G., Hanany, S., \& Mesinger, D.E. 1995, ApJ, 453, 238
\bibitem{} Roberge, W.G., DeGraff, T.A., \& Flaherty, J.E., 1993, ApJ,
 418, 287-306
\bibitem{} Roberge, W.G., \& Lazarian, A. 1999, MNRAS, 305, 615-630
\bibitem{} Roberge, W.G. \& Desch S.J. 1990, BAAS, 22, 1256
\bibitem{} Robinson, J.W., 1991, Practical handbook of Spectroscopy, Boca Raton: CRC Press
\bibitem{} Rosenbush, V.K., Kolokolova, L., Lazarian, A., Shakhovskoy, N.,
Kiselev, N. 2006, Icarus, 186, 317-330
\bibitem{} Rosenbush, V.K., Rosenbush, A.E., Dement'ev, M. S. 1994, Icarus,
108, 81-91
\bibitem{} Rosenbush, V. K., Shakhovskoj, N. M., \& Rosenbush, A. E
1999, Earth, Moon, and Planets, 78, 381-386
\bibitem{} Rouan, D., Leger, A., Omont, A., Giard, M.
1992, A \& A,  253, 498-514.
\bibitem{} Safier, P.N., McKee, C.F.; Stahler, S.W. 1997, ApJ, 485, 660-679
\bibitem{} Salpeter, E.E., \&  Wickramasinche, N.C. 1969, Nature, 222, 442-443
\bibitem{} Schmidt, Th. 1971, A\&A, 12, 456-463
\bibitem{} Sellgren, K., Rouan, D., Leger, A. 1988,
A\&A, 196, 252-254.
\bibitem{} Serezhkin, Y. 2000, in Instruments, Methods, and Missions
for Astrobiology III, ed. R. Hoover, Proc. SPIE Vol 4137, p. 1-12
\bibitem{} Serkowski, K. 1973, in IAU Symp. 52, Interstellar Dust
and Related Topics, ed. J.M. Greenberg \& H.C. van de Hulst
(Dordrecht: Kluwer), 145-160
\bibitem{} Serkowski, K., Mathewson, D.S. \& Ford, V.L. 1975, ApJ,
196, 261-290
\bibitem{} Shukla, P. \& Stenflo, L. 2005, ApJL, 629, L93-L95
\bibitem{} Smith, G.H., Wright, C.M., Aitken, D.K., Roche, P.F.
\& Hough J.H. 2000, MNRAS, 312, 327-361
\bibitem{} Sorrell, W.H., 1995a, MNRAS, 273, 169-186
\bibitem{}  \rule{1.2cm}{0.2mm} 1995b, MNRAS, 273, 187-200
\bibitem{} Spitzer, L., Jr, Tukey, J.W. 1950, ApJ, 187-192
\bibitem{} Spitzer, L.,Jr \& McGlynn T.A. 1979, ApJ, 231, 417-424, (SM79)
\bibitem{} Tadhunter, C. N., Sparks, W.,
Axon, D. J., Bergeron, L., Jackson, N. J.,
Packham, C., Hough, J. H., Robinson, A.,
Young, S. 2000, MNRAS, 313, L52-L56
\bibitem{} Tamura, M., Hough, J.H., Greaves, J.S.,
Morino, J-I,
Chrysostomou, A., Holland, W.S., \& Momose, M. 1999,
ApJ, 525, 832-836
\bibitem{} Vaillancourt, J. 2006, astro-ph/060933
\bibitem{} van Vleck, J.H., 1937, J. Chem. Phys. 5, 320-325
\bibitem{} Velikov, S.J. 1959, J. Exper. Theoret. Phys. (USSR), 36, 1398-1405
\bibitem{} Ward-Thompson, D., Kirk, J.M., Crutcher, R.M., Greaves, J.S.,
Holland, W.S., \& Andre, P. 2000, ApJ, 537, L135-L138
\bibitem{} Ward-Thompson, D., Andre, P., \& Kirk, J. 2002, MNRAS, 329, 257-276
\bibitem{}Weidenschilling, S.J. \& Ruzmaikina, T.V. 1994, ApJ, 430, 713
\bibitem{} Weingartner, J. 2006, ApJ, 647, 390-396
\bibitem{} Weingartner, J. \& Draine, B. 2001, ApJ, 548, 296
\bibitem{} Weingartner, J., \& Draine, B. 2003, ApJ, 589, 289 (DW03)
\bibitem{} Whittet, D.C.B. 1992, Dust in Galactic Environment,
Bristol:IPP
\bibitem{} Whittet, D.C.B., \& van Brenda, I.G. 1978, A\&A, 66, 57-63
\bibitem{} Whittet, D.C.B., Gerakines, P.A., Hough, J.H. \& Snenoy
2001, ApJ, 547, 872-884
\bibitem{} Wiebe, D.S, \& Watson, W.D. 2001, ApJ, 549, L115-118
\bibitem{} Wolff, M. J., Clayton, G.C., Kim, S-H; Martin, P.G.,
Anderson, C.M. 1997, ApJ, 478, 395-402
\bibitem{} Wolstencroft, R.D. 1985, in Cosmical Gas Dynamics, ed. F.D. Kahn
(Utrecht: VNU Science Press), p. 251-261
\bibitem{} Yan, H. \& Lazarian, A. 2003, ApJ, 592, L33-L36
\bibitem{} Yan, H. \& Lazarian, A. 2006a, ApJ, in press, astro-ph/0611281
\bibitem{} Yan, H. \& Lazarian, A. 2007, ApJ, 657, 618-640
\bibitem{} Yan, H., Lazarian, A. \& Draine, B. 2004, ApJ, 616, 895-911
\end{thebibliography}
\end{document}